\documentclass[12pt]{article}
\usepackage{amsfonts}
\usepackage{amsmath}
\usepackage{epsfig,epstopdf}
\usepackage{cite}
\begin{document}

\begin{center}
{\LARGE \bf Symbolic expressions for fully differential single top quark production cross section and decay width of polarized top quark in the presence of anomalous \textit{Wtb} couplings.} \\	

\vspace{1.5cm}
	
{\large \bf Edward Boos\footnote{boos@theory.sinp.msu.ru} and Viacheslav Bunichev\footnote{bunichev@theory.sinp.msu.ru}}\\
\vspace{1.5cm}

Skobeltsyn Institute of Nuclear Physics,
Moscow State University\\ 
119991 Moscow, Russia \\	
\end{center}
\vspace{2cm}

\begin{abstract}
Spin correlations in the t-channel single top quark production and its subsequent decay are investigated
for the case of contributions involving anomalous \textit{Wtb} couplings. 
We obtain analytical expressions for the differential widths for the three-particle decay of a polarized t quark in its rest frame and also expressions for the differential cross sections of the full process of production and decay of the t quark ($2\to 4$) as a function of the energy of a charged lepton and two angles of orientation of the quantization axis of the t-quark spin. The expression is presented in the most general form for the case of real and imaginary vector and tensor anomalous \textit{Wtb} couplings. We show that shapes of certain multidimensional kinematic distributions of final state particles are significantly different for the contributions proportional to different combinations of the anomalous couplings. 
The most noticeable differences appear in the shape of the surfaces of two-dimensional distributions, where one of the variables is the energy of a charged lepton and the other is one of the t-quark spin orientation angles.
Observed properties are confirmed by two methods of computations either from the obtained symbolic expression for the differential cross sections of the full process of the polarized single top quark production with its subsequent decay or by means of the CompHEP program for the complete process involving the t-channel single top. 
In addition, using the obtained analytical expressions, we estimate the statistical accuracy of extracting values of the anomalous \textit{Wtb} couplings for different levels of the expected integral luminosity at the LHC.  
\end{abstract}

\newpage

\tableofcontents 

\newpage

\section{Introduction}
\label{intro}
With the Higgs boson discovery at the LHC the Standard Model (SM) is completed in a sense that all predicted particles are found and all the interactions between particles are fixed. However, most likely, the SM is a kind of effective field theory working (and working amazingly well) at the energy range determined by the electroweak (EW) energy scale. The top quark being the heaviest found fundamental particle with closest to the EW scale mass is a promising place to search for possible deviations from the SM(see, recent reviews on the top quark \cite{Deliot:2014uua,Boos:2015bta,Gerber:2015upa,Bernreuther:2015wqa,Cristinziani:2016vif,Husemann:2017eka}). In particular such deviations may be related to the presence of the top quark anomalous couplings which are usually parametrized in terms of a number of the gauge-invariant dimension-six operators 
given in \cite{buchmuller} and in the Warsaw basis \cite{Grzadkowski:2010es} following the notations from \cite{AguilarSaavedra:2018nen}
\begin{align}
& O_{\phi q}^{(3,33)} = \frac{i}{2} \, \left[ \phi^\dagger \tau^I (D_\mu \phi)
  - (D_\mu \phi^\dagger) \tau^I  \phi \right] (\bar q_{L3} \gamma^{\mu} \tau^I q_{L3}) \,,
&& {O_{\phi u d}^{(33)}} = i (\tilde \phi^\dagger D_\mu \phi)
        (\bar t_{R} \gamma^{\mu} b_{R}) \,,  \\
& O_{dW}^{(33)} = (\bar q_{L3} \sigma^{\mu \nu} \tau^I b_{R}) \phi \, W_{\mu \nu}^I \,,
&& O_{uW}^{(33)} = (\bar q_{L3} \sigma^{\mu \nu} \tau^I t_{R}) \tilde \phi \, W_{\mu \nu}^I \,. \notag
\end{align}
These operators lead to the effective Lagrangian ~\cite{Kane:1991bg} allowed by the Lorentz invariance parametrizing the anomalous terms in the \textit{Wtb} vertex 

\begin{equation}
\label{anom_wtb_eq_lagrangian}\begin{split}
 {\cal L} =& -\frac{g}{\sqrt{2}} \overline{b}{\gamma}^{\mu} \big( f_{LV} P_L + f_{RV} P_R \big) t W^{-}_{\mu} - \frac{g}{\sqrt{2}}  \overline{b} ~\frac{{i\sigma}^{\mu\nu}  }{2 M_W} \big( f_{LT} P_L + f_{RT} P_R \big)tW^{-}_{\mu\nu} 
 + H.c. \end{split},
\end{equation}

where $M_W$ is the W-boson mass, $P_{L,R} = (1 \mp \gamma_5)/2 $ is 
the left-(right-) handed  projection operator, 
$W^{-}_{\mu\nu} = \partial_{\mu}W^{-}_{\nu} - \partial_{\nu}W^{-}_{\mu}$,
$g$ is the weak isospin gauge coupling, and parameters 
 $f_{LV(T)}$ and  $f_{RV(T)}$ are the dimensionless coefficients 
that parametrize the strengths of the left-vector (tensor) and 
the right-vector (tensor) structures in the Lagrangian. 

The couplings in the Lagrangian~(\ref{anom_wtb_eq_lagrangian}) are related in the following way to constants
$C_{\phi q}^{(3,33)}$, $C_{\phi u d}^{(33)}$,  $C_{dW}^{(33)}$, $C_{uW}^{(33)}$
in front of the gauge-invariant dimension-six operators
~\cite{Whisnant:1997qu,Boos:1999ca,AguilarSaavedra:2008gt,Birman:2016jhg} :

\begin{equation}
\label{anom_constants}
f_{LV} = V_{tb} + C_{\phi q}^{(3,33)} \frac{v^2}{\Lambda^2},~~ f_{RV} = \frac{1}{2} C_{\phi u d}^{(33)} \frac{v^2}{\Lambda^2},~~ f_{LT} = \sqrt 2 C_{dW}^{(33)} \frac{v^2}{\Lambda^2},~~ f_{RT} = \sqrt 2 C_{uW}^{(33)} \frac{v^2}{\Lambda^2}. 
\end{equation}

If one assumes that naturally the constants in front of the operators are of the order of unity, the natural values for the anomalous couplings are of the order of $\frac{v^2}{\Lambda^2}$ and therefore rather small.
This is confirmed by the recent most stringent experimental limits \cite{Khachatryan:2016sib}.

In the SM all fermions interact through the left-handed currents  and all constants (\ref{anom_constants}) are equal 
to zero, except $f_{LV}=V_{tb}$ (Cabibbo-Kobayashi-Maskawa-matrix element). Note that the anomalous coupling 
parameters could be complex in the most general case.

In this paper we discuss a simple idea allowing us to improve further the limits on anomalous couplings. The idea is based on the top quark spin correlation properties in the t- and s-channel single top production with its subsequent decay. Spin correlations in the presence of anomalous couplings have been studied previously in \cite{Jezabek:1994zv, Jezabek:1994qs, Mahlon:1996pn, Mahlon:1999gz, Grzadkowski:1999iq, Grzadkowski:2001tq, Boos:2002xw, Zhang:2010dr, Prasath:2014mfa, Aguilar-Saavedra:2014eqa, Aguilar-Saavedra:2014ywa, Aguilar-Saavedra:2017nik}. The effect of decoupling of anomalous top decay vertices in angular distributions has been studied in \cite{Grzadkowski:2002gt, Hioki:2015moa, Godbole:2018wfy}. $T$-odd correlations in polarized top quark decays have been studied in \cite{Fischer:2018lme}. 
Decays of polarized top quarks in the frame of the SM at next-to-Leading and next-to-next-to-leading order in QCD have been studied in papers \cite{Gao:2012ja, Bernreuther:2014dla, Berger:2017zof, Czarnecki:2018}. 
The next-to-leading order calculation in QCD for single-top-quark production processes in frame of Standard model effective field theory (SMEFT) is presented in \cite{Neumann:2019kvk}.

\begin{figure}
\begin{center}
\begin{minipage}[t]{.45\linewidth}
\centering
\includegraphics[width=6cm,height=5cm]{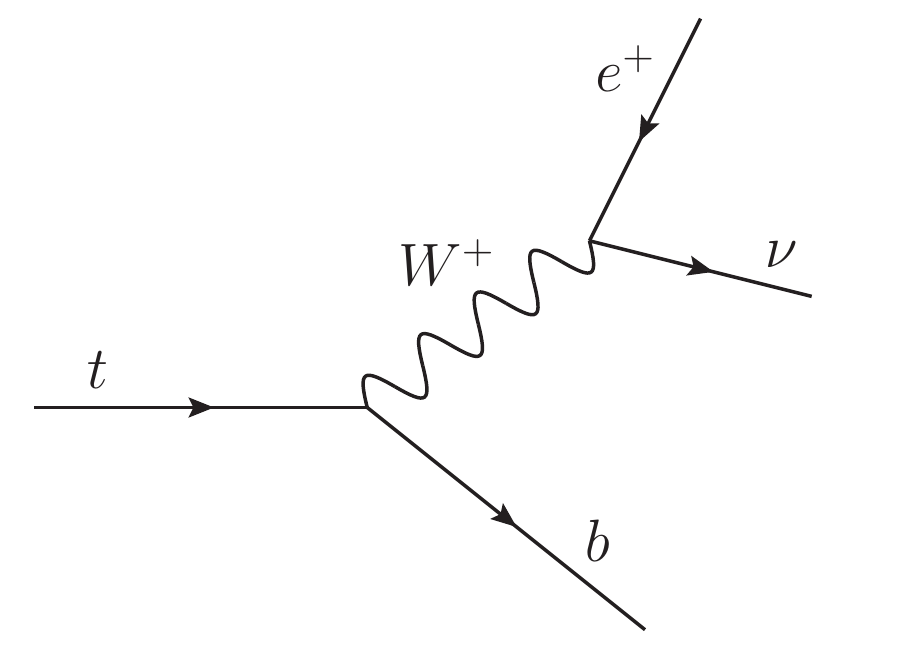}
\caption{\label{pic1} \footnotesize Top quark leptonic decay process. }
\end{minipage}
\hfill
\begin{minipage}[t]{.45\linewidth}
\centering
\includegraphics[width=6cm,height=5cm]{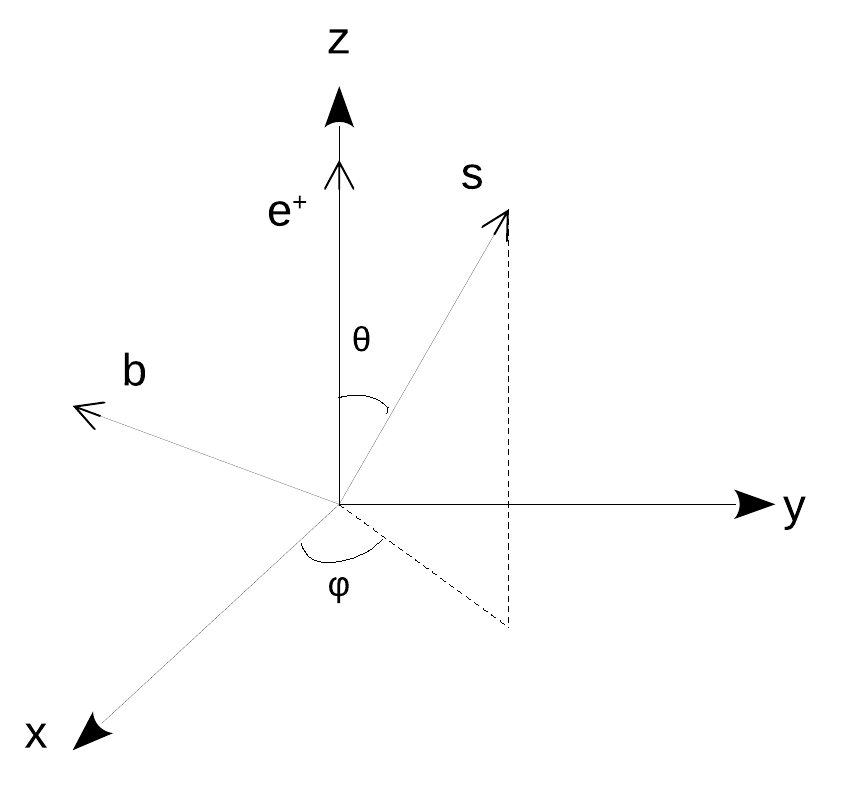}
\caption{ \label{pic2} \footnotesize Top quark spin quantization axis.}
\end{minipage}
\end{center}
\end{figure}

As well known in the SM, the positively charged lepton from the top quark decay in its rest frame tends to follow the top quark spin direction \cite{Jezabek:1994zv, Jezabek:1994qs}. In the t- and s-channel single top production the direction of the top quark spin in the top rest frame is highly correlated with the d-quark momentum for the t channel (outgoing light jet) and anti-d-quark momentum (incoming parton corresponding to the beam axis) for the s channel \cite{Mahlon:1996pn, Mahlon:1999gz}. One can understand this very simply by considering the single top production as a decay back in time \cite{Boos:2002xw}. The direction of the preferred spin configuration in the single top production in the presence of anomalous couplings is changed insignificantly compared to the SM due to the smallness of the anomalous couplings, so one can chose the direction related to the mentioned d-quark momentum as a top spin quantization axis and make sure that this will be the preferred spin direction of the top quark in its rest frame similar to the SM. 
After production, the highly polarized top quark decays (Fig.~\ref{pic1}), and one can use the properties of such a polarized decay.

%=========================================================================

%=============================================================================
\begin{figure}
\begin{center}
\begin{minipage}[t]{.45\linewidth}
\centering
\includegraphics[width=6cm,height=5cm]{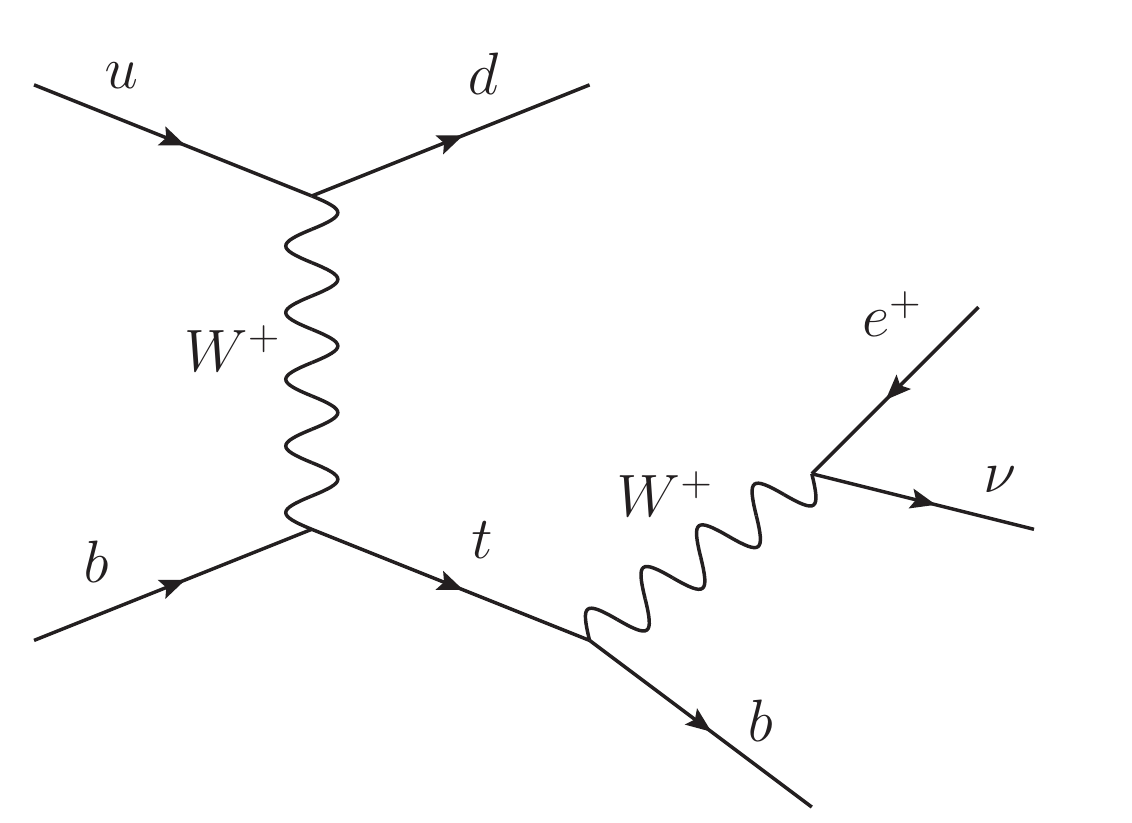}
\caption{\label{pic3} \footnotesize t-channel single top quark production and its leptonic decay processes. }
\end{minipage}
\hfill
\begin{minipage}[t]{.45\linewidth}
\centering
\includegraphics[width=6cm,height=5cm]{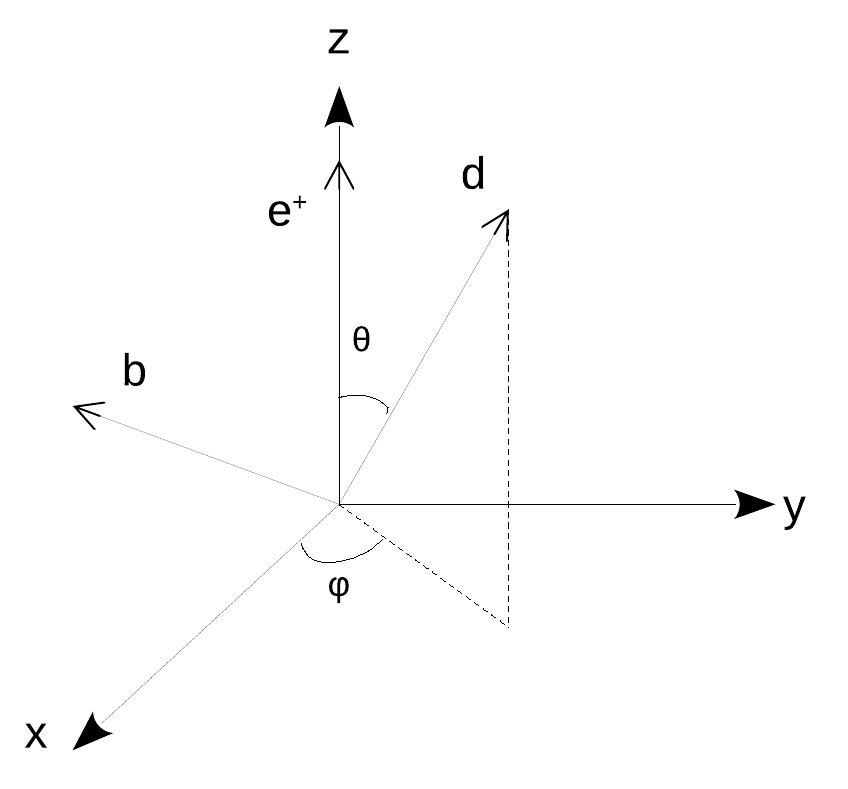}
\caption{ \label{pic4} \footnotesize Top quark spin quantization axis in the top rest frame is chosen as the direction of the d-quark momentum for the t-channel (outgoing light jet).}
\end{minipage}
\end{center}
\end{figure}
%=============================================================================

\section{Top decay}
\label{Top decay}
To investigate the effect of anomalous parameters on the decay of a polarized top quark, we will calculate the differential width of the three-body t-quark decay. At first, using Feynman rules from Lagrangian (\ref{anom_wtb_eq_lagrangian}), with the help of the symbolic manipulation system FORM \cite{Vermaseren:2000nd}, we calculate the matrix element squared of the polarized t-quark decay into a charged lepton, d quark and b quark:  

\begin{align}\label{melement}
&|M_{t \to b \nu e^+}|^2~~=~~\frac{g^4}{\left(2(p_{\nu} p_{e^+})-m_W^2\right)^2 + \Gamma_W^2 m_W^2}\cdot \big[\\ \nonumber \\
\nonumber &
+ |f_{LV}|^2~~\cdot~~(p_b p_{\nu})\cdot\big(~(p_{e^+} p_t) - (p_{e^+} s)\cdot m_t~\big) \\
\nonumber &
+ |f_{LT}|^2~~\cdot~~\frac{2}{m_W^2} \cdot (p_b p_{e^+})(p_{\nu} p_{e^+})\cdot\big(~(p_{e^+} p_t) - (p_{e^+} s)\cdot m_t~\big) \\
\nonumber &
+ |f_{RT}|^2~~\cdot~~\frac{2}{m_W^2} \cdot (p_b p_{\nu})(p_{\nu} p_{e^+})\cdot\big(~(p_{\nu} p_t) + (p_{\nu} s)\cdot m_t~\big) \\
\nonumber &
+ |f_{RV}|^2~~\cdot~~(p_b p_{e^+})\cdot\big(~(p_{\nu} p_t) + (p_{\nu} s)\cdot m_t~\big) \\
\nonumber &
+ (Re f_{LV}\cdot Re f_{RT} + Im f_{LV}\cdot Im f_{RT})~~\cdot~~\frac{2}{m_W}\cdot (p_b p_{\nu})\cdot\big(~(p_{\nu} p_{e^+})\cdot m_t + (p_{\nu} s)(p_{e^+} p_t) - (p_{\nu} p_t)(p_{e^+} s)~\big) \\
\nonumber &
+ (Re f_{LT} \cdot Re f_{RV} + Im f_{LT}\cdot Im f_{RV})~~\cdot~~\frac{2}{m_W} \cdot (p_b p_{e^+})\cdot\big(~(p_{\nu} p_{e^+})\cdot m_t + (p_{\nu} s)(p_{e^+} p_t) - (p_{\nu} p_t)(p_{e^+} s)~\big) \\
\nonumber &
+ (Re f_{LV}\cdot Im f_{RT} - Im f_{LV}\cdot Re f_{RT})~~\cdot~~ \frac{-2}{m_W}\cdot\epsilon_{\alpha\beta\rho\sigma}p_t^{\alpha} p_b^{\beta} p_{e^+}^{\rho} s^{\sigma} \cdot\left( \frac{m_t^2}{2}-(p_{e^+} p_t) \right) \\
\nonumber &
+ (Re f_{LT}\cdot Im f_{RV}  - Im f_{LT}\cdot Re f_{RV}  )~~\cdot~~\frac{-2}{m_W}\cdot\epsilon_{\alpha\beta\rho\sigma}p_t^{\alpha} p_b^{\beta} p_{e^+}^{\rho} s^{\sigma}  \cdot\left((p_b p_t) + (p_{e^+} p_t) - \frac{m_t^2}{2}\right)~~~\big]
\end{align}

where $s$ is the spin vector of the top quark.
\\

For further analytic calculations, we use the coordinate system that shown in Fig.~\ref{pic2}. Here, the angle $\theta$ is the angle between the charged lepton momentum and the direction of the top quark spin quantization axis, and $\phi$ is the angle in the plane perpendicular to the lepton momenta counted from the decay plane formed by the top quark decay products. 
Therefore, in the top quark rest frame, we have the following parametrization for the direction of the quantization axis of the top quark and for the 3-momentum of the positron and b quark:
\begin{align}\label{param}
\bold s=(\sin\theta\cos\phi,\sin\theta\sin\phi,\cos\theta),~~~~~
\bold p_{e^+}=|\bold p_{e^+}|\cdot(0,0,1),~~~~~
\bold p_b=|\bold p_b|\cdot(\sin\theta_{be},0,\cos\theta_{be})
\end{align}
The cosine of the angle between the b quark and the top quark spin quantization axis is:~
$\cos\theta_{bs} = \sin\theta_{be}\cdot\sin\theta\cdot\cos\phi + \cos\theta_{be}\cdot\cos\theta$.
One can also express the $\cos\phi$ in terms of other angles:~
$\cos\phi=(\cos\theta_{bs}-\cos\theta_{be}\cdot\cos\theta)/(\sin\theta_{be}\cdot\sin\theta)$. 
It leads to:
\begin{align}\label{cosphi}
\phi= \arccos\left( \frac{\cos\theta_{bs}-\cos\theta_{be}\cdot\cos\theta}{\sin\theta_{be}\cdot\sin\theta}\right)
\end{align}
The expression (\ref{cosphi}) is used 
 to reconstruct the $\phi$ angle in a numerical Monte Carlo simulation.

We substitute parametrization (\ref{param}) to the matrix element squared (\ref{melement}).
Using the 4-momentum conservation law in the t-quark rest system, one can express the neutrino momentum through the momentum of the b quark and positron. Then, we integrate the matrix element squared of the polarized t-quark decay (\ref{melement}) over the b-quark energy using the narrow-width approximation for the W-boson decay and neglecting the b-quark mass in comparison to the top quark and W-boson masses. Symbolic computation gives the following expression for the fully differential partial decay width of the top quark in its rest frame:  

\begin{align}\label{twidth0}
&\frac{d\Gamma_{t \to b \nu e^+}}{d\epsilon\cdot d\cos\theta\cdot d\phi}~~=~~\frac{\alpha^2\cdot m_t^3}{128\cdot \pi\cdot \sin^4{\Theta_W}\cdot \Gamma_W\cdot m_W}\cdot \big[\\ \nonumber \\ \nonumber &
+ |f_{LV}|^2~~\cdot~~(1-\epsilon)\cdot \epsilon \cdot(1 + \cos\theta)\\ \nonumber &
+ |f_{LT}|^2~~\cdot~~(\epsilon-r^2)\cdot \epsilon \cdot(1 + \cos\theta)\\ \nonumber &
+ |f_{RT}|^2~~\cdot~~(1-\epsilon)\cdot\left(1+r^2-\epsilon + \frac{2r\cdot c(\epsilon)}{\epsilon}\cdot\sin\theta\cos\phi + \left(\frac{2r^2}{\epsilon}+\epsilon-r^2-1\right)\cdot\cos\theta\right)\\ \nonumber &
+ |f_{RV}|^2~~\cdot~~(\epsilon-r^2)\cdot\left(1+r^2-\epsilon + \frac{2r\cdot\c(\epsilon)}{\epsilon}\cdot\sin\theta\cos\phi + \left(\frac{2r^2}{\epsilon}+\epsilon-r^2-1\right)\cdot\cos\theta\right)\\ \nonumber &
+ (Re f_{LV}\cdot Re f_{RT} + Im f_{LV}\cdot Im f_{RT})~~\cdot~~(1-\epsilon)\cdot 2\cdot\left(c(\epsilon)\cdot\sin\theta\cos\phi + r\cdot(1+\cos\theta)\right) \\ \nonumber &
+ (Re f_{LT}\cdot Re f_{RV} + Im f_{LT}\cdot Im f_{RV})~~\cdot~~(\epsilon-r^2)\cdot 2\cdot\left(c(\epsilon)\cdot\sin\theta\cos\phi + r\cdot(1+\cos\theta)\right)\\ \nonumber \\ \nonumber &
+ (Re f_{LV}\cdot Im f_{RT} - Im f_{LV}\cdot Re f_{RT})~~\cdot~~(1-\epsilon)\cdot (-2\cdot c(\epsilon)\cdot\sin\theta\sin\phi) \\ \nonumber &
+ (Re f_{LT}\cdot Im f_{RV} - Im f_{LT}\cdot Re f_{RV})~~\cdot~~(\epsilon-r^2)\cdot (-2\cdot c(\epsilon)\cdot\sin\theta\sin\phi) ~~~\big]
\end{align}
\\

where:~~
$c(\epsilon) = \sqrt{(1-\epsilon)(\epsilon-r^2)},~~~\epsilon = 2E_{e^+}/m_t,~~~\epsilon_{max} = 1,~~~\epsilon_{min} = r^2,~~~r=m_W/m_t$.
\\

This expression was obtained for the first time in such a complete form for such a parameterization.
The expression (\ref{twidth0}) contains of eight terms corresponding to different possible quadratic combinations of the coupling products. Each term contains a multiplier $(1-\epsilon)$ or $(\epsilon-r^2)$, as well as a polynomial multiplier being a function of kinematic variables $\epsilon$, $\cos\theta$, $\sin\theta$, $\cos\phi$ or $\sin\phi$. 
One should note that all the eight terms are different functions of the variables resulting in different shapes of multidimensional distributions.
Therefore, such multidimensional distribution shapes (multidimensional surfaces) can be used to separate contributions from different anomalous coupling combinations.

Let us integrate the expression of the fully differential partial decay width of the top quark (\ref{twidth0}) over the angle $\phi$ 
from 0 to $2\pi$ and obtain the doubly differential partial width $\frac{d\Gamma_{t \to b \nu e^+}}{d\epsilon~\cdot~d\cos\theta}$. Since the first six terms in (\ref{twidth0}) are even functions of $\phi$, their integral over $\phi$ in the range of $\pi$ to $2\pi$ is equal to the integral over $\phi$ in the range of $0$ to $\pi$.
At the same time, the last two terms in (\ref{twidth0}) proportional to $\sin\phi$ are odd functions of $\phi$, and their integral over $\phi$ in the range of $\pi$ to $2\pi$
equals the integral over $\phi$ in the range of $0$ to $\pi$ taken with the opposite sign, 

\begin{align}\label{twidth1}
&\frac{d\Gamma_{t \to b \nu e^+}}{d\epsilon \cdot d\cos\theta}=\int\limits^{\pi}_{0}\frac{d\Gamma_{t \to b \nu e^+}}{d\epsilon \cdot d\cos\theta\cdot d\phi}d\phi+\int\limits^{2\pi}_{\pi}\frac{d\Gamma_{t \to b \nu e^+}}{d\epsilon \cdot d\cos\theta\cdot d\phi}d\phi =
\frac{\alpha^2\cdot m_t^3}{64\cdot\sin^4{\Theta_W}\cdot \Gamma_W\cdot m_W}\cdot \big[\\ \nonumber &
+ |f_{LV}|^2~~\cdot~~(1-\epsilon)\cdot\epsilon \cdot(1 + \cos\theta)\\ \nonumber &
+ |f_{LT}|^2~~\cdot~~(\epsilon-r^2)\cdot\epsilon\cdot(1 + \cos\theta)\\ \nonumber &
+ |f_{RT}|^2~~\cdot~~(1-\epsilon)\cdot\left(1+r^2-\epsilon + \left(\frac{2r^2}{\epsilon}+\epsilon-r^2-1\right)\cdot\cos\theta\right)\\ \nonumber &
+ |f_{RV}|^2~~\cdot~~(\epsilon-r^2)\cdot\left(1+r^2-\epsilon + \left(\frac{2r^2}{\epsilon}+\epsilon-r^2-1\right)\cdot\cos\theta\right)\\ \nonumber &
+ (Re f_{LV}\cdot Re f_{RT} + Im f_{LV}\cdot Im f_{RT})~~\cdot~~(1-\epsilon)\cdot 2r\cdot(1+\cos\theta)\\ \nonumber &
+ (Re f_{LT}\cdot Re f_{RV} + Im f_{LT}\cdot Im f_{RV})~~\cdot~~(\epsilon-r^2)\cdot 2r\cdot(1+\cos\theta)\\ \nonumber &
+ (Re f_{LV}\cdot Im f_{RT} - Im f_{LV}\cdot Re f_{RT})~~\cdot~~(1-\epsilon)\cdot \left(-\frac{2}{\pi}\right)\cdot c(\epsilon)\cdot\sin\theta~~\cdot~~(1~-1) \\ \nonumber &
+ (Re f_{LT}\cdot Im f_{RV} - Im f_{LT}\cdot Re f_{RV})~~\cdot~~(\epsilon-r^2)\cdot \left(-\frac{2}{\pi}\right)\cdot c(\epsilon)\cdot\sin\theta~~\cdot~~(1~-1) ~~~\big]
\end{align}

The angular dependence of the expression was simplified after integration, but the energy-dependent factors $(1-\epsilon)$ and $(\epsilon-r^2)$ did not change and the differences between the various terms of the expression remained. The first six terms of formula (\ref{twidth1}) agree with the expression previously obtained in \cite{Hioki:2015moa}. The last two terms give zero contribution to the 2D distribution on the energy and $\cos\theta$, but we kept them in (\ref{twidth1}) as a factor (1-1) stressing that the term consists of two equal contributions with different signs at $\phi$ intervals from 0 to $\pi$ and from $\pi$ to $2\pi$. This fact allows one to extract corresponding anomalous couplings looking at the differences (or asymmetry in $\phi$) of the two distributions.

Also, integrating the expression (\ref{twidth0}) over $\cos\theta$ from -1 to 1, one can obtain the doubly differential partial width $\frac{d\Gamma_{t \to b \nu e^+}}{d\epsilon~\cdot~d\phi}$:
\begin{align}\label{twidth2}
&\frac{d\Gamma_{t \to b \nu e^+}}{d\epsilon \cdot d\phi}~~=~~\frac{\alpha^2\cdot m_t^3}{64\cdot\pi\cdot\sin^4{\Theta_W}\cdot \Gamma_W\cdot m_W}\cdot \big[\\ \nonumber &
+ |f_{LV}|^2~~\cdot~~(1-\epsilon)\cdot\epsilon\\ \nonumber &
+ |f_{LT}|^2~~\cdot~~(\epsilon-r^2)\cdot\epsilon\\ \nonumber &
+ |f_{RT}|^2~~\cdot~~(1-\epsilon)\cdot\left(1+r^2-\epsilon + \frac{\pi r}{2\epsilon}\cdot c(\epsilon)\cdot\cos\phi\right)\\ \nonumber &
+ |f_{RV}|^2~~\cdot~~(\epsilon-r^2)\cdot\left(1+r^2-\epsilon + \frac{\pi r}{2\epsilon}\cdot c(\epsilon)\cdot\cos\phi\right)\\ \nonumber &
+ (Re f_{LV}\cdot Re f_{RT} + Im f_{LV}\cdot Im f_{RT})~~\cdot~~(1-\epsilon)\cdot\left(2r+\frac{\pi}{2}\cdot c(\epsilon)\cdot\cos\phi\right) \\ \nonumber &
+ (Re f_{LT}\cdot Re f_{RV} + Im f_{LT}\cdot Im f_{RV})~~\cdot~~(\epsilon-r^2)\cdot\left(2r+\frac{\pi}{2}\cdot c(\epsilon)\cdot\cos\phi\right)\\ \nonumber &
+ (Re f_{LV}\cdot Im f_{RT} - Im f_{LV}\cdot Re f_{RT})~~\cdot~~(1-\epsilon)\cdot\left(-\frac{\pi}{2}\right)\cdot c(\epsilon)\cdot\sin\phi \\ \nonumber &
+ (Re f_{LT}\cdot Im f_{RV} - Im f_{LT}\cdot Re f_{RV})~~\cdot~~(\epsilon-r^2)\cdot\left(-\frac{\pi}{2}\right)\cdot c(\epsilon)\cdot\sin\phi ~~~\big]
\end{align}
\\

Parts of polynomials containing the $\cos\theta$ function disappear after integrating the expression (\ref{twidth0}) over $\cos\theta$. The function $\sin\theta$ in the remaining parts of (\ref{twidth0}) is replaced by the factor $\pi/2$. 
However, despite these changes, the differences between the eight terms of formula (\ref{twidth2}), proportional to anomalous couplings, remain.

Now, integrating the expression of the fully differential top quark partial decay width (\ref{twidth0}) over $\epsilon$ from $r^2$ to 1, we obtain the doubly differential partial width $\frac{d\Gamma_{t\to b\nu e^+}}{d\cos\theta~\cdot~d\phi}$ as a function of the two orientation angles: 

\begin{align}\label{twidth3}
&\frac{d\Gamma_{t\to b \nu e^+}}{d\cos\theta\cdot d\phi}~~=~~\frac{\alpha^2\cdot m_t^3}{256\cdot 3\cdot \pi\cdot\sin^4{\Theta_W}\cdot \Gamma_W\cdot m_W}\cdot \big[\\ \nonumber \\ \nonumber &
+ |f_{LV}|^2~~\cdot~~(1 - r^2)^2(1 + 2r^2) \cdot(1 + \cos\theta)\\ \nonumber &
+ |f_{LT}|^2~~\cdot~~(1 - r^2)^2(r^2 + 2) \cdot(1 + \cos\theta)\\ \nonumber &
+ |f_{RT}|^2~~\cdot~~\big(~~(1 - r^2)^2(r^2 + 2)+ \frac{3\pi r}{2}(-3 + 8r - 6r^2 + r^4)\cdot\sin\theta\cos\phi \\ \nonumber &
~~~~~~~~~~~~~~~~~~~~~~~~~~~~-\big(~~(1-r^2)(2 - r^4 + 11r^2) + 24r^2\cdot\ln(r)~~\big)\cdot \cos\theta~~\big)\\ \nonumber\\ \nonumber &
+ |f_{RV}|^2~~\cdot~~\big(~~(1 - r^2)^2(1 + 2r^2) 
+ \frac{3\pi r}{2}(1 - 6r^2 + 8r^3 - 3r^4)\cdot\sin\theta\cos\phi \\ \nonumber & 
~~~~~~~~~~~~~~~~~~~~~~~~~~~~~~-\big(~~(1-r^2)(1 - 2r^4 - 11r^2) - 24r^4\cdot \ln(r)~~\big)\cdot\cos\theta~~\big)\\ \nonumber\\ \nonumber &
+ (Re f_{LV}\cdot Re f_{RT} + Im f_{LV}\cdot Im f_{RT})~~\cdot~~(1 - r^2)^2\left(\frac{3\pi}{4}(1 - r^2)\cdot\sin\theta\cos\phi + 6r\cdot(1+\cos\theta)\right) \\ \nonumber &
+ (Re f_{LT}\cdot Re f_{RV} + Im f_{LT}\cdot Im f_{RV})~~\cdot~~(1 - r^2)^2\left(\frac{3\pi}{4}(1 - r^2)\cdot\sin\theta\cos\phi + 6r\cdot(1+\cos\theta)\right)\\ \nonumber &
+ (Re f_{LV}\cdot Im f_{RT} - Im f_{LV}\cdot Re f_{RT})~~\cdot~~\left(-\frac{3\pi}{4}\right)(1 - r^2)^3\cdot\sin\theta\sin\phi \\ \nonumber &
+ (Re f_{LT}\cdot Im f_{RV} - Im f_{LT}\cdot Re f_{RV})~~\cdot~~\left(-\frac{3\pi}{4}\right)(1 - r^2)^3\cdot\sin\theta\sin\phi ~~~\big]
\end{align}

The first term in the expression (\ref{twidth3}) is the well-known SM-like contribution \cite{Jezabek:1994zv, Jezabek:1994qs} corresponding to 100\% spin correlation behavior $(1 + \cos\theta)$. 
The interference between the left-vector term with real coupling and the right-tensor term with the imaginary coupling corresponding to the charge-parity-violating part coincides exactly with the expression presented in \cite{Zhang:2010dr}. As one can see, integration (\ref{twidth0}) over $\epsilon$ eliminates the factors $(1-\epsilon)$ and $(\epsilon-r^2)$ and makes the terms proportional to anomalous couplings in the expression (\ref{twidth3}) more similar to each other.

Finally, integrating (\ref{twidth0}) over $\epsilon$, $\cos\theta$ and $\phi$ (ranging from 0 to $2\pi$), we obtain the expression for the t-quark three-body leptonic partial decay width $\Gamma_{t \to b \nu e^+}$,

\begin{align}\label{twidth4}
\Gamma_{t \to b \nu e^+}~~=~~\frac{\alpha^2\cdot m_t^3}{64\cdot 3\cdot\sin^4{\Theta_W}\cdot \Gamma_W\cdot m_W}&\cdot (1 - r^2)^2\cdot\big[\\ \nonumber &
+ |f_{LV}|^2~~\cdot~~(1 + 2r^2)\\ \nonumber &
+ |f_{LT}|^2~~\cdot~~(r^2 + 2)\\ \nonumber &
+ |f_{RT}|^2~~\cdot~~(r^2 + 2)\\ \nonumber &
+ |f_{RV}|^2~~\cdot~~(1 + 2r^2)\\ \nonumber &
+ (Re f_{LV}\cdot Re f_{RT} + Im f_{LV}\cdot Im f_{RT})~~\cdot~~6r \\ \nonumber &
+ (Re f_{LT}\cdot Re f_{RV} + Im f_{LT}\cdot Im f_{RV})~~\cdot~~6r
~~~\big]
\end{align}

For the case of purely real anomalous couplings, this expression is consistent with the previously obtained \cite{MohammadiNajafabadi:2006tfl}.

\section{Top production and decay}
\label{sec:Numreal}

Now we use the expression for the differential decay width of the t quark
to derive a differential cross section for the complete process of the single top production with its subsequent decay.

The dominant process of single top production at the LHC collider is the t-channel process shown in Fig.~\ref{pic3}.
It is known that in the framework of the SM in a t-channel process $ub\to td$, in its rest frame the t quark is produced polarized in the direction of the d quark. Therefore, we set the components of the spin vector of the t quark along this direction (Fig.~\ref{pic4}) in amplitudes of the production and decay and square the amplitude of the complete process $u b \to d, b, \nu, e^+$.
Going into the t-quark rest frame and writing explicit scalar products through the components of the 4-momentum and angles, summing over the spin components one gets an expression for the matrix element of the complete process $u b \to d, b, \nu, e^+$.

To calculate the cross section of the complete process, we use the formula for the t-channel anomalous production of the top quark \cite{Boos:2016zmp},

\begin{align}\label{prodcross}
	&\sigma(\hat{s})_{ub\to td}~~=~~\frac{\pi\cdot\alpha^2\cdot V_{ud}^2}{4\cdot\sin^4{\Theta_W}\cdot m_W^2}~\cdot\big[\\ \nonumber &
	+ |f_{LV}|^2~~\cdot~~c_2 \beta^4 \\ \nonumber &
	+ |f_{LT}|^2~~\cdot~~\beta^2\left( c_1\cdot ln(a)~ - ~2\beta^2\right)\\ \nonumber &
	+ |f_{RV}|^2~~\cdot~~\left( - (1 + c_1)r_s^2\cdot ln(a)~ + ~(1+2r_s^2)\cdot\beta^2\right)\\ \nonumber &
	+ |f_{RT}|^2~~\cdot~~\big(~~~~(1+2r_s^2)\cdot ln(a)~ - ~(1 + c_1)\cdot c_2\beta^2\big)\\ \nonumber &
	+ (Re f_{LV}\cdot Re f_{RT} + Im f_{LV}\cdot Im f_{RT})~~\cdot~~2r_s\sqrt{1-\beta^2}\left(ln(a)~ - ~c_2\beta^2\right)  \\ \nonumber &
	+ (Re f_{LT}\cdot Re f_{RV} + Im f_{LT}\cdot Im f_{RV})~~\cdot~~2r_s\sqrt{1-\beta^2}\left(- c_1\cdot ln(a)~ + ~2\beta^2\right)~~\big]  \nonumber &
\end{align}
where
\begin{align}
	\nonumber a = 1+\frac{\beta^2}{r_s^2},~~~c_1 = \beta^2+2r_s^2,~~~c_2 = \frac{1}{\beta^2 +  r_s^2},~~~\beta^2 = 1-\frac{m_t^2}{\hat{s}},~~~r_s=\frac{m_W}{\sqrt{\hat{s}}}
\end{align}

Integrating the matrix element of the complete process $u b \to d, b, \nu, e^+$ over all variables of the phase space, except ($\epsilon$, $\cos\theta$, $\phi$), neglecting the b-quark mass in comparison to the top quark and W-boson masses, using formulas (\ref{twidth0}) and (\ref{prodcross}), and keeping only terms up to the second order of magnitude of anomalous parameters in the numerator, we obtain the differential cross section of the polarized single top quark production with its subsequent decay:

\begin{align}\label{totalcrossec}
&\frac{d\sigma(\hat{s})_{u b \to d b \nu e^+}}{d\epsilon\cdot d\cos\theta\cdot d\phi}~~=~~\frac{1}{\Gamma_t}~\cdot\big[~~
	\sigma(\hat{s})_{ub\to td}\cdot\frac{d\Gamma_{t \to b \nu e^+}}{d\epsilon\cdot d\cos\theta\cdot d\phi}\\ \nonumber&
	- \sigma_{R}(\hat{s})_{ub\to td}\cdot\frac{\alpha^2\cdot m_t^3\cdot V_{tb}^2}{64\cdot \pi\cdot \sin^4{\Theta_W}\cdot \Gamma_W\cdot m_W}\cdot(1-\epsilon)\cdot\epsilon \cdot\cos\theta~~\big]\\ \nonumber&
\end{align}

where $\Gamma_t$ is the total decay width of the t quark, taking into account the anomalous couplings and all decay modes,   $\sigma(\hat{s})_{ub\to td}$ is the cross section of the unpolarized t-quark production (\ref{prodcross}), $\sigma_R(\hat{s})_{ub\to td}$ is part of the cross section of the unpolarized t-quark production that is proportional to $|f_{RV}|^2$ and $|f_{RT}|^2$, and $\frac{d\Gamma_{t \to b \nu e^+}}{d\epsilon~\cdot~d\cos\theta~\cdot~d\phi}$ is the differential partial width of the polarized t-quark decay (\ref{twidth0}), and $\theta$ and $\phi$ are orientation angles of the positron with respect to the direction of the d-quark momentum. We mean that we omit terms of the third and fourth order of magnitude of anomalous couplings in the numerator of the first part of (\ref{totalcrossec}). 
\\

The expression (\ref{totalcrossec}) was obtained for the first time. The detailed derivation of the expression (\ref{totalcrossec}) as well as the full expression, which includes terms of all orders by anomalous parameters is given in the Appendix of the article. Expression (\ref{totalcrossec}) contains the contribution of the Standard Model, which has the form

\begin{align}\label{totalcrossecsm}
&\frac{d\sigma_{SM}(\hat{s})_{u b \to d b \nu e^+}}{d\epsilon\cdot d\cos\theta\cdot d\phi}~=~\\ \nonumber&
~~~~~~~~~~~~~~~~\frac{\alpha^2\cdot V_{ud}^2\cdot V_{tb}^2}{8\cdot3\cdot\sin^4{\Theta_W}\cdot m_W^2\cdot (1 - r^2)^2(1+2r^2)}\cdot\frac{(\hat{s}-m_t^2)^2}{\hat{s}(\hat{s}-m_t^2+m_W^2)}\cdot (1-\epsilon)\cdot\epsilon\cdot(1+\cos\theta)
\end{align}
\\

The expression (\ref{totalcrossecsm}) was also obtained for the first time. After integrating it over all variables, we have

\begin{align}\label{fullcrossecsm}
&\sigma_{SM}(\hat{s})_{u b \to d b \nu e^+}~=~\frac{\pi\cdot\alpha^2\cdot V_{ud}^2\cdot V_{tb}^2}{4\cdot9\cdot\sin^4{\Theta_W}\cdot m_W^2}\cdot\frac{(\hat{s}-m_t^2)^2}{\hat{s}(\hat{s}-m_t^2+m_W^2)}
\end{align}
\\

Integrating (\ref{totalcrossec}) over one of the variables, we obtain expressions for all possible double-differential scattering cross sections:

\begin{align}\label{totalcrossec1}
	&\frac{d\sigma(\hat{s})_{u b \to d b \nu e^+}}{d\epsilon\cdot d\cos\theta}~~=~~\frac{1}{\Gamma_t}~\cdot\big[~~
	\sigma(\hat{s})_{ub\to td}\cdot\frac{d\Gamma_{t \to b \nu e^+}}{d\epsilon\cdot d\cos\theta}\\ \nonumber&
	- \sigma_{R}(\hat{s})_{ub\to td}\cdot\frac{\alpha^2\cdot m_t^3\cdot V_{tb}^2}{32\cdot \sin^4{\Theta_W}\cdot \Gamma_W\cdot m_W}\cdot(1-\epsilon)\cdot\epsilon \cdot\cos\theta~~\big]\\ \nonumber&
\end{align}
\begin{align}\label{totalcrossec2}
	&\frac{d\sigma(\hat{s})_{u b \to d b \nu e^+}}{d\epsilon\cdot d\phi}~~=~~\sigma(\hat{s})_{ub\to td}\cdot\frac{1}{\Gamma_t}~\cdot\frac{d\Gamma_{t \to b \nu e^+}}{d\epsilon\cdot d\phi}&
\end{align}

\begin{align}\label{totalcrossec3}
	&\frac{d\sigma(\hat{s})_{u b \to d b \nu e^+}}{d\cos\theta\cdot d\phi}~~=~~\frac{1}{\Gamma_t}~\cdot\big[~~
	\sigma(\hat{s})_{ub\to td}\cdot\frac{d\Gamma_{t \to b \nu e^+}}{d\cos\theta\cdot d\phi}\\ \nonumber&
	- \sigma_{R}(\hat{s})_{ub\to td}\cdot\frac{\alpha^2\cdot m_t^3\cdot V_{tb}^2\cdot (1 - r^2)^2 (1+2r^2)}{128\cdot 3\cdot \pi\cdot \sin^4{\Theta_W}\cdot \Gamma_W\cdot m_W}\cdot\cos\theta~~\big]\\ \nonumber&
\end{align}
After integration over all variables, the cross section takes on a well-known simple form:
\begin{align}\label{totalcrossec4}
	&\sigma(\hat{s})_{u b \to d b \nu e^+}~~=~~\sigma(\hat{s})_{ub\to td}\cdot Br_{t \to b \nu e^+}&
\end{align}

where $Br_{t \to b \nu e^+}=\Gamma_{t \to b \nu e^+}/\Gamma_t$.
	 
\section{Numerical illustration, real couplings}
\label{sec:Numreal}

\begin{figure}
\begin{center}
\begin{minipage}[t]{.325\linewidth}
\centering
\includegraphics[width=6cm,height=6cm]{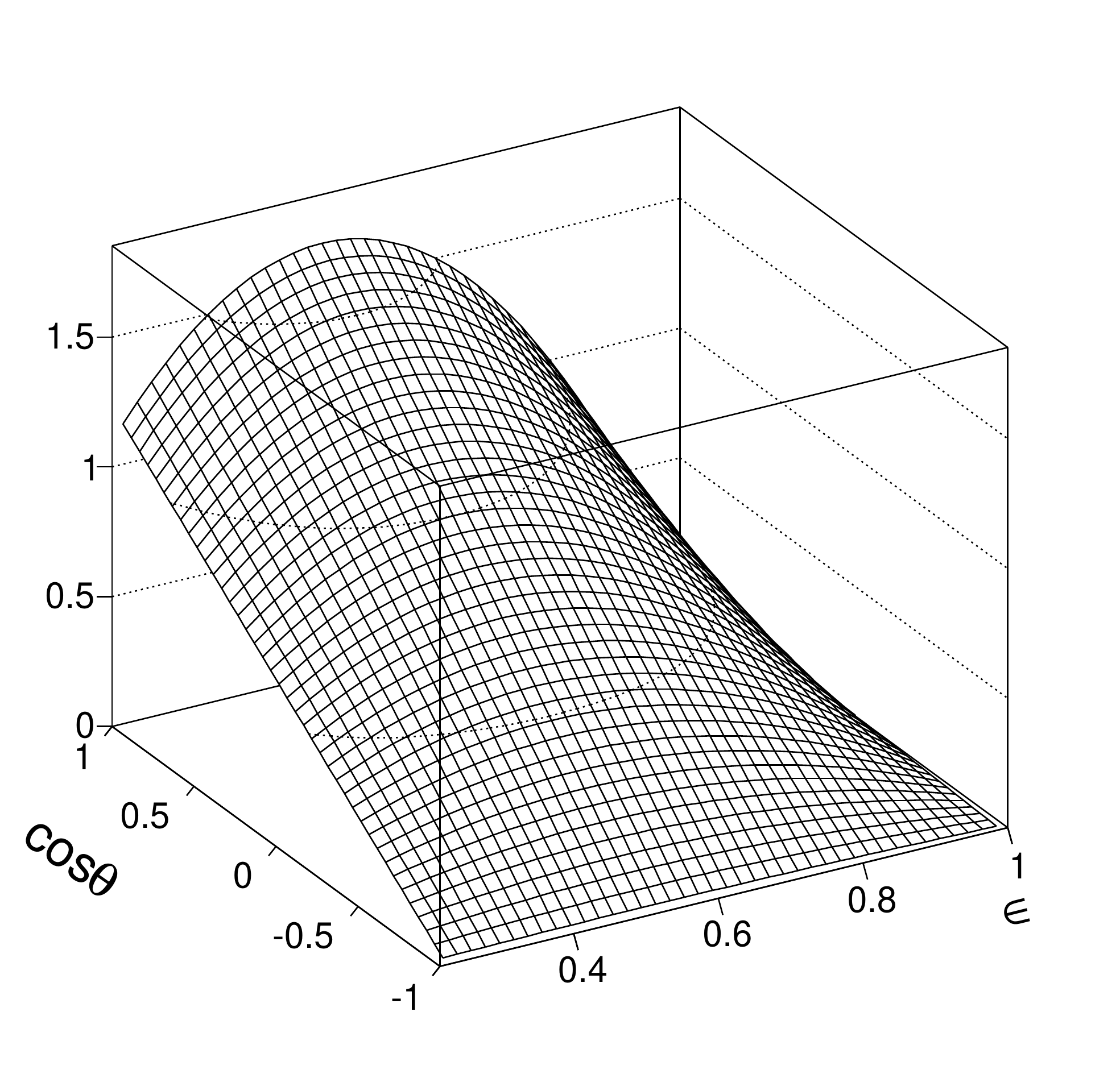}
\begin{center}
{\it $|f_{LV}|^2$}
\end{center}
\end{minipage}
%\hfill
\begin{minipage}[t]{.325\linewidth}
\centering
\includegraphics[width=6cm,height=6cm]{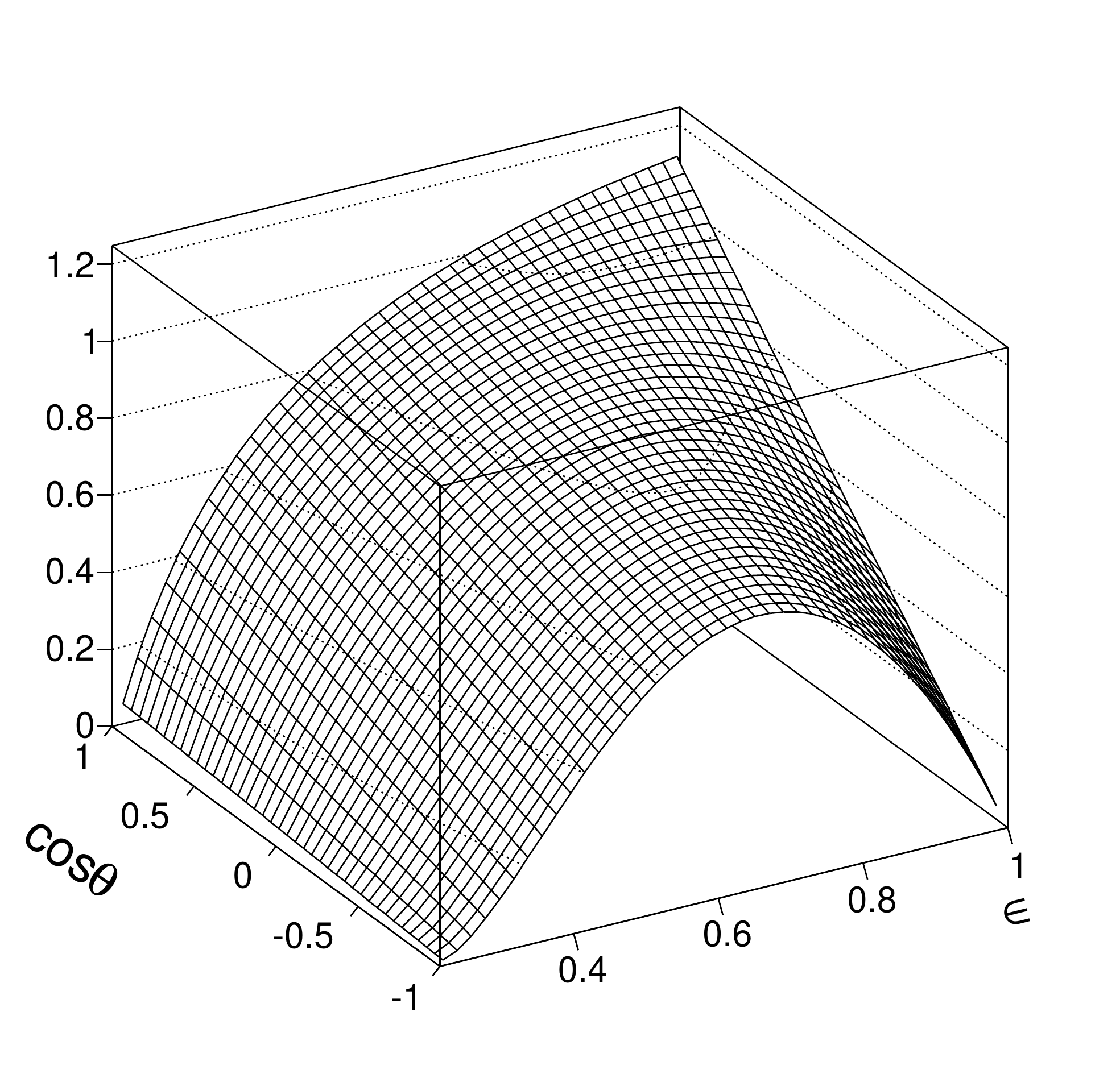}
\begin{center}
{\it $|f_{RV}|^2$}
\end{center}
\end{minipage}
\begin{minipage}[t]{.325\linewidth}
\centering
\includegraphics[width=6cm,height=6cm]{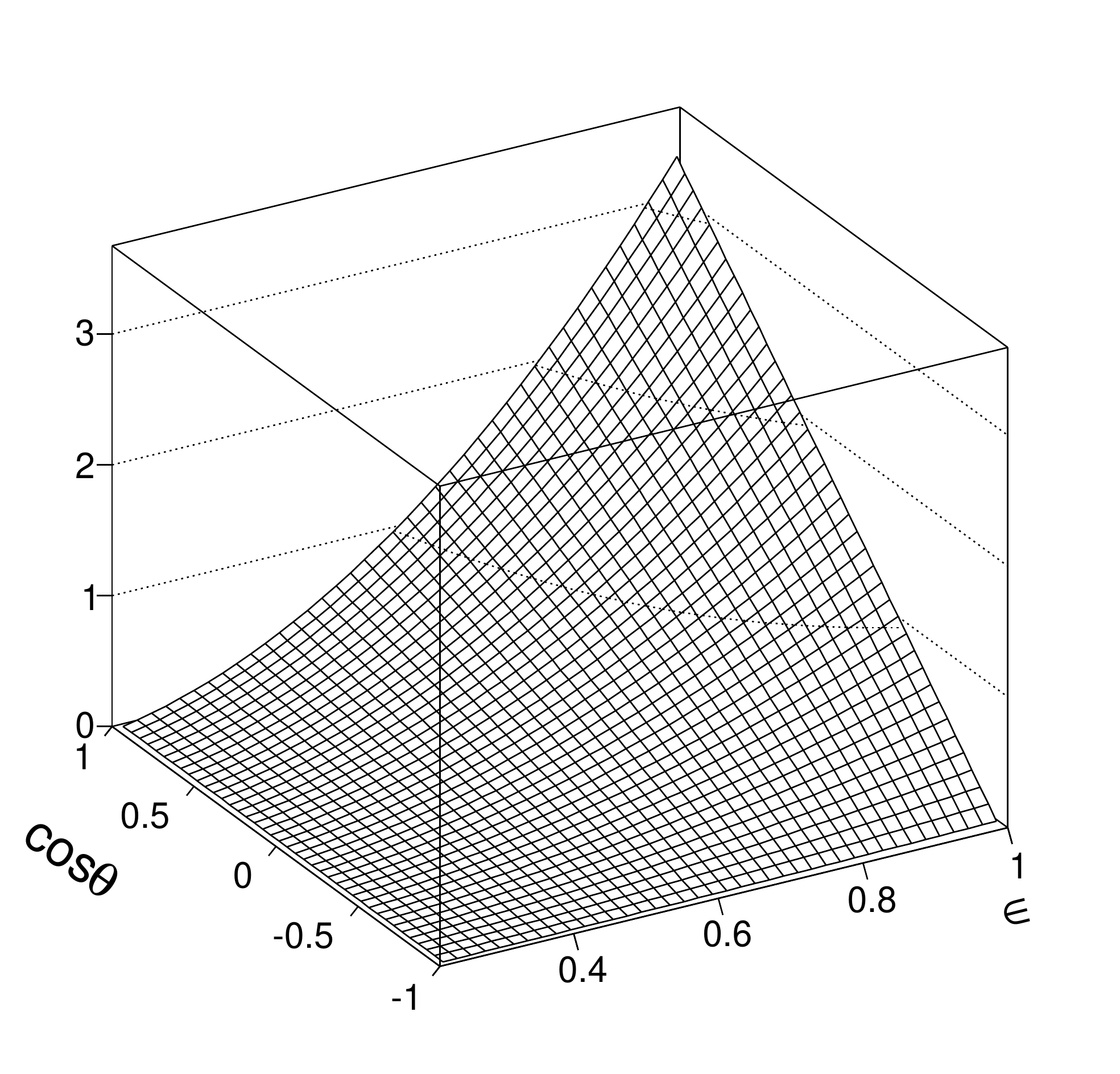}
\begin{center}
{\it $|f_{LT}|^2$}
\end{center}
\end{minipage}
\begin{minipage}[t]{.325\linewidth}
\centering
\includegraphics[width=6cm,height=6cm]{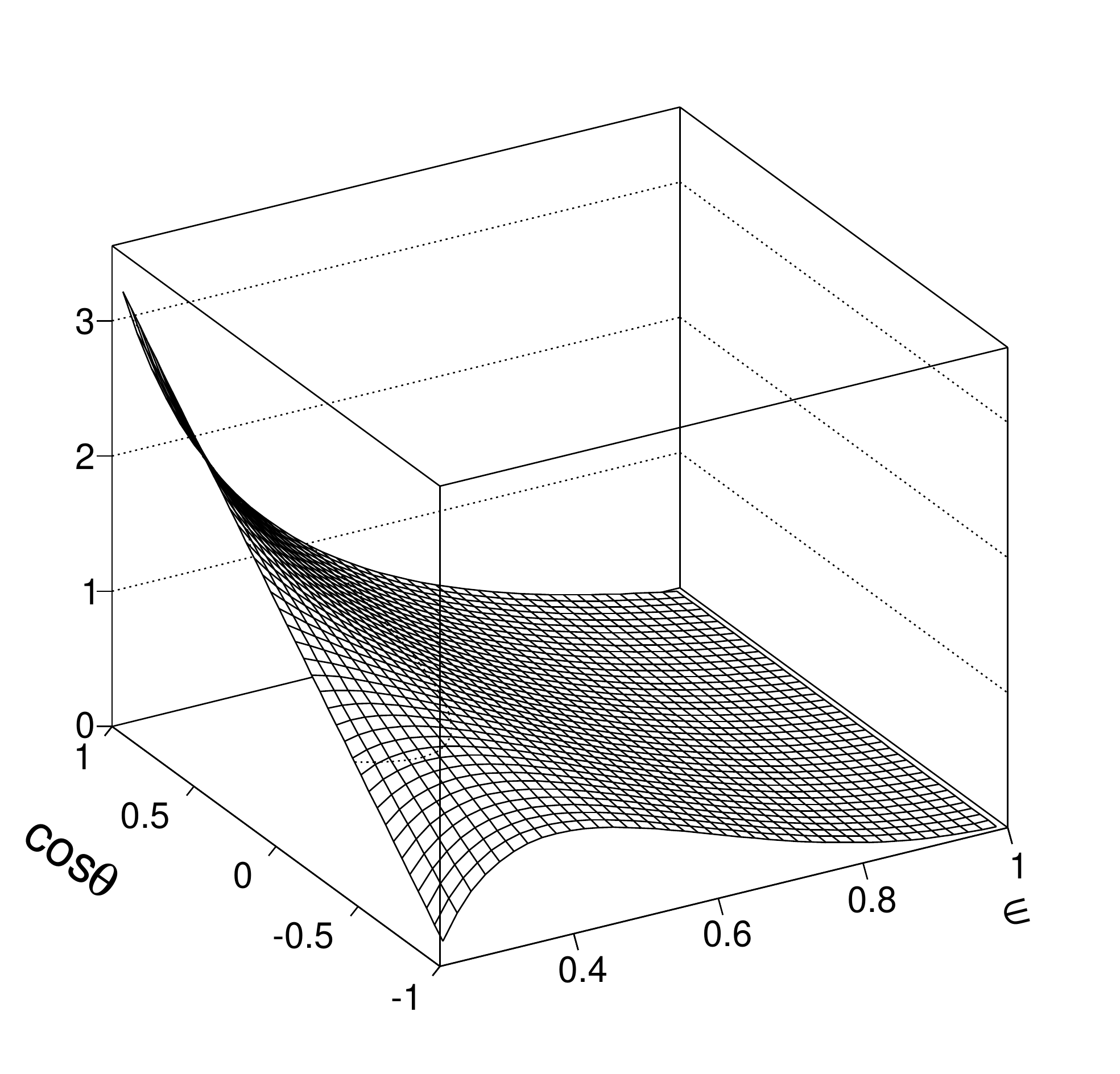}
\begin{center}
{\it $|f_{RT}|^2$}
\end{center}
\end{minipage}
\begin{minipage}[t]{.325\linewidth}
\centering
\includegraphics[width=6cm,height=6cm]{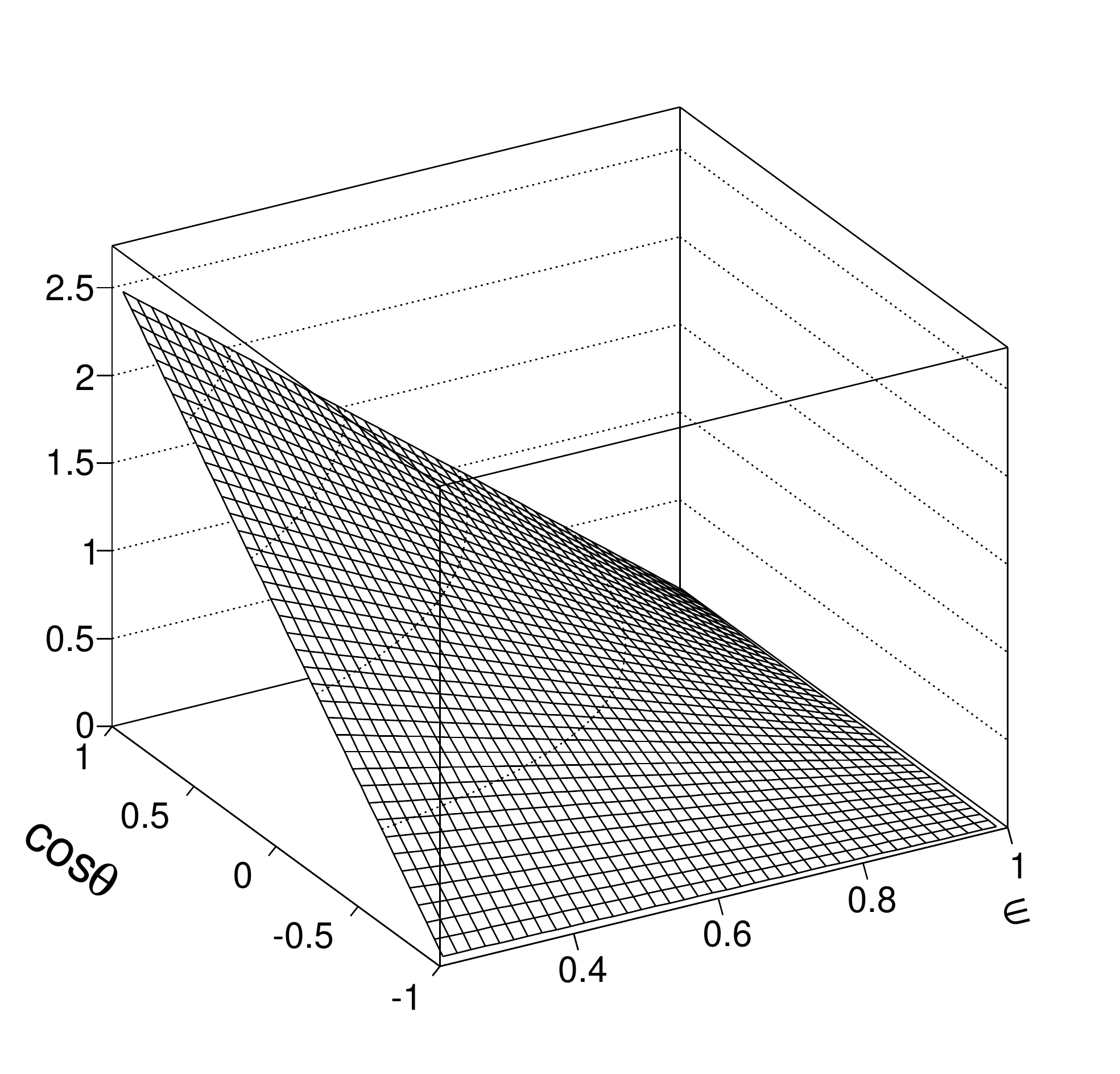}
\begin{center}
{\it $Re f_{LV}Re f_{RT}+Im f_{LV}Im f_{RT}$}
\end{center}
\end{minipage}
\begin{minipage}[t]{.325\linewidth}
\centering
\includegraphics[width=6cm,height=6cm]{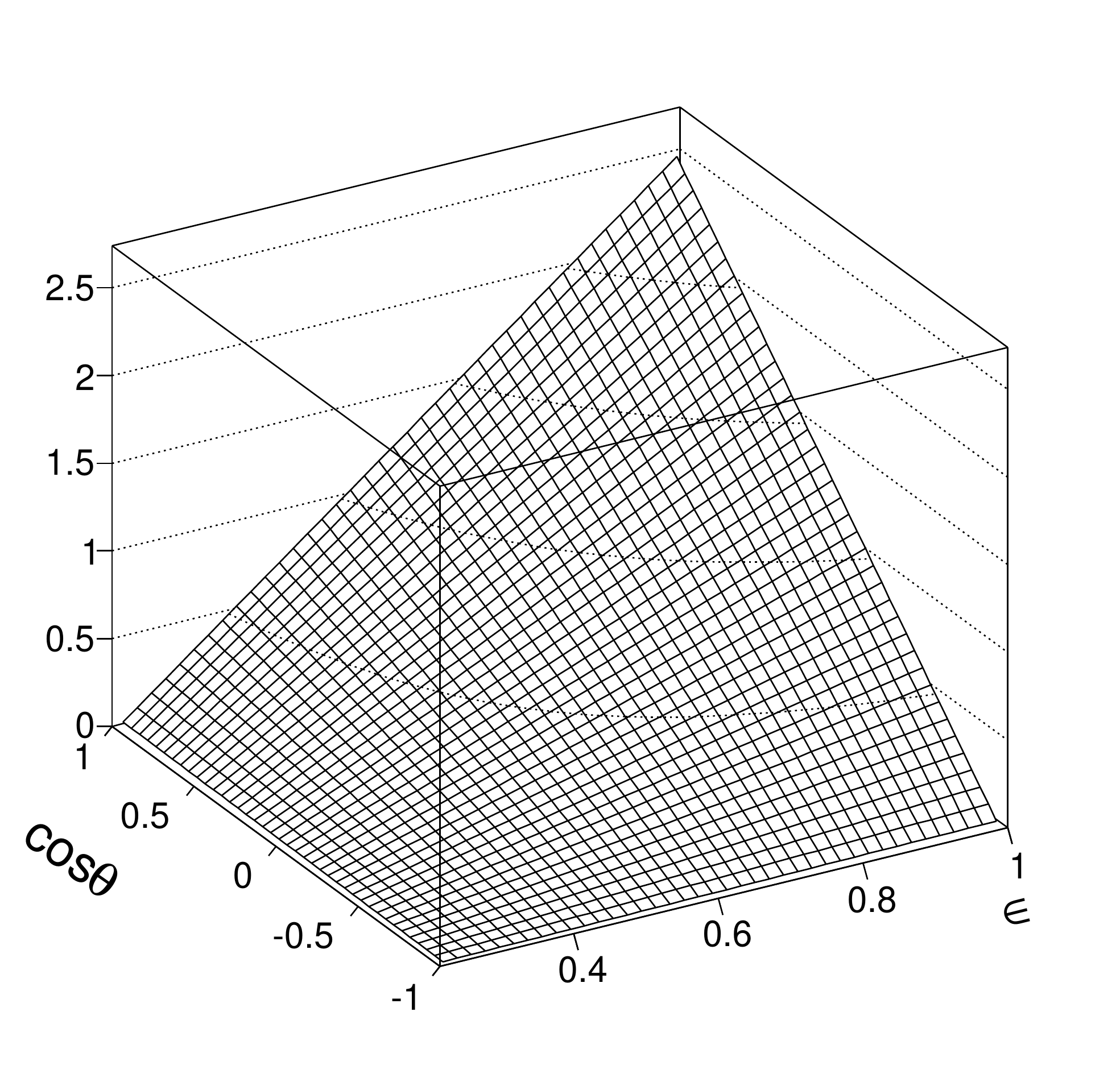}
\begin{center}
{\it $Re f_{LT}Re f_{RV}+Im f_{LT}Im f_{RV}$}
\end{center}
\end{minipage}
\begin{minipage}[t]{.325\linewidth}
	\centering
	\includegraphics[width=6cm,height=6cm]{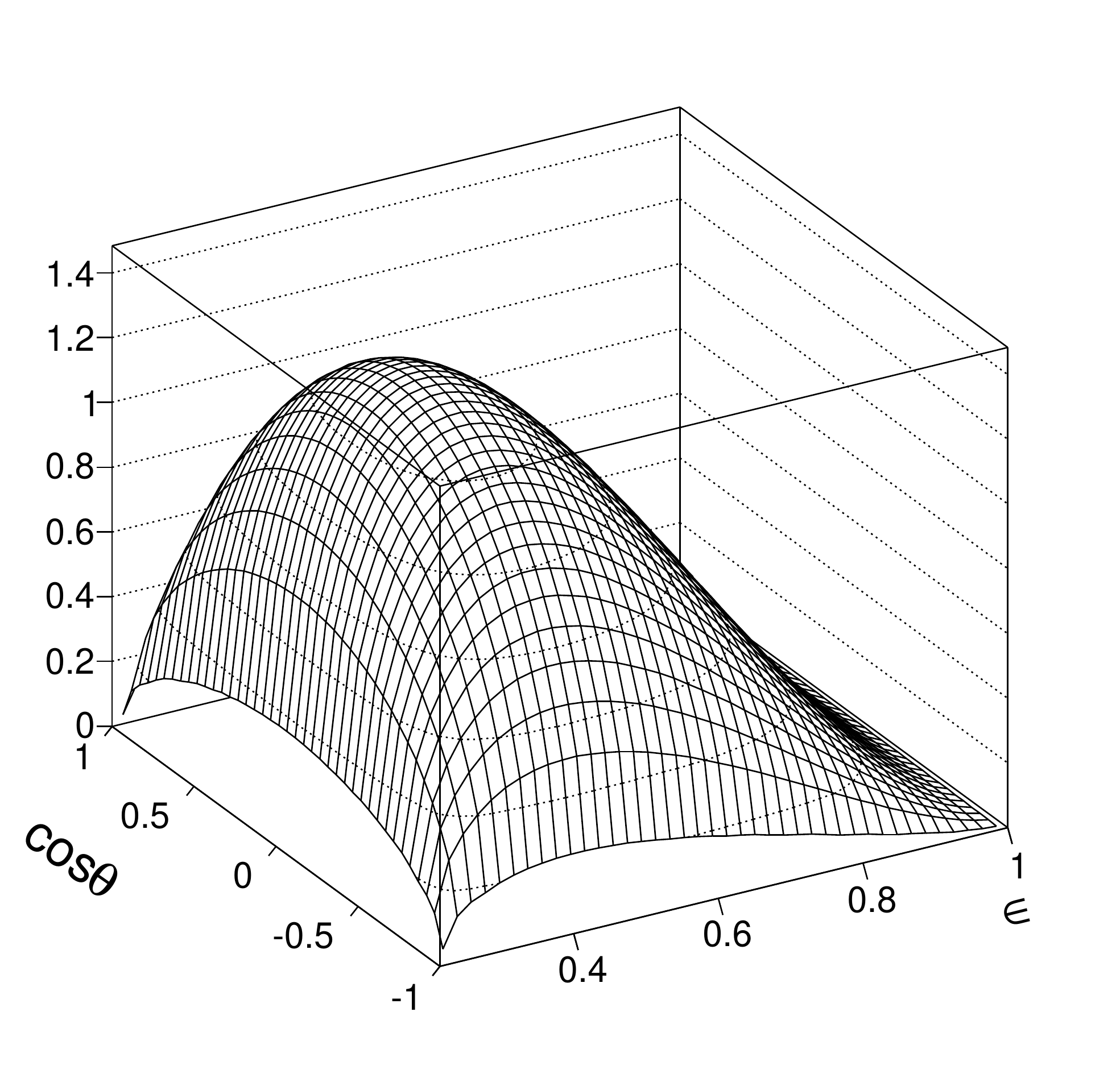}
	\begin{center}
		{\it $Re f_{LV}Im f_{RT}-Im f_{LV}Re f_{RT}$}
	\end{center}
\end{minipage}
\begin{minipage}[t]{.325\linewidth}
	\centering
	\includegraphics[width=6cm,height=6cm]{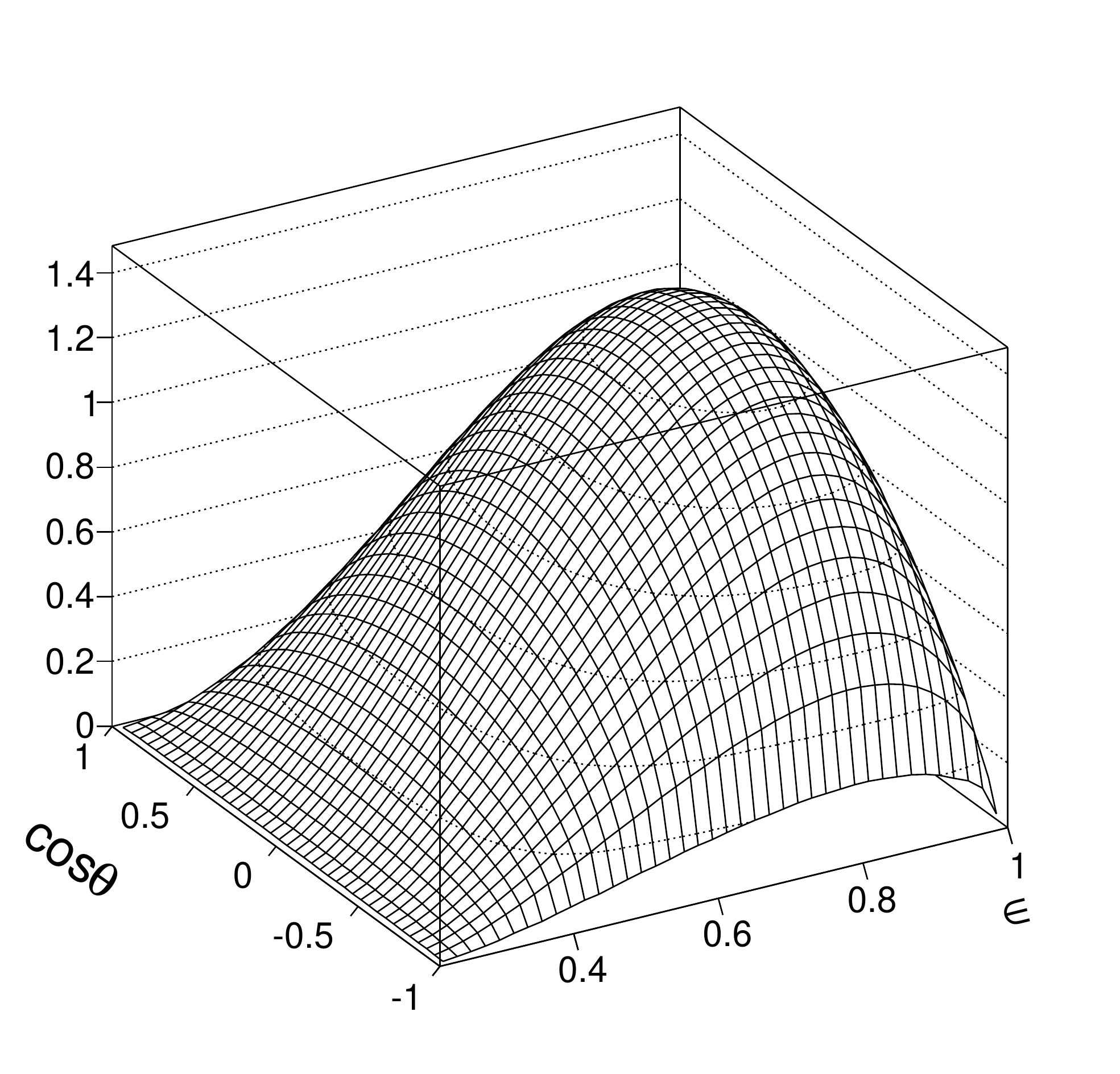}
	\begin{center}
		{\it $Re f_{LT}Im f_{RV}-Im f_{LT}Re f_{RV}$}
	\end{center}
\end{minipage}		
\end{center}
\caption{ \label{pic5} \footnotesize  Plots of normalized double-differential  t-quark decay partial width $\frac{d(\Gamma-\Gamma_{SM})}{d\epsilon~\cdot~d\cos\theta}$ for eight terms of formula (\ref{twidth1}) corresponding to possible combinations of anomalous couplings.}
\end{figure}

As the first numerical illustration, we draw normalized plots (Fig.~\ref{pic5}) of differential partial width $d\Gamma_{t \to b \nu e^+}/(d\epsilon\cdot dcos\theta)$ at the $\phi$ interval from 0 to $\pi$ for all eight terms of formula (\ref{twidth1}), which correspond to different combinations of anomalous couplings.  
As one can see, the shapes of the surfaces corresponding to the various terms of (\ref{twidth1}) are very different from each other. This set of eight different surfaces can be used as a basis for a multiparametric fitting function to get limits on anomalous couplings.

%=========================================================================
%=========================================================================

To verify this statement, we present the following numerical illustration for a few simple scenarios where we alternately set one of the anomalous couplings to a nonzero value. 
In each case, we present surfaces corresponding to formulas (\ref{twidth1}), (\ref{twidth2}), (\ref{twidth3}) for the  differential t-quark width. Then, we show surfaces for the same kinematic variables for differential cross sections as follows from formulas (\ref{totalcrossec1}), (\ref{totalcrossec2}), (\ref{totalcrossec3}) after integration with the parton distribution functions \footnote{The CTEQ6l parton distribution functions (PDFs) are used for definiteness. We have checked that various sets of PDFs do not influence the shapes of the surfaces.} at 14 TeV energy at the LHC collider. To validate the correctness of the results, we perform Monte Carlo event generation for the $2 \rightarrow 4$ process for the dominating t-channel single top production with subsequent three-body decay of t quark $u b \to d, b, \nu, l^+$ (Fig.~\ref{pic3}) using the CompHEP program \cite{Boos:2004kh} and show the same multidimensional distributions in the top rest frame taking all angles with respect to the d-quark momentum (Fig.~\ref{pic4}) and restore angle $\phi$ using the expression (\ref{cosphi}). Note that the anomalous couplings are included both in the top quark production and decay process. Values for anomalous couplings are taken according to the latest experimental upper limits \cite{Khachatryan:2016sib}. To make deviations from the SM more pronounced, we subtract the SM contribution. To correctly compare the shapes of different distributions, we normalized them to the value of the full integral. Writing explicitly from (\ref{totalcrossec}) the leading terms up to the second order of magnitude on anomalous couplings, one can write

\begin{align}\label{diffcrossec}
	\frac{1}{I_{norm}}\frac{d(\sigma-\sigma_{SM})_{u b \to d b \nu e^+}}{d\epsilon\cdot d\cos\theta\cdot d\phi}&=
	\frac{1}{(\sigma_{ub\to td}\Gamma_{t \to b \nu e^+}-r_{\Gamma}\cdot(\sigma_{SM})_{ub\to td}(\Gamma_{SM})_{t \to b \nu e^+})}\cdot\big[~~\\ \nonumber \\ \nonumber&
	+ (\sigma_{SM})_{ub\to td}\cdot\frac{d(\Gamma-\Gamma_{SM})_{t \to b \nu e^+}}{d\epsilon\cdot d\cos\theta\cdot d\phi}\\ \nonumber&
	+ \left(\sigma_{ub\to td}-r_{\Gamma}\cdot(\sigma_{SM})_{ub\to td}\right)\cdot \frac{d(\Gamma_{SM})_{t \to b \nu e^+}}{d\epsilon\cdot d\cos\theta\cdot d\phi}\\ \nonumber&
	+(\sigma_{RT})_{ub\to td}\cdot \frac{d(\Gamma_{RT})_{t \to b \nu e^+}}{d\epsilon\cdot d\cos\theta\cdot d\phi}\\ \nonumber&
	- (\sigma_{R})_{ub\to td}\cdot \frac{\alpha^2\cdot m_t^3\cdot V_{tb}^2}{64\cdot \pi\cdot \sin^4{\Theta_W}\cdot \Gamma_W\cdot m_W}\cdot(1-\epsilon)\cdot\epsilon \cdot\cos\theta~~\big]
\end{align}
where $\sigma_{ub\to td}$ is the cross section of the unpolarized t-quark production (\ref{prodcross}) and $(\sigma_R)_{ub\to td}$ is a part of this cross section that is proportional to $|f_{RV}|^2$ and $|f_{RT}|^2$, $(\sigma_{RT})_{ub\to td}$ is a part that is proportional to $V_{tb}\cdot Ref_{RT}$, $(\Gamma_{RT})_{t \to b \nu e^+}$ is a part of the top partial width that is proportional to $V_{tb}\cdot Ref_{RT}$ and $V_{tb}\cdot Imf_{RT}$, $(\sigma_{SM})_{ub\to td}$ is the SM cross section of the unpolarized t-quark production, $r_{\Gamma}=\frac{\Gamma_t}{(\Gamma_{SM})_t}$, $\Gamma_t$ is the total decay width of the t quark, taking into account the anomalous couplings and all decay modes, and $(\Gamma_{SM})_t$ is the SM total decay width of the t quark.
\\

The detailed derivation of the expression is given in the Appendix of this article.
The first term of formula (\ref{diffcrossec}) reproduces shapes of the distributions as follows from the formula for the differential width (\ref{twidth0}), whereas the second, third and fourth terms give additional contributions.

\begin{figure}
	\begin{center}
		\begin{minipage}[t]{.325\linewidth}
			\centering
			\includegraphics[width=6cm,height=6cm]{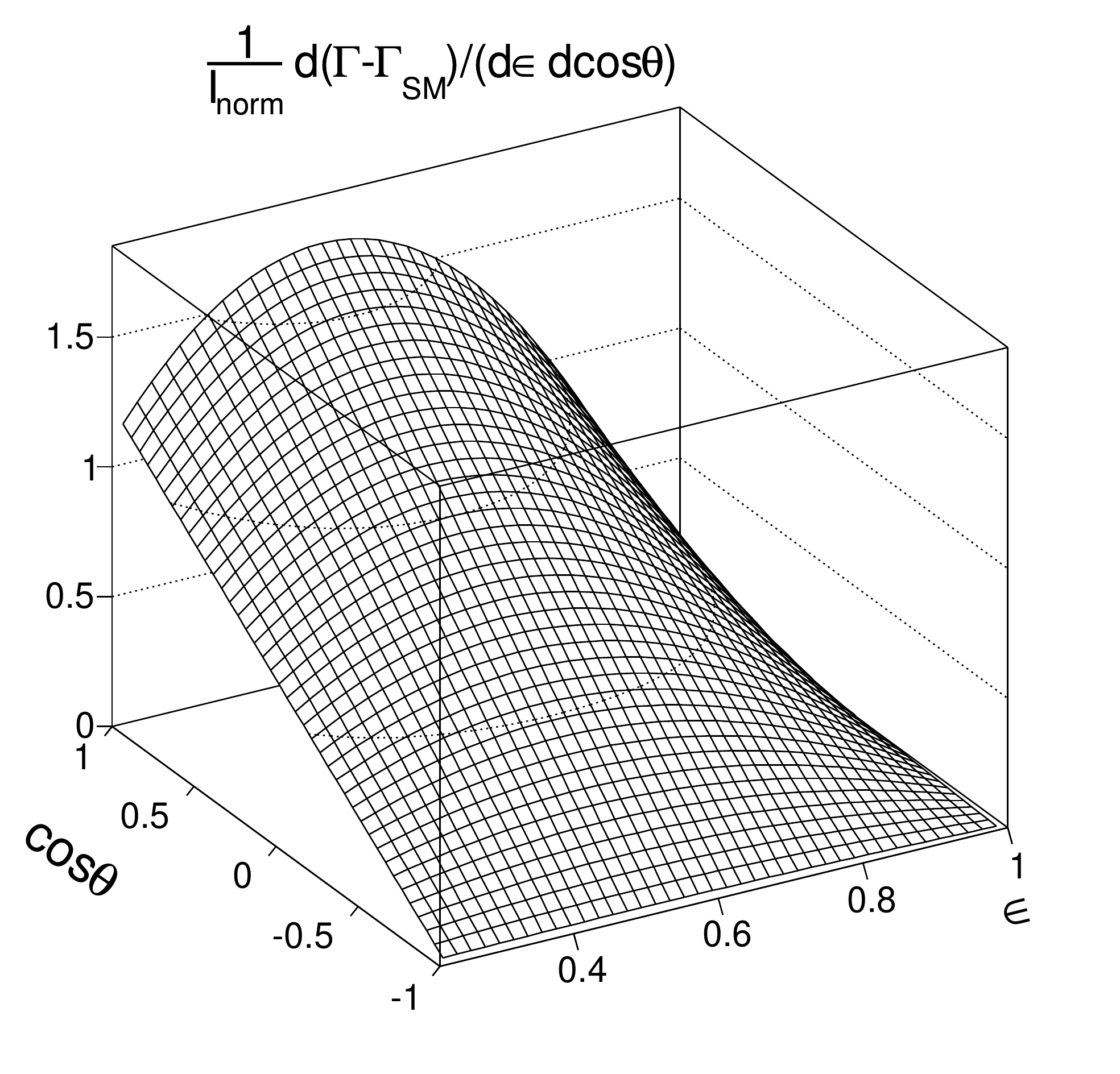}
		\end{minipage}
		\begin{minipage}[t]{.325\linewidth}
			\centering
			\includegraphics[width=6cm,height=6cm]{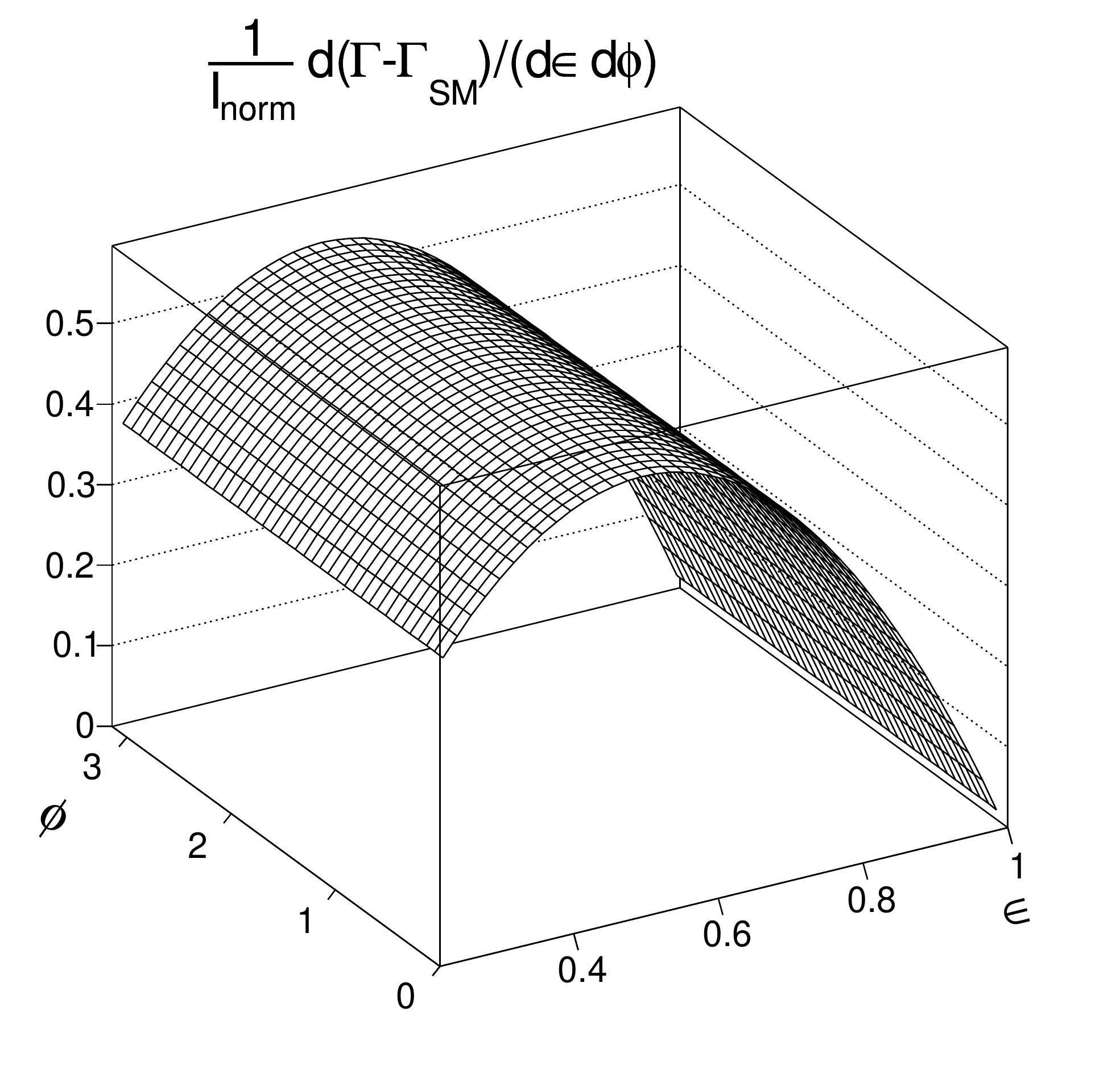}
		\end{minipage}
		\begin{minipage}[t]{.325\linewidth}
			\centering
			\includegraphics[width=6cm,height=6cm]{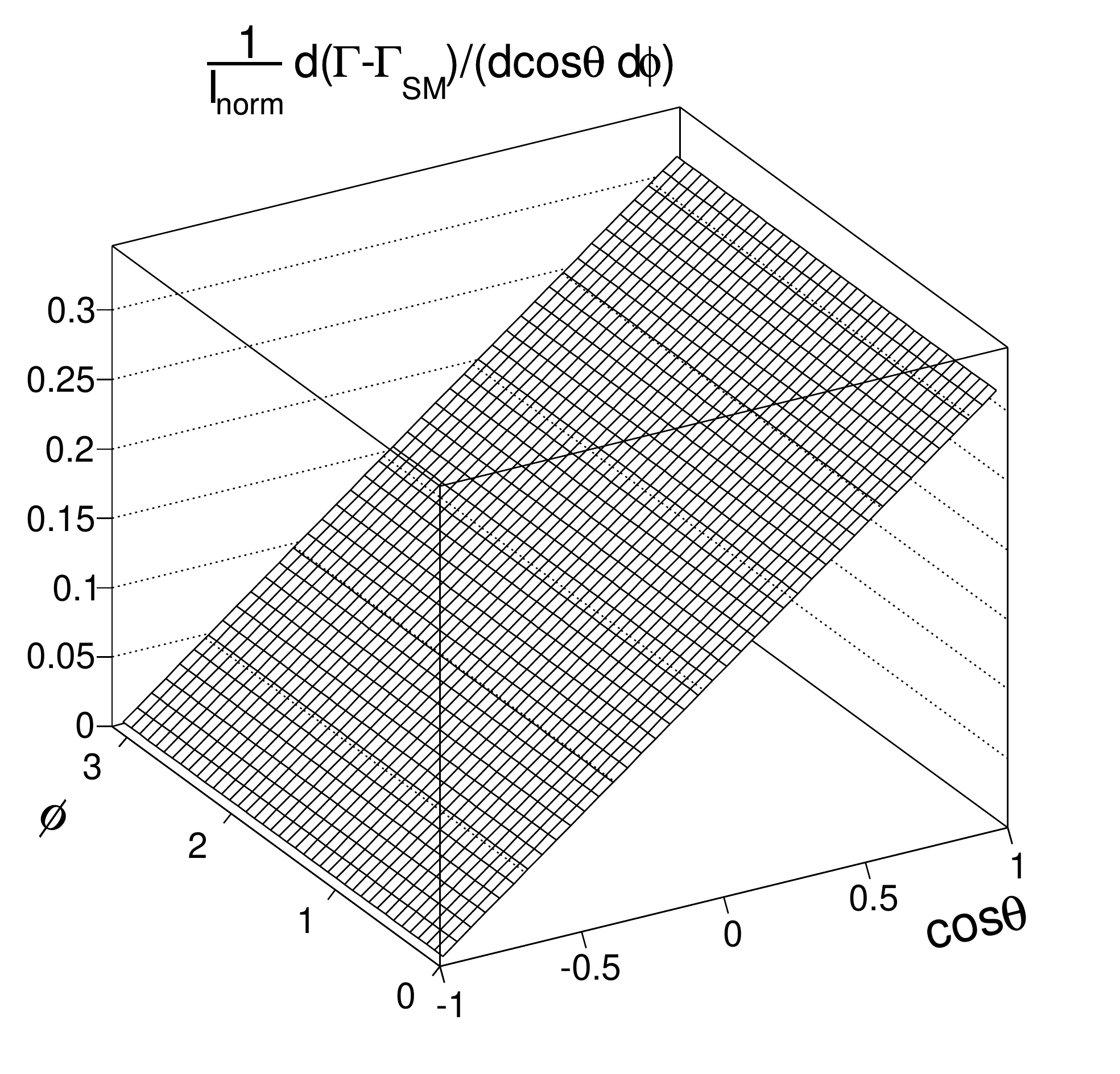}
		\end{minipage}
	\\
		\begin{minipage}[t]{.325\linewidth}
			\centering
			\includegraphics[width=6cm,height=6cm]{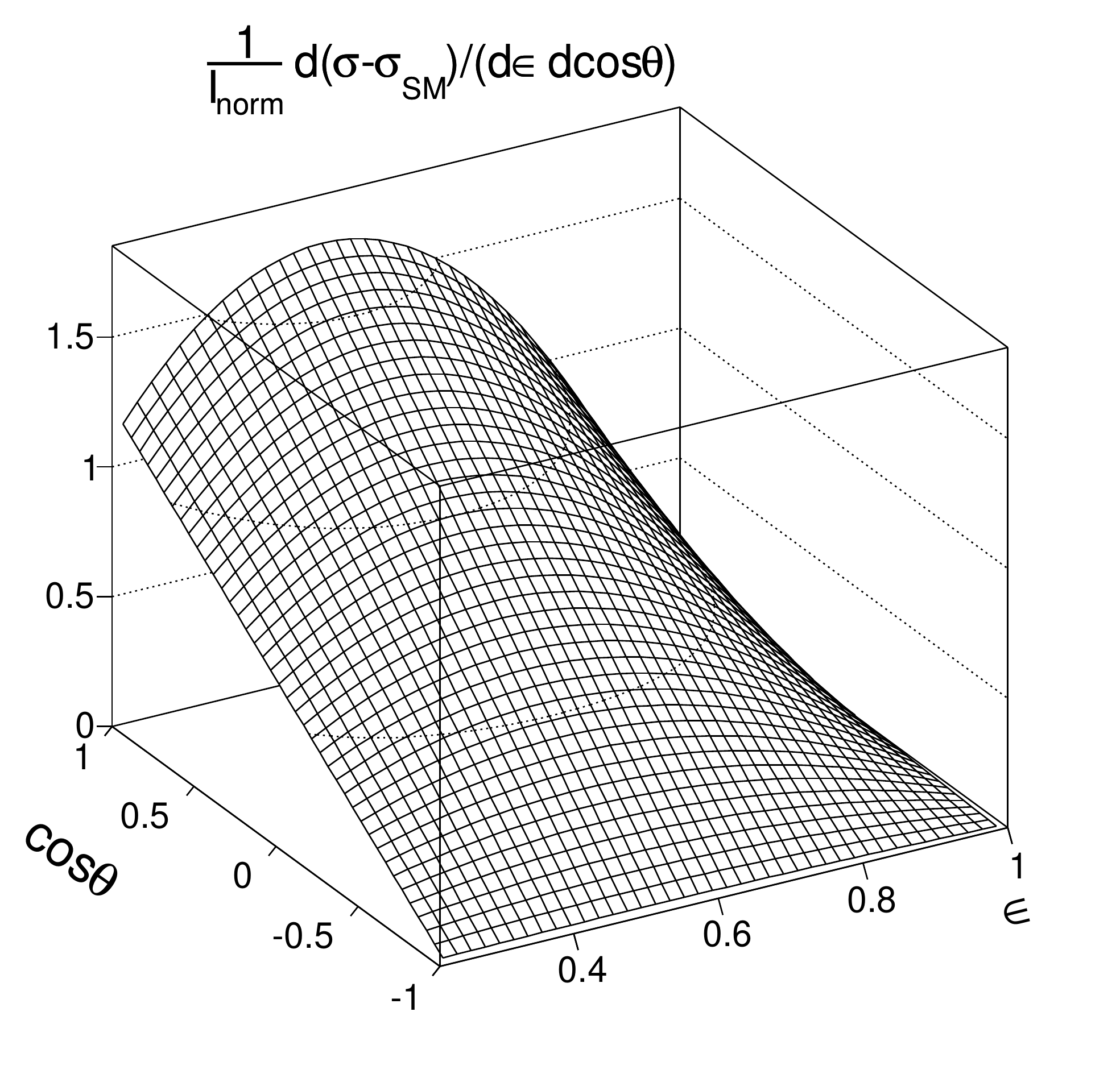}
		\end{minipage}
		\begin{minipage}[t]{.325\linewidth}
			\centering
			\includegraphics[width=6cm,height=6cm]{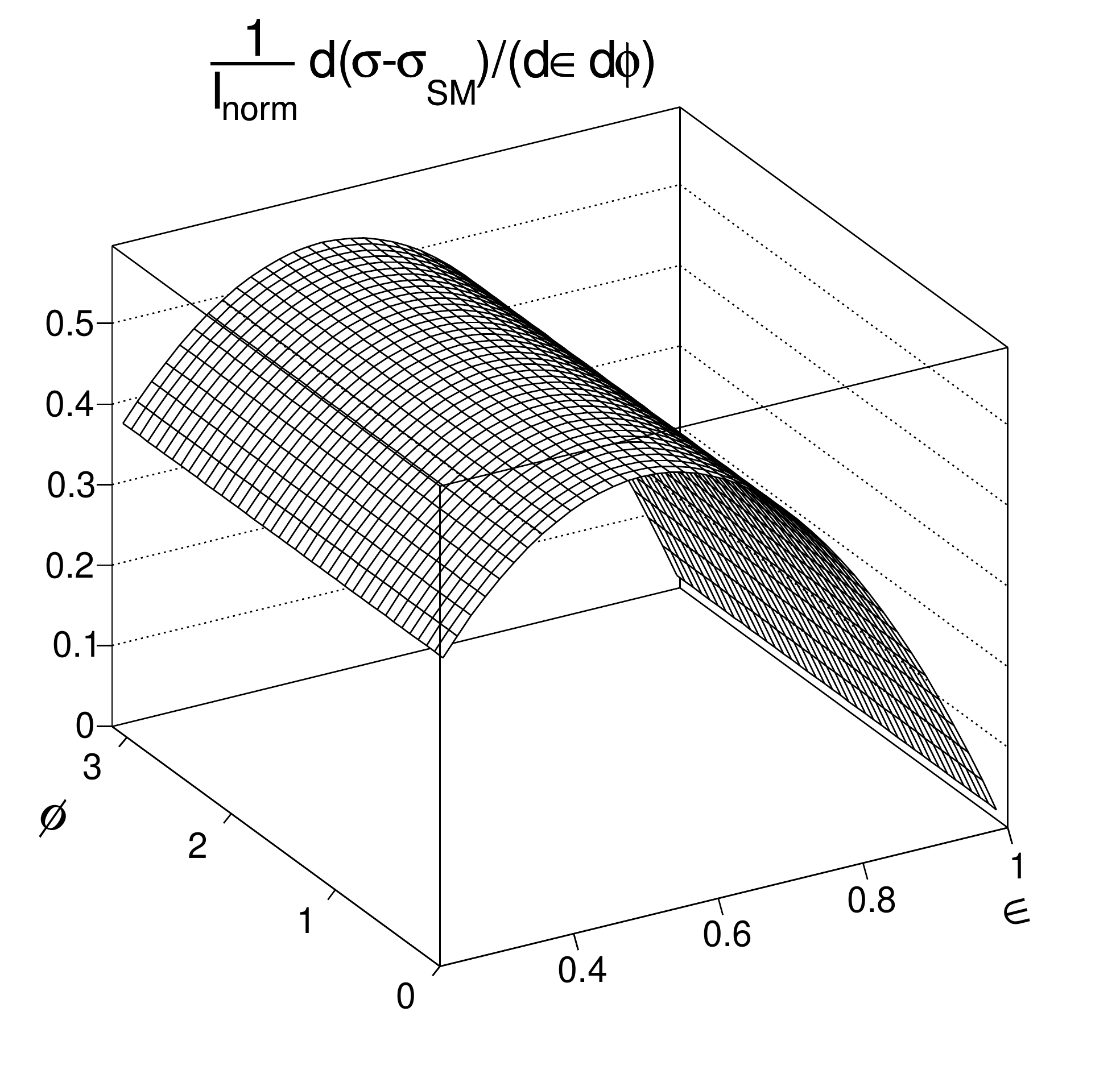}
		\end{minipage}
		\begin{minipage}[t]{.325\linewidth}
			\centering
			\includegraphics[width=6cm,height=6cm]{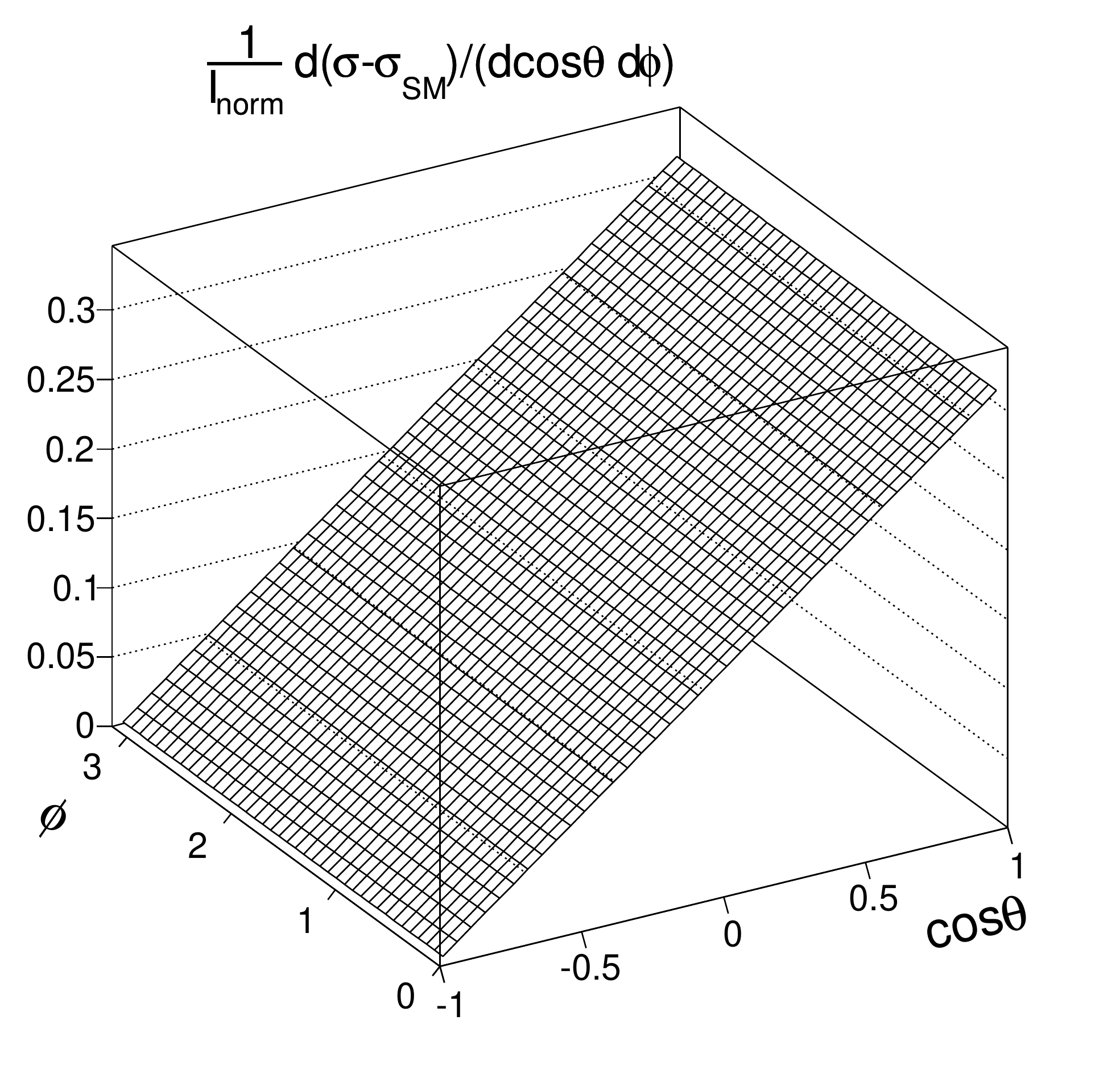}
		\end{minipage}
	\\	
		\begin{minipage}[t]{.325\linewidth}
			\centering
			\includegraphics[width=6cm,height=6cm]{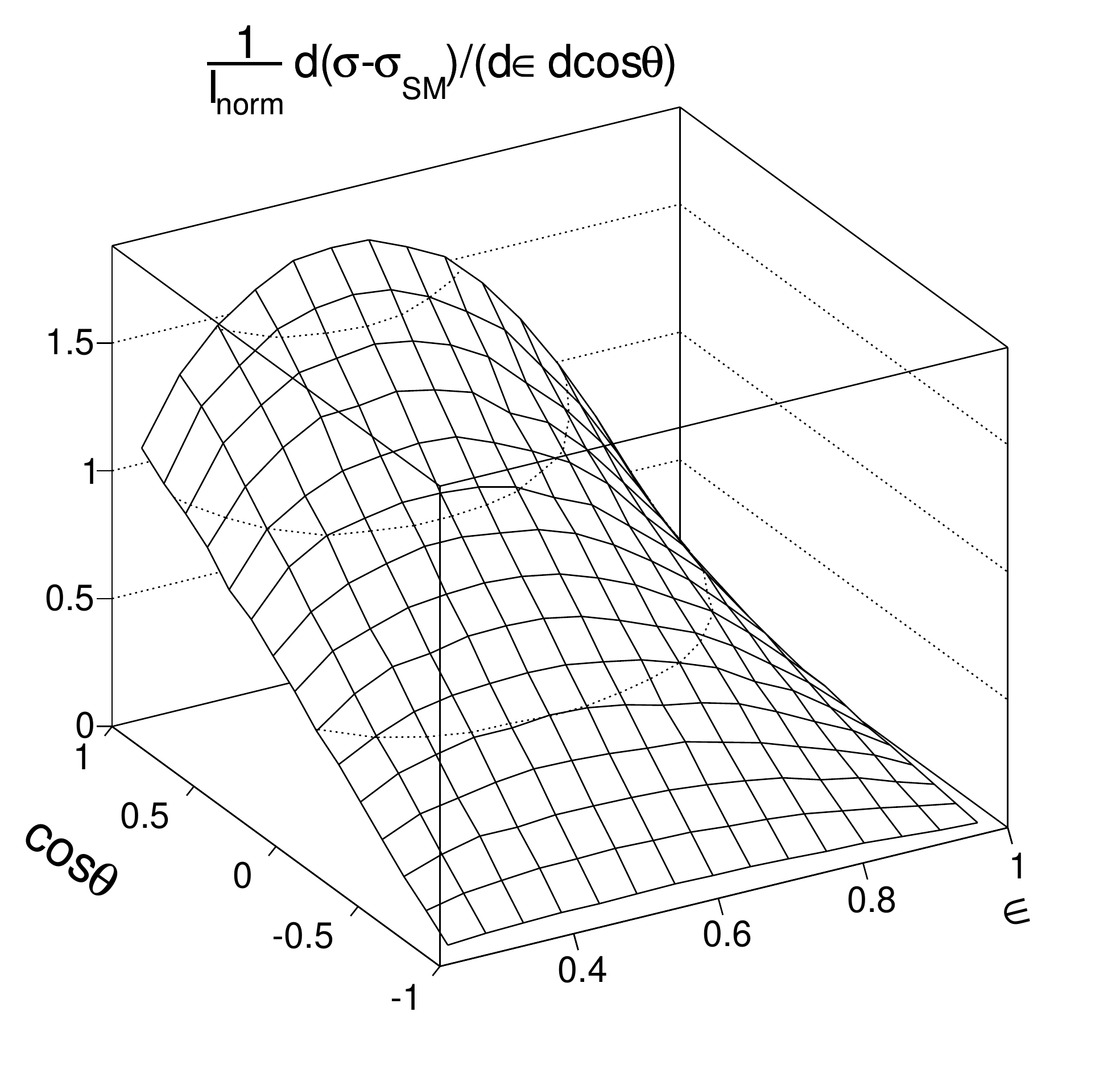}
		\end{minipage}
		\begin{minipage}[t]{.325\linewidth}
			\centering
			\includegraphics[width=6cm,height=6cm]{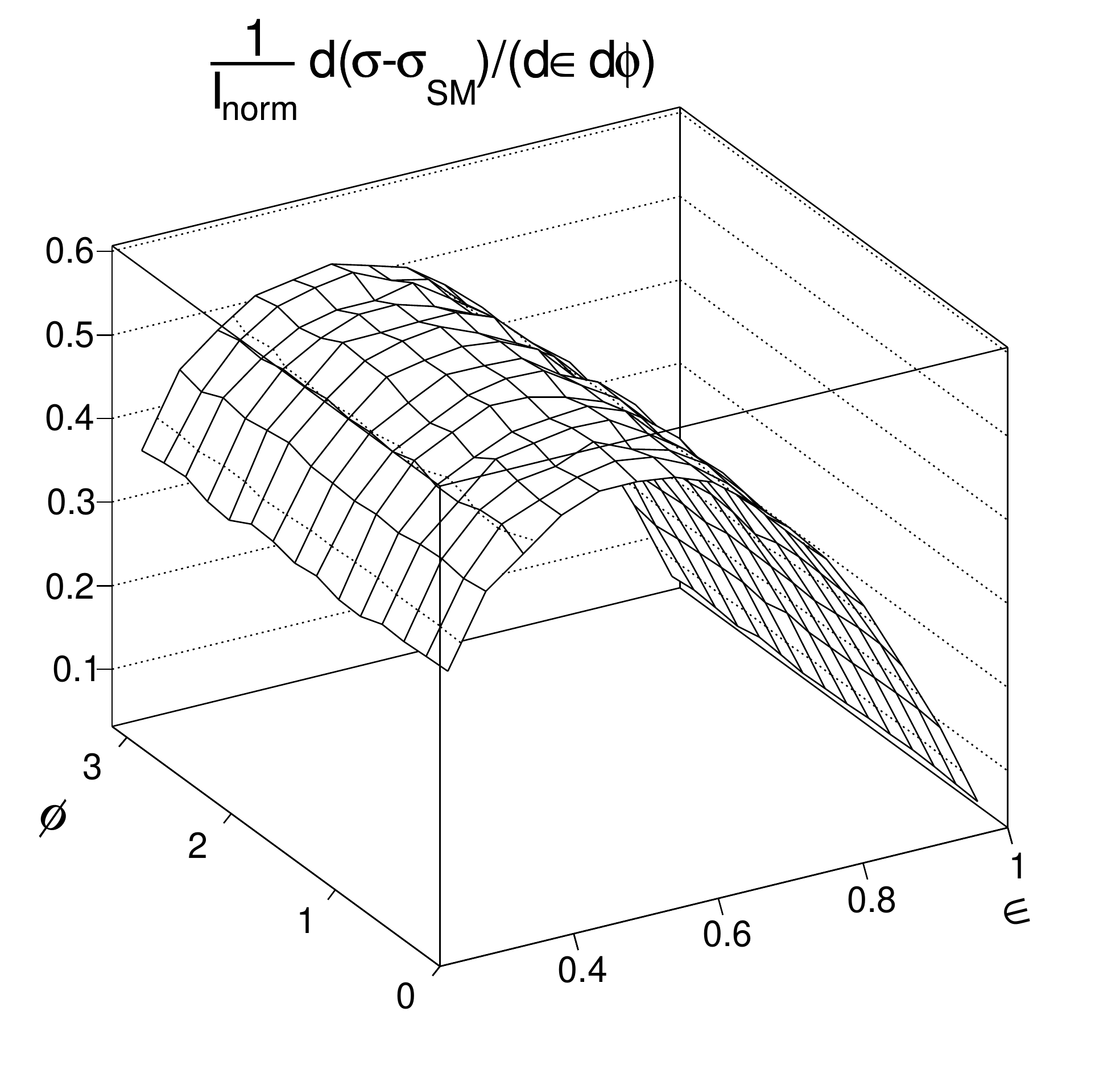}
		\end{minipage}
		\begin{minipage}[t]{.325\linewidth}
		    \centering
		    \includegraphics[width=6cm,height=6cm]{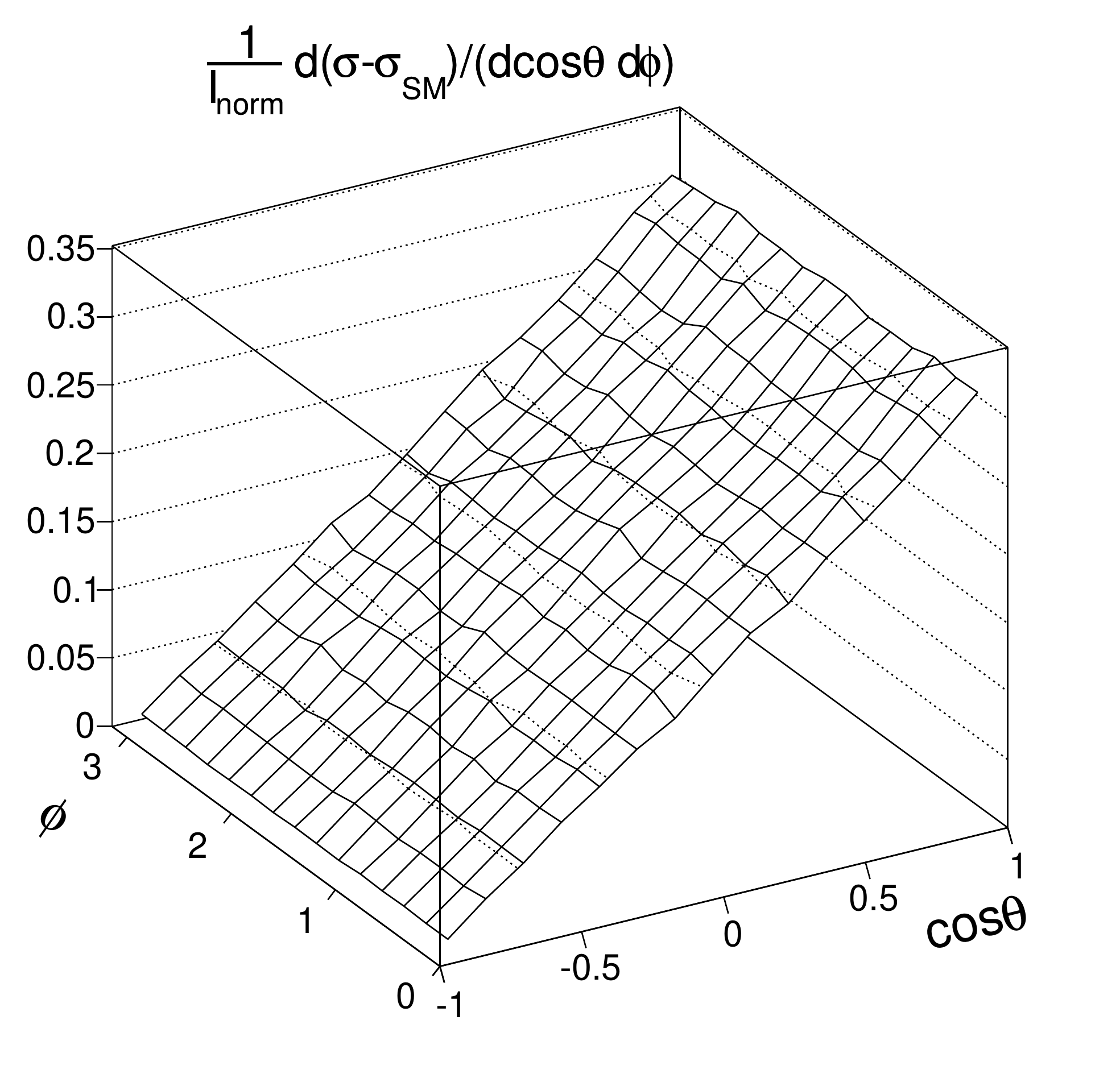}
	    \end{minipage}
  	\\			
	\end{center}
	\caption{ \label{pic6} \footnotesize  Scenario $Ref_{LV}$ = 1.03. The upper figures show plots of the normalized double-differential t-quark decay partial width $\frac{d(\Gamma-\Gamma_{SM})}{d\epsilon~\cdot~d\cos\theta}$, ~$\frac{d(\Gamma-\Gamma_{SM})}{d\epsilon~\cdot~d\phi}$, and $\frac{d(\Gamma-\Gamma_{SM})}{d\cos\theta~\cdot~d\phi}$ built from formulas (\ref{twidth1}), (\ref{twidth2}), (\ref{twidth3}). The middle figures show plots of the normalized double-differential cross sections $\frac{d(\sigma-\sigma_{SM})}{d\epsilon~\cdot~d\cos\theta}$, ~$\frac{d(\sigma-\sigma_{SM})}{d\epsilon~\cdot~d\phi}$, and $\frac{d(\sigma-\sigma_{SM})}{d\cos\theta~\cdot~d\phi}$ built from formulas (\ref{totalcrossec1}), (\ref{totalcrossec2}), (\ref{totalcrossec3}). The lower figures show plots of the normalized double-differential cross sections built from Monte Carlo events.}
\end{figure}

As the first scenario, we consider the case where the left-vector anomalous coupling ($Ref_{LV}-V_{tb}$) is not equal to 0, and the remaining anomalous couplings are equal to 0. In this case, the last two terms of the expression (\ref{diffcrossec}) are zero, and the first two terms reproduce the same shapes of the surfaces as those for the Standard Model with 100\% spin correlation behavior $(1 + \cos\theta)$, the maximum of the energy distribution at $E_{e^+}=m_t/4\approx43$ GeV, and no dependence on the $\phi$ angle. Figure \ref{pic6} shows normalized distributions corresponding to this scenario. 
The upper figures show plots of the normalized double-differential t-quark decay partial width. The middle figures show plots of the normalized double-differential cross sections and the lower figures show plots of the normalized double-differential cross sections built from Monte Carlo events. 
 
\begin{figure}
	\begin{center}
		\begin{minipage}[t]{.325\linewidth}
			\centering
			\includegraphics[width=6cm,height=6cm]{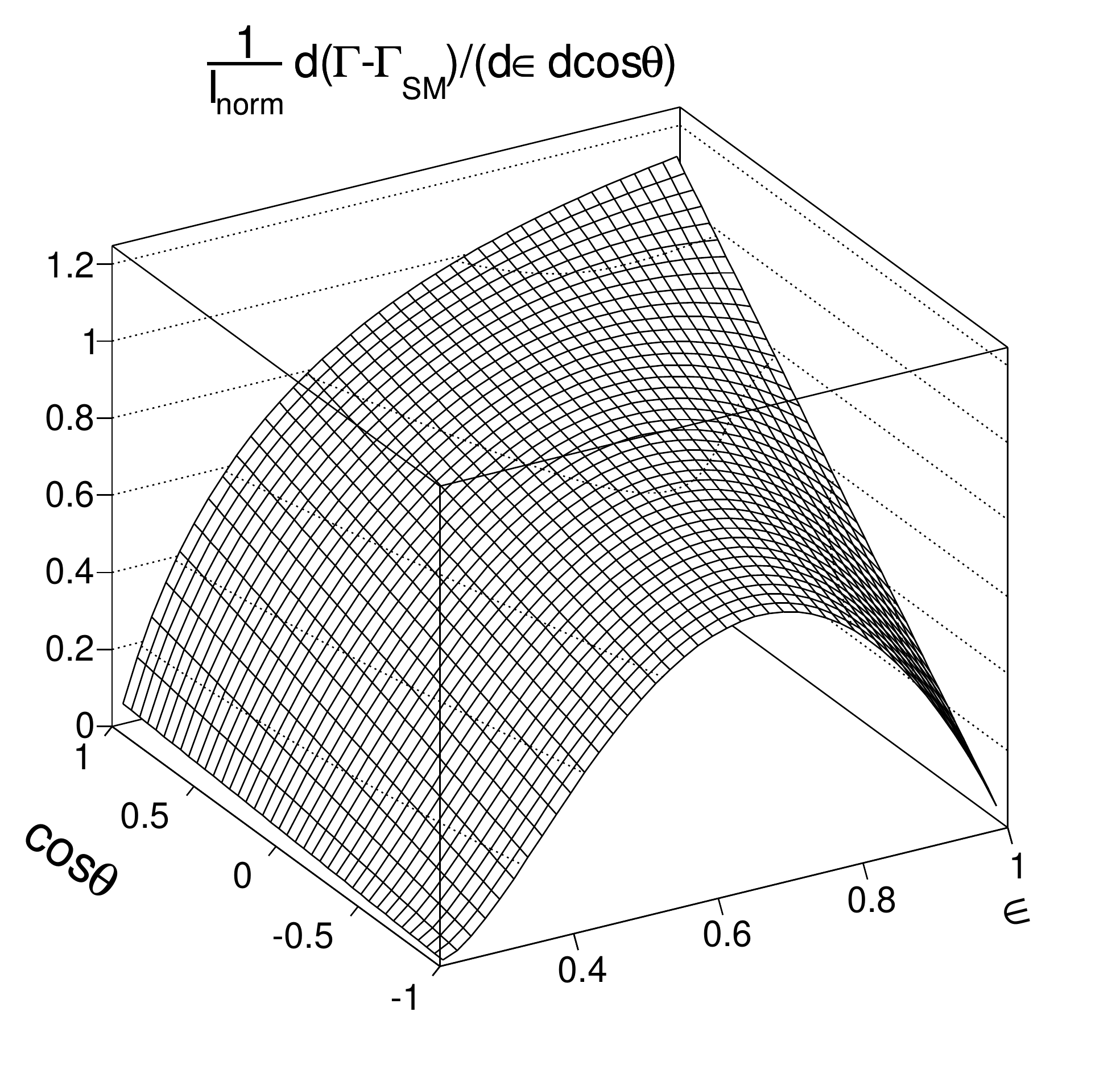}
		\end{minipage}
		\begin{minipage}[t]{.325\linewidth}
			\centering
			\includegraphics[width=6cm,height=6cm]{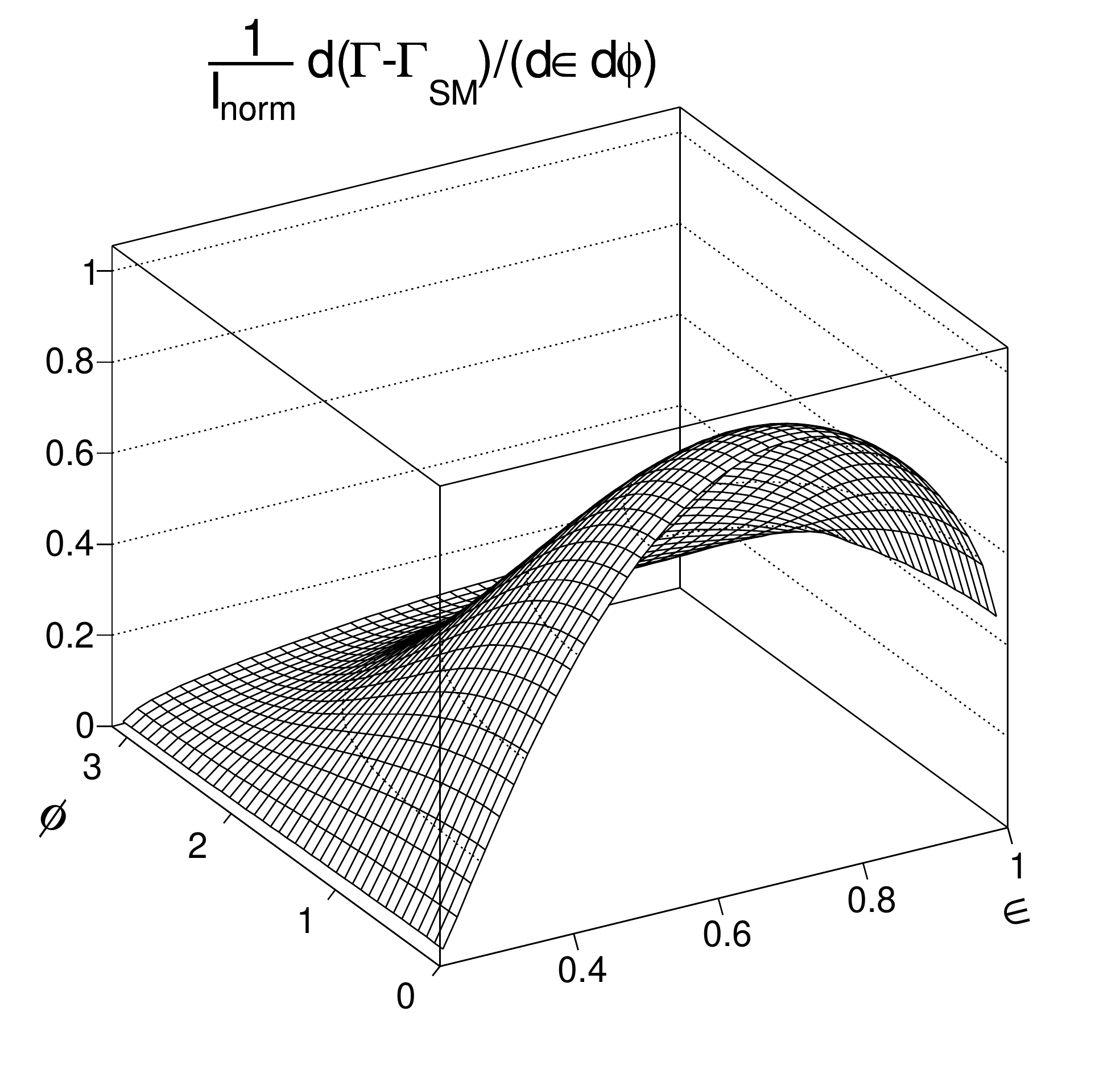}
		\end{minipage}
		\begin{minipage}[t]{.325\linewidth}
			\centering
			\includegraphics[width=6cm,height=6cm]{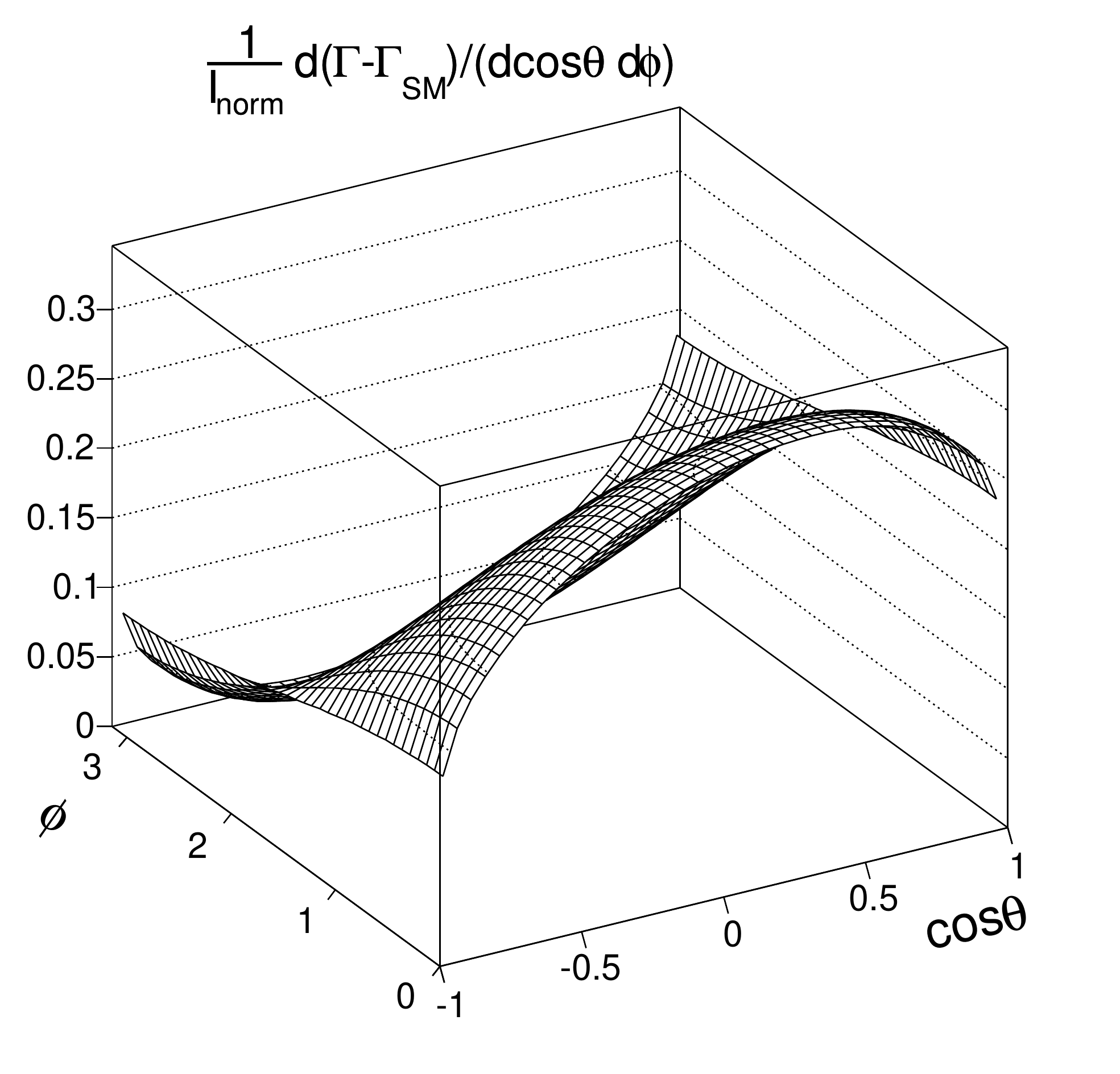}
		\end{minipage}
	\\
		\begin{minipage}[t]{.325\linewidth}
			\centering
			\includegraphics[width=6cm,height=6cm]{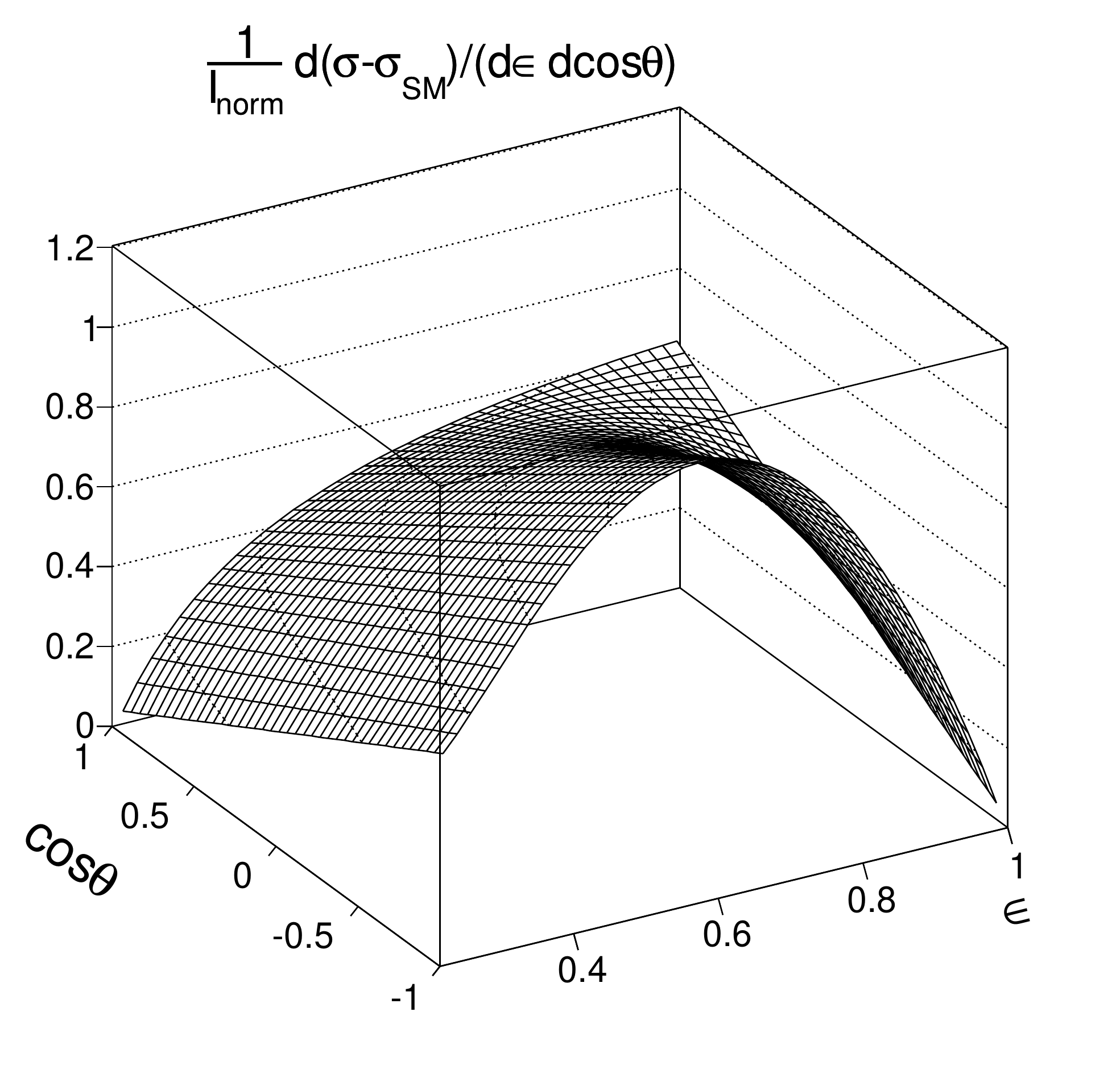}
		\end{minipage}
		\begin{minipage}[t]{.325\linewidth}
			\centering
			\includegraphics[width=6cm,height=6cm]{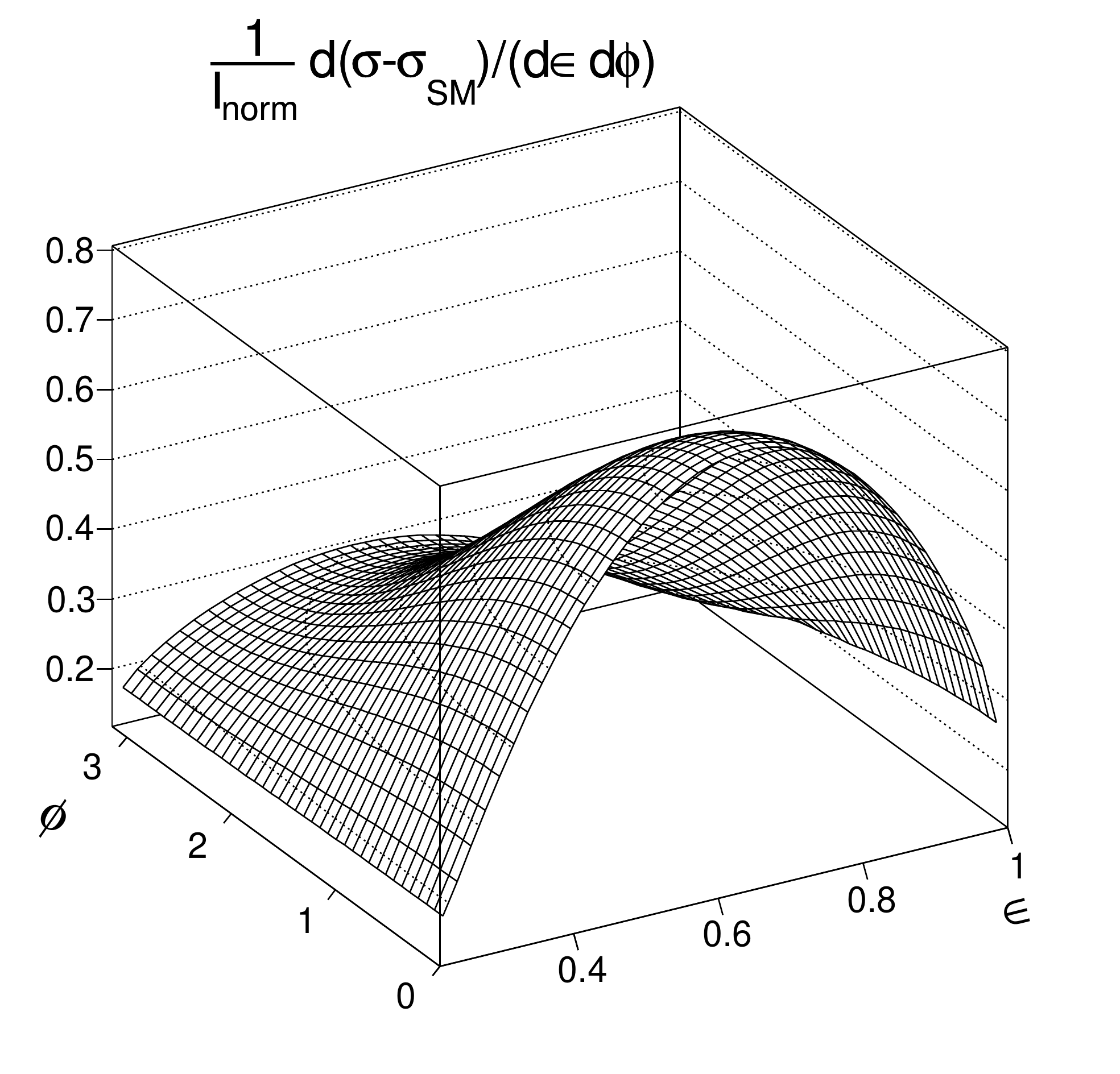}
		\end{minipage}
		\begin{minipage}[t]{.325\linewidth}
			\centering
			\includegraphics[width=6cm,height=6cm]{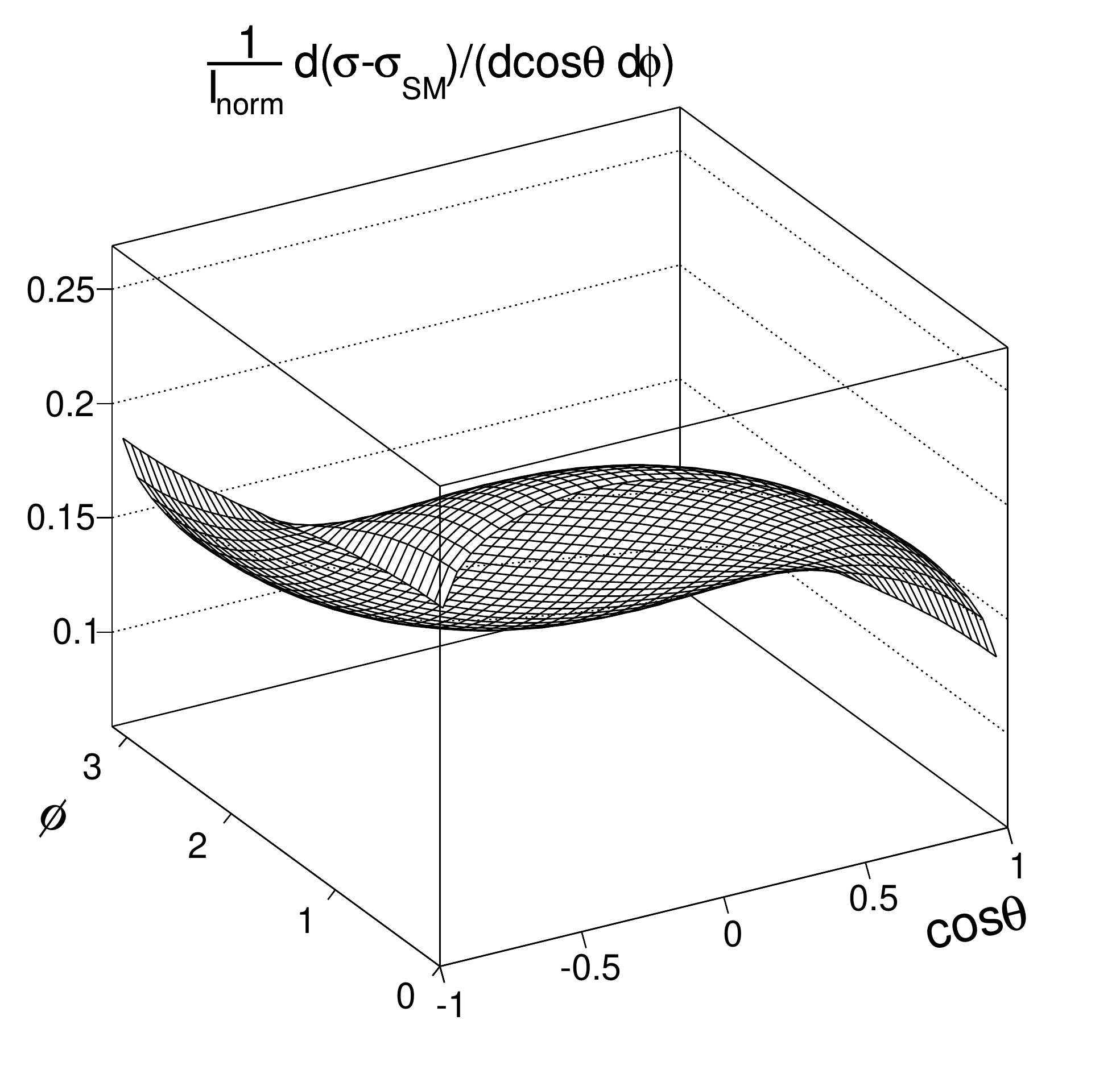}
		\end{minipage}
	\\
		\begin{minipage}[t]{.325\linewidth}
			\centering
			\includegraphics[width=6cm,height=6cm]{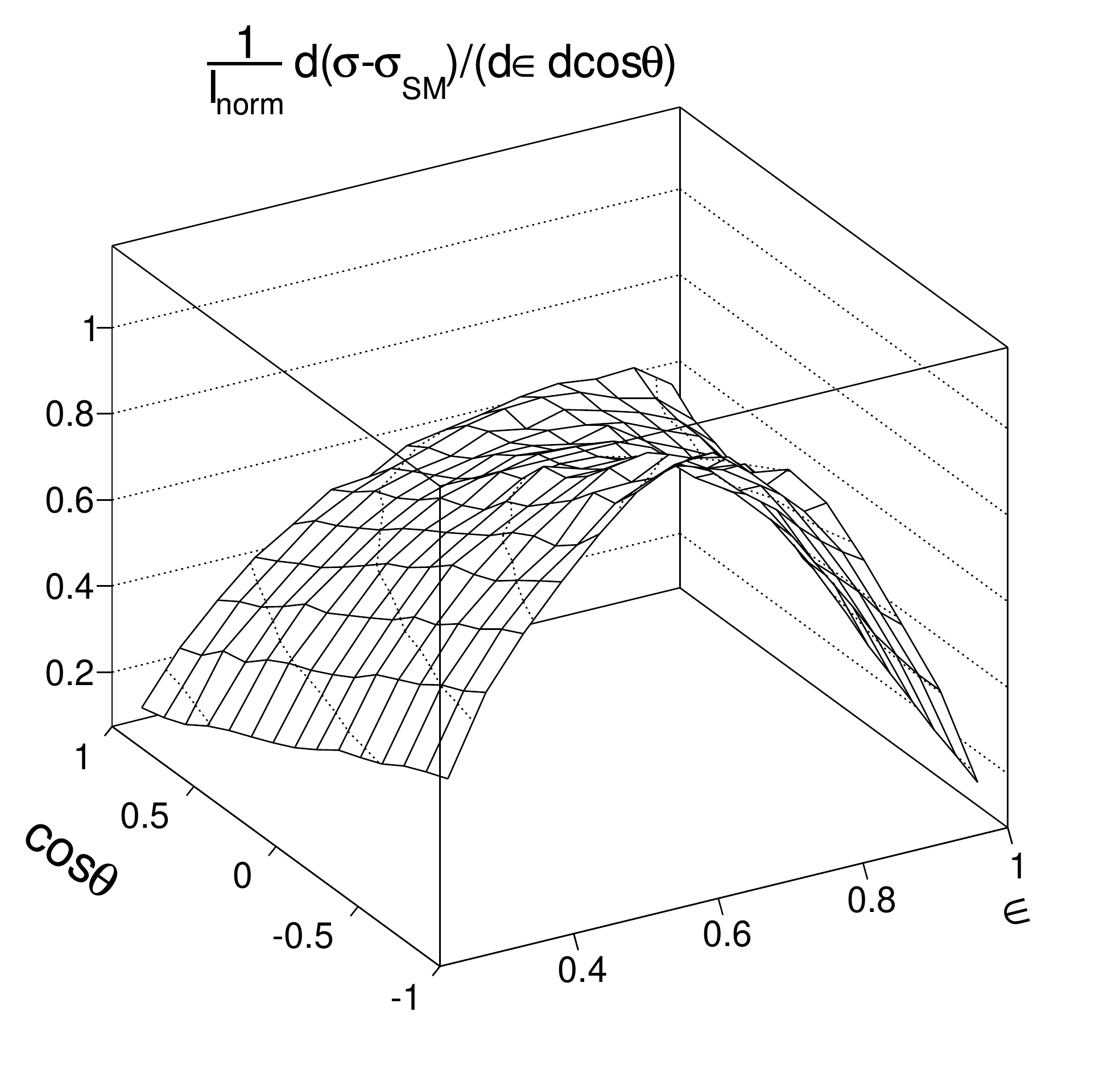}
		\end{minipage}
		\begin{minipage}[t]{.325\linewidth}
			\centering
			\includegraphics[width=6cm,height=6cm]{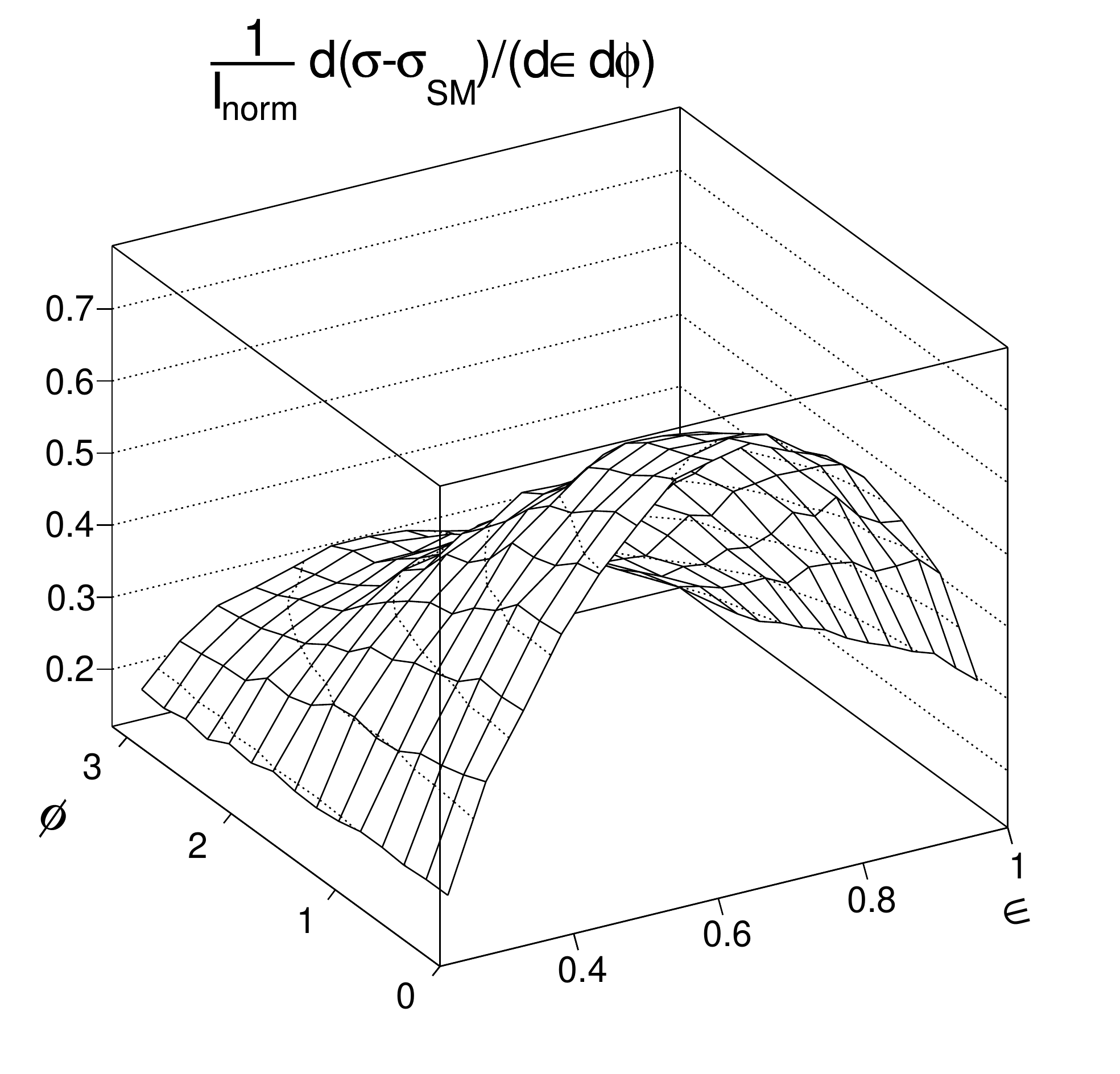}
		\end{minipage}
		\begin{minipage}[t]{.325\linewidth}
			\centering
			\includegraphics[width=6cm,height=6cm]{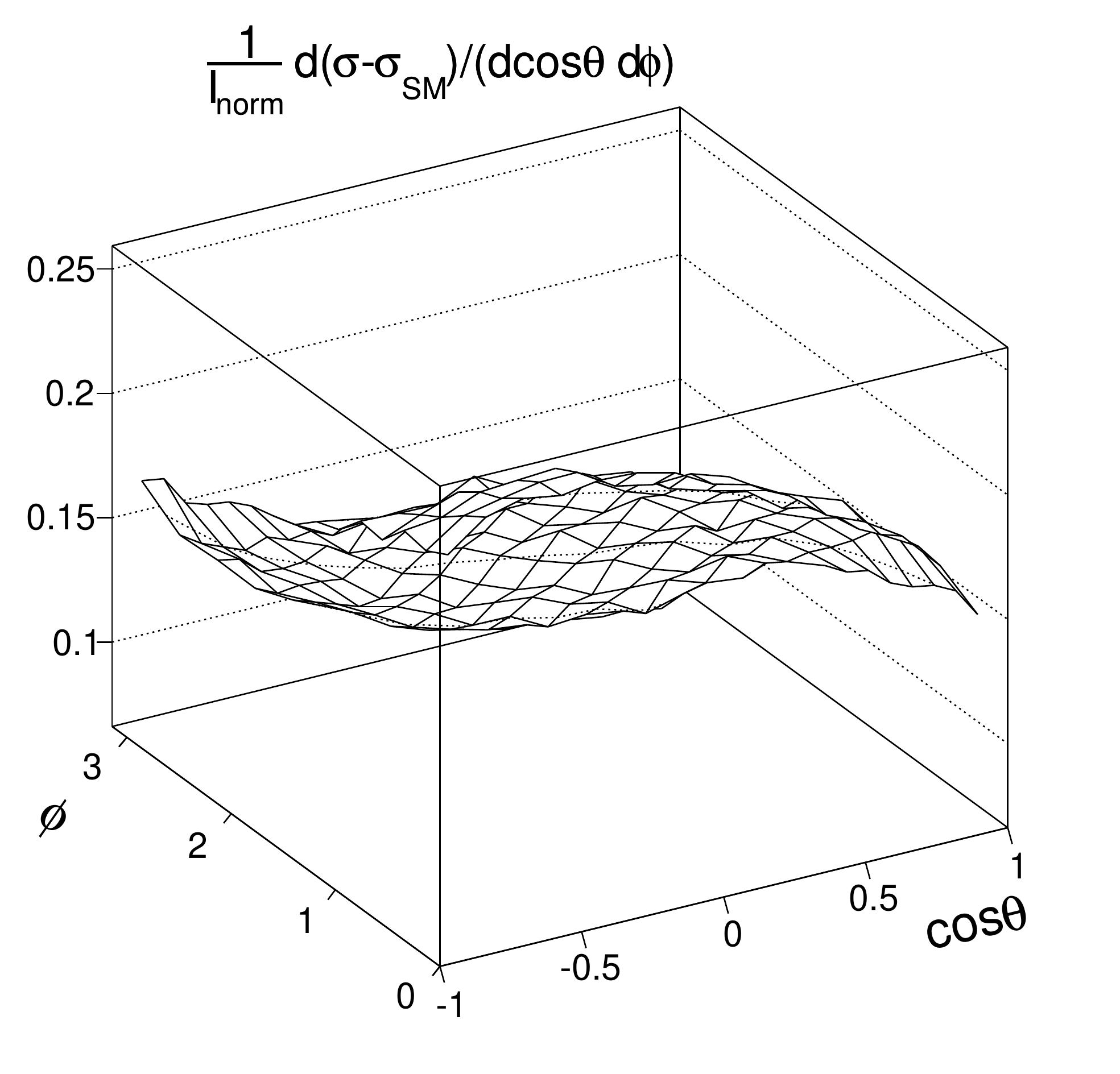}
		\end{minipage}
	\\				
	\end{center}
	\caption{ \label{pic7} \footnotesize  Scenario $Ref_{RV}$ = 0.16. The upper figures show plots of the normalized double-differential t-quark decay partial width $\frac{d(\Gamma-\Gamma_{SM})}{d\epsilon~\cdot~d\cos\theta}$, ~$\frac{d(\Gamma-\Gamma_{SM})}{d\epsilon~\cdot~d\phi}$, and $\frac{d(\Gamma-\Gamma_{SM})}{d\cos\theta~\cdot~d\phi}$ built from formulas (\ref{twidth1}), (\ref{twidth2}), (\ref{twidth3}). The middle figures show plots of the normalized double-differential cross sections $\frac{d(\sigma-\sigma_{SM})}{d\epsilon~\cdot~d\cos\theta}$, ~$\frac{d(\sigma-\sigma_{SM})}{d\epsilon~\cdot~d\phi}$, and $\frac{d(\sigma-\sigma_{SM})}{d\cos\theta~\cdot~d\phi}$ built from formulas (\ref{totalcrossec1}), (\ref{totalcrossec2}), (\ref{totalcrossec3}). The lower figures show plots of the normalized double-differential cross sections built from Monte Carlo events.}
\end{figure}

For the second scenario, $Ref_{RV}$ is set to be nonzero while the remaining anomalous couplings are taken to be zero.
Unlike the previous one, in this scenario, all terms of formula (\ref{diffcrossec}) have different behavior with respect to the kinematic variables and give contributions to two-dimensional surfaces shown in Fig.~\ref{pic7}. The first term (\ref{diffcrossec}) depends on used variables 
$\epsilon$, $\cos\theta$, and the $\phi$ angle and reproduces the same shapes as those from the differential top width (upper plot in Fig.~\ref{pic7}).
The second term in (\ref{diffcrossec}) gives the shapes as for the SM slightly affecting the common shapes.
The third term is zero. The fourth term does not depend on the $\phi$ angle, but being proportional to the $|f_{RV}|^2$ coupling changes the overall shapes significantly.  Since this term is proportional to $\cos\theta$, its influence is most noticeable in $\cos\theta$ distributions (Fig.~\ref{pic7} middle left and middle right).

\begin{figure}
	\begin{center}
		\begin{minipage}[t]{.325\linewidth}
			\centering
			\includegraphics[width=6cm,height=6cm]{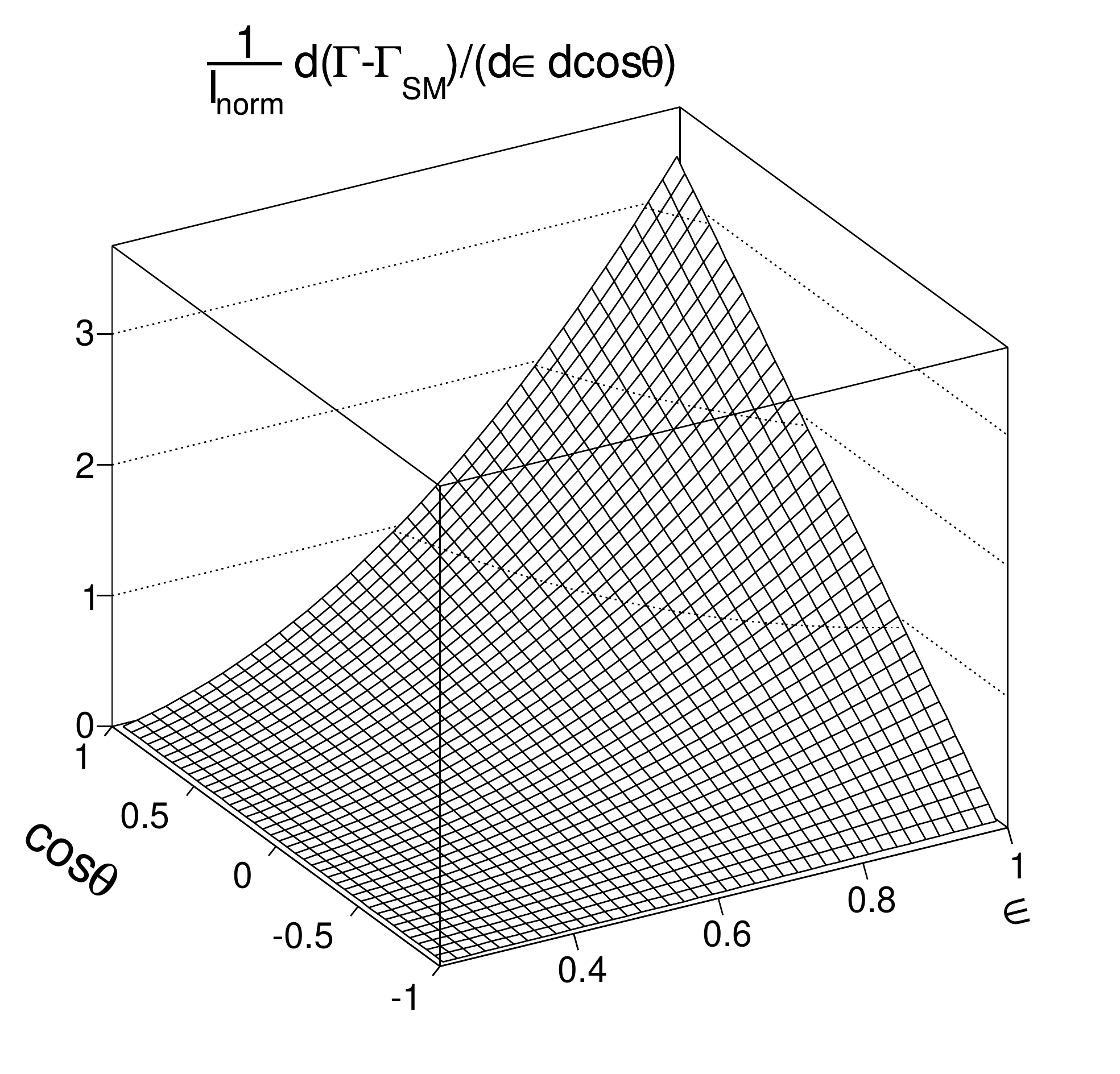}
		\end{minipage}
		\begin{minipage}[t]{.325\linewidth}
			\centering
			\includegraphics[width=6cm,height=6cm]{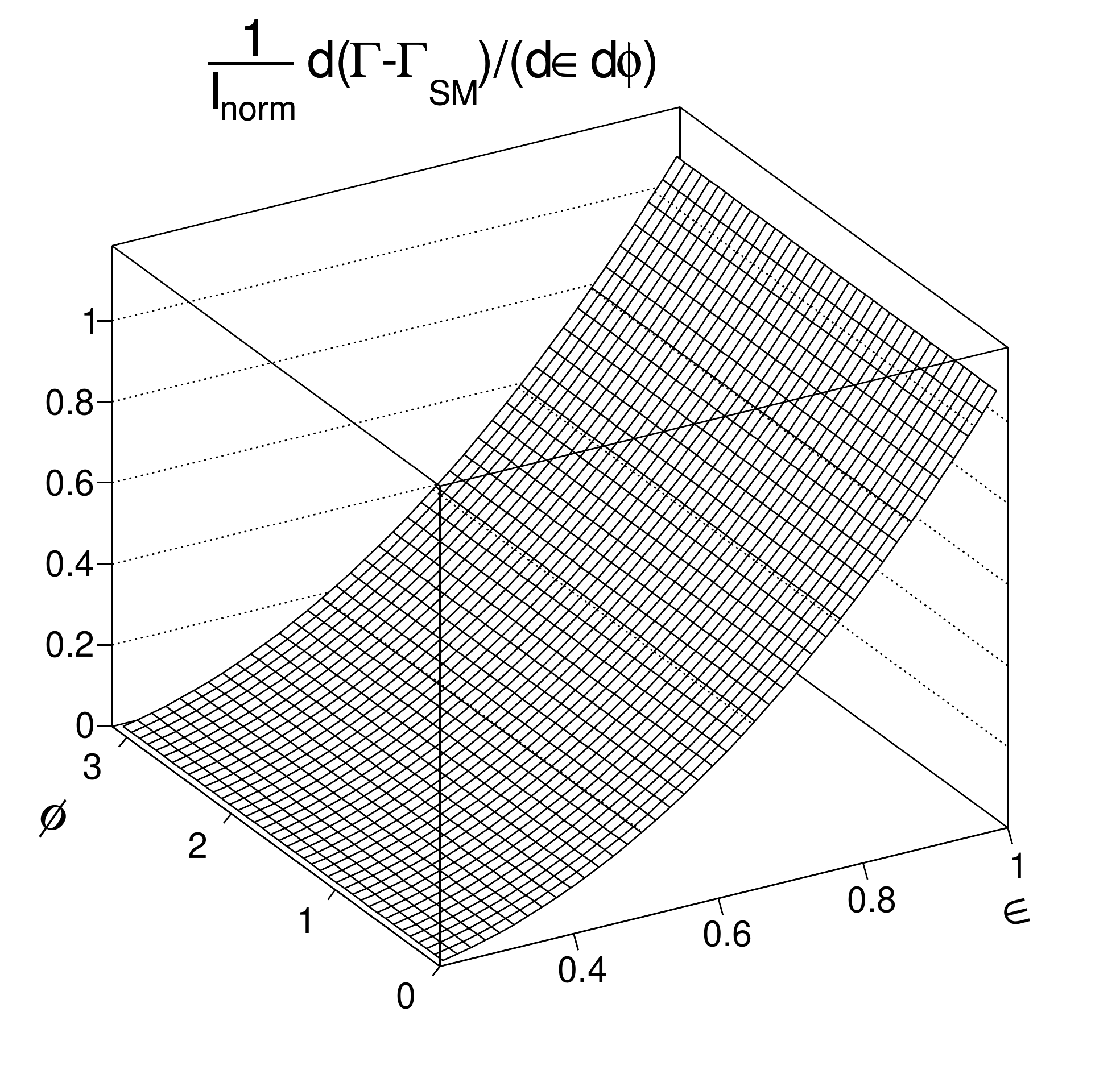}
		\end{minipage}
		\begin{minipage}[t]{.325\linewidth}
			\centering
			\includegraphics[width=6cm,height=6cm]{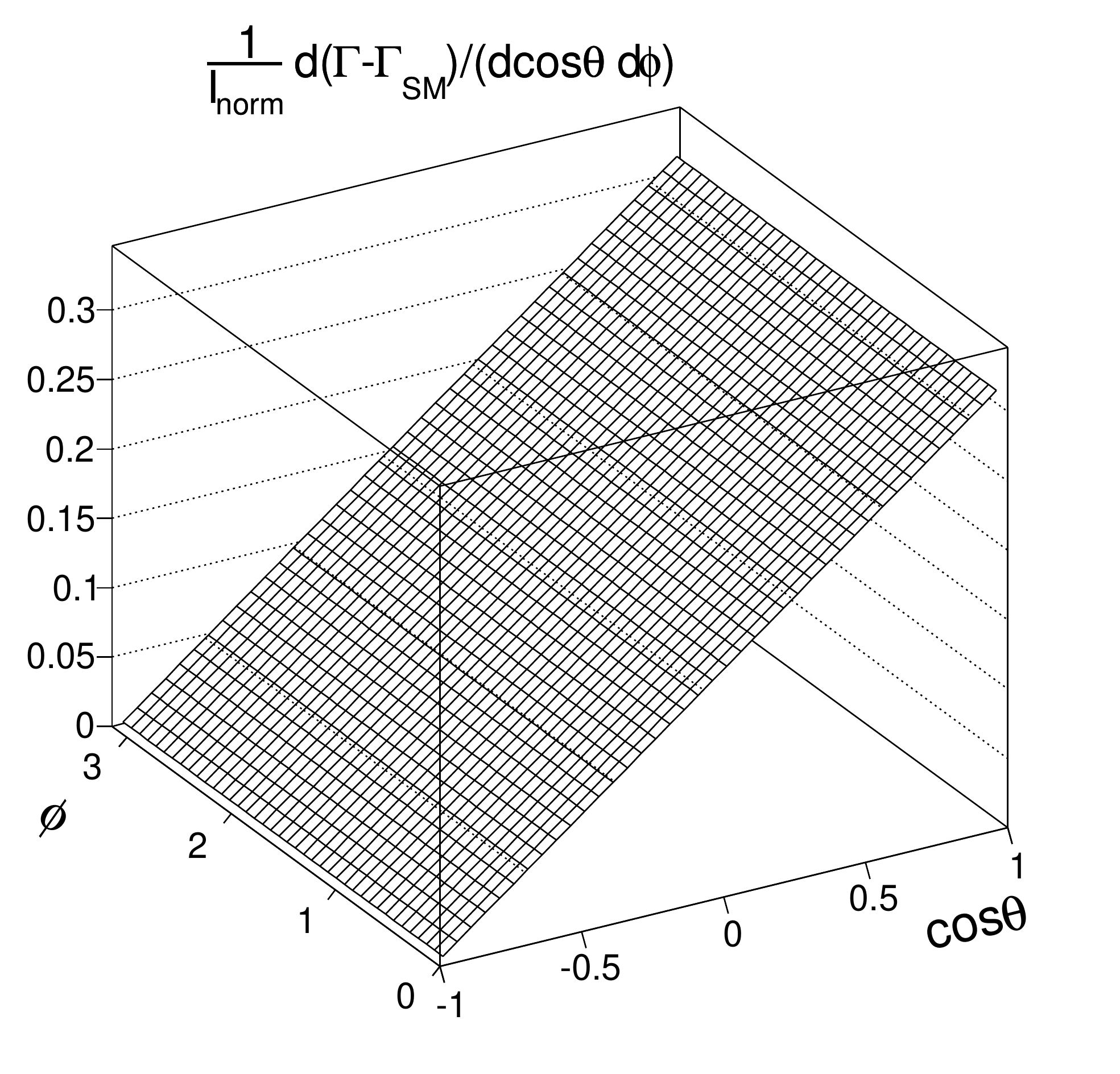}
		\end{minipage}
	\\
		\begin{minipage}[t]{.325\linewidth}
			\centering
			\includegraphics[width=6cm,height=6cm]{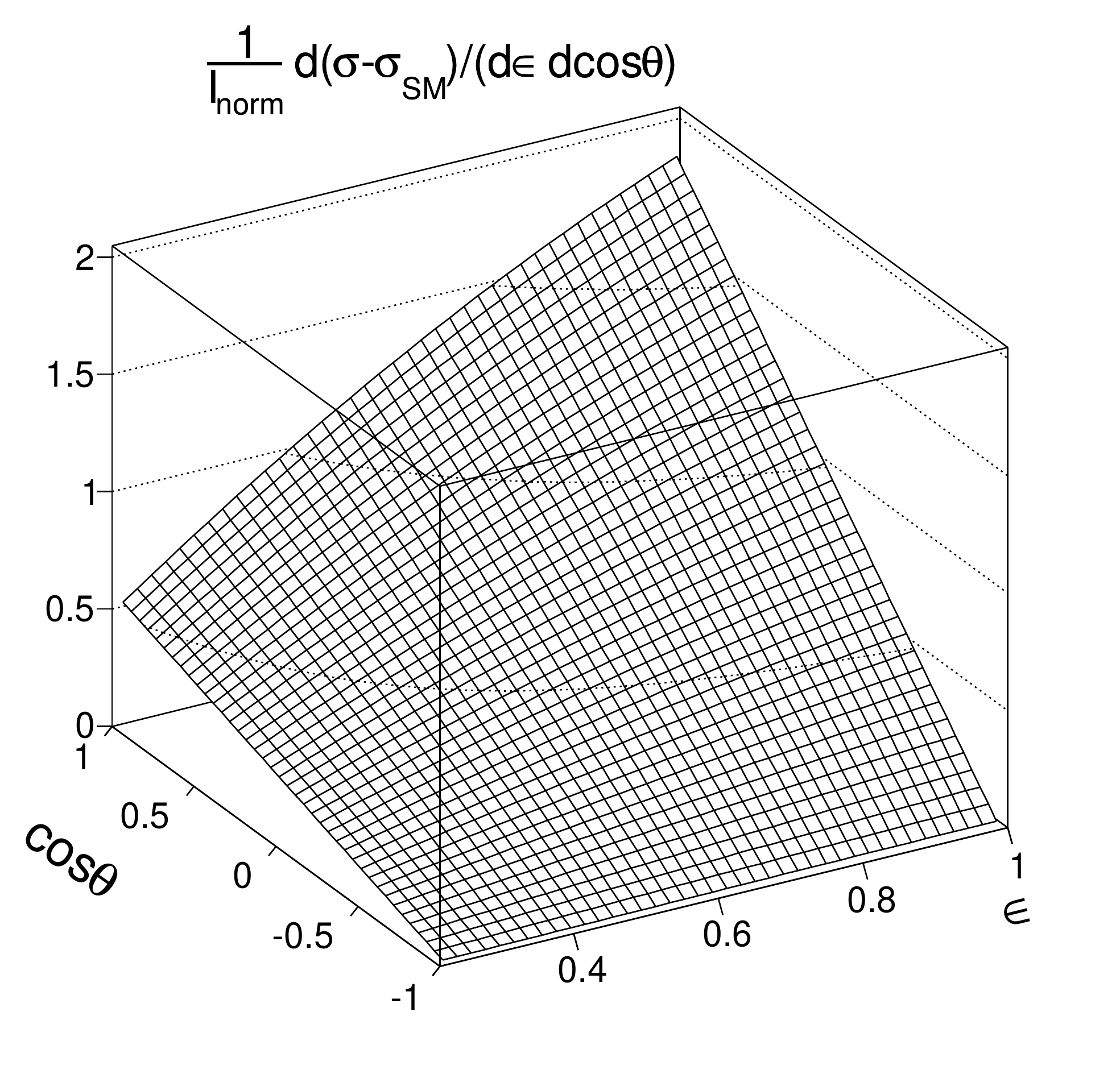}
		\end{minipage}
		\begin{minipage}[t]{.325\linewidth}
			\centering
			\includegraphics[width=6cm,height=6cm]{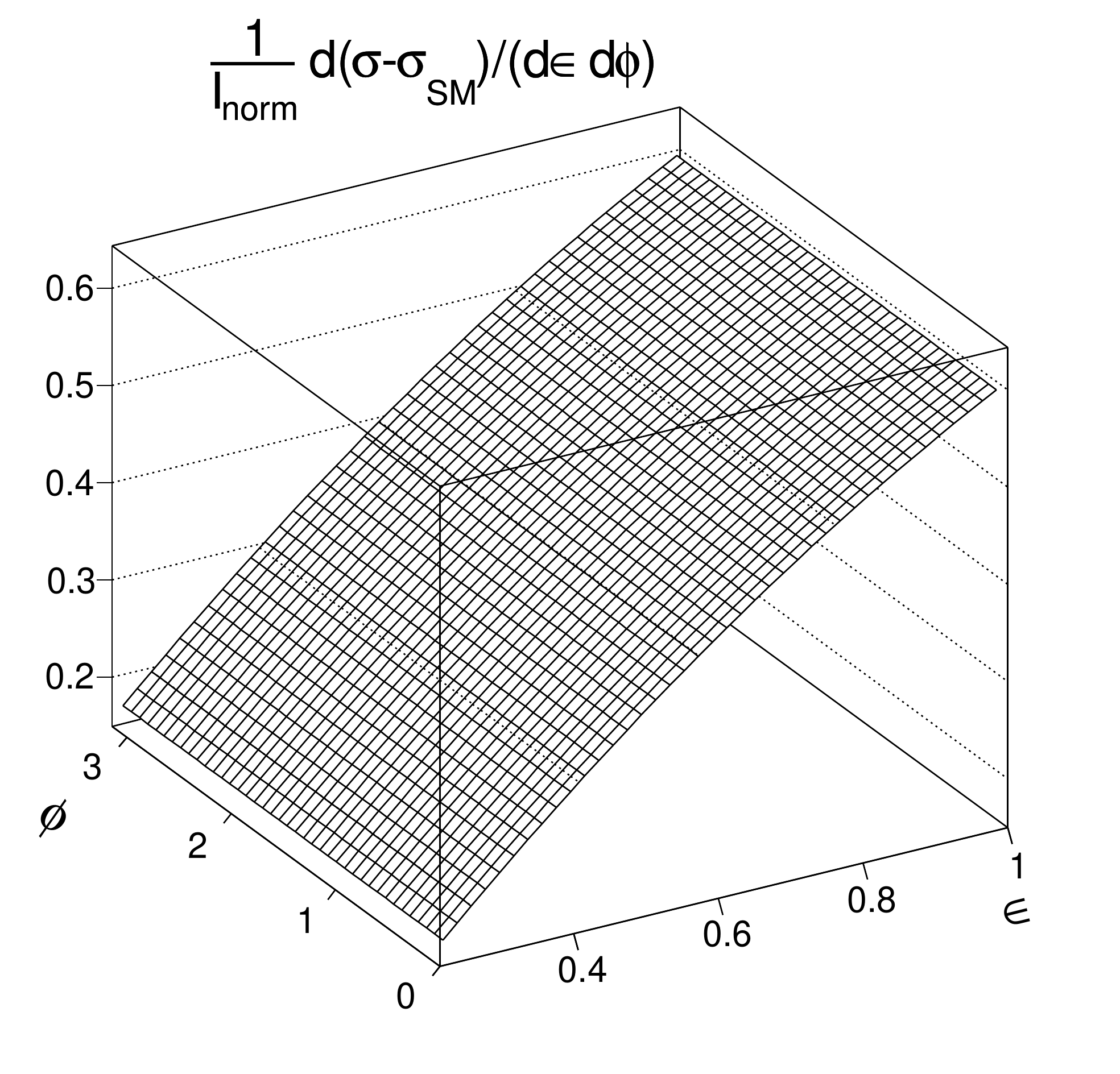}
		\end{minipage}
		\begin{minipage}[t]{.325\linewidth}
			\centering
			\includegraphics[width=6cm,height=6cm]{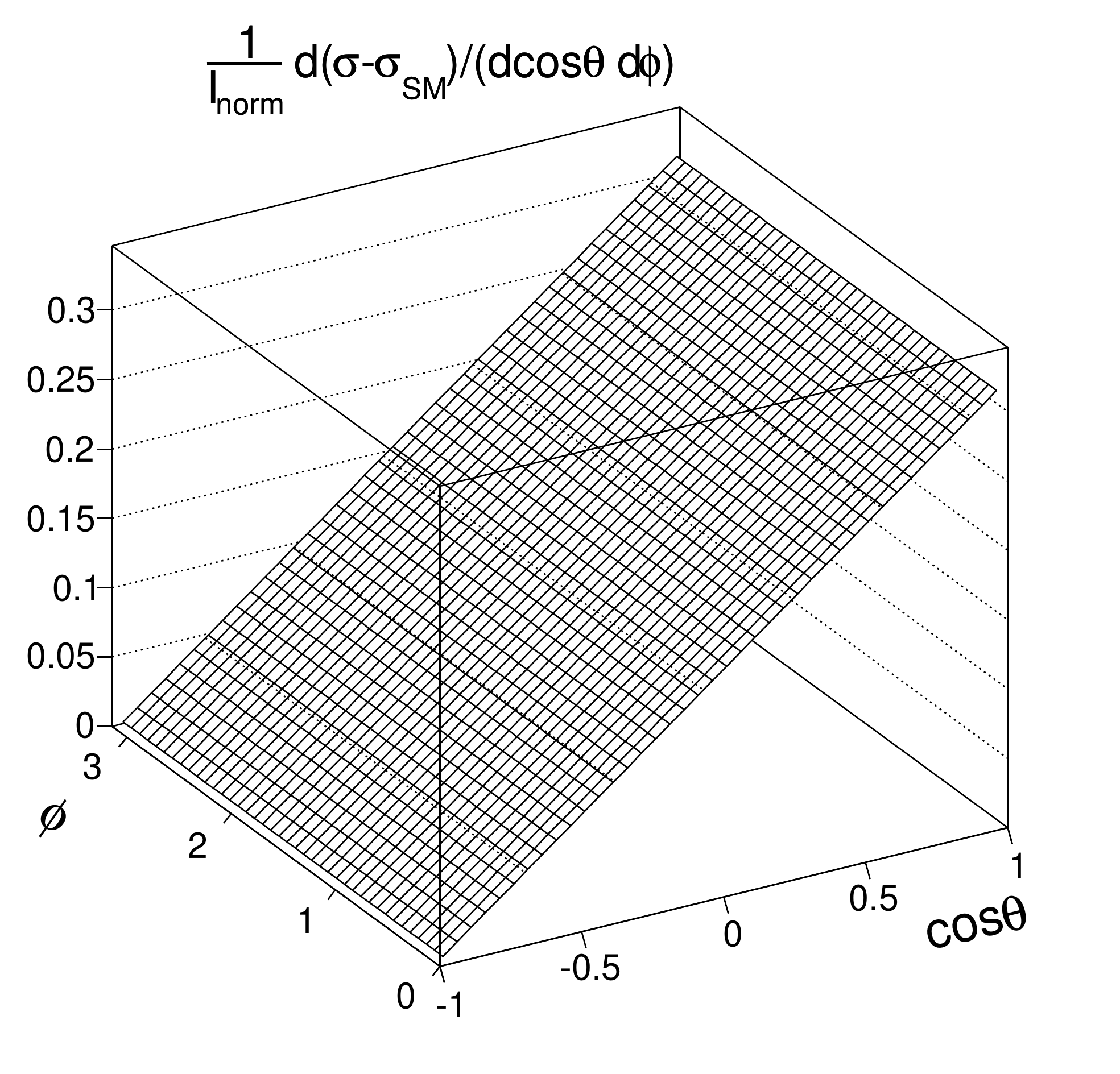}
		\end{minipage}
	\\
		\begin{minipage}[t]{.325\linewidth}
			\centering
			\includegraphics[width=6cm,height=6cm]{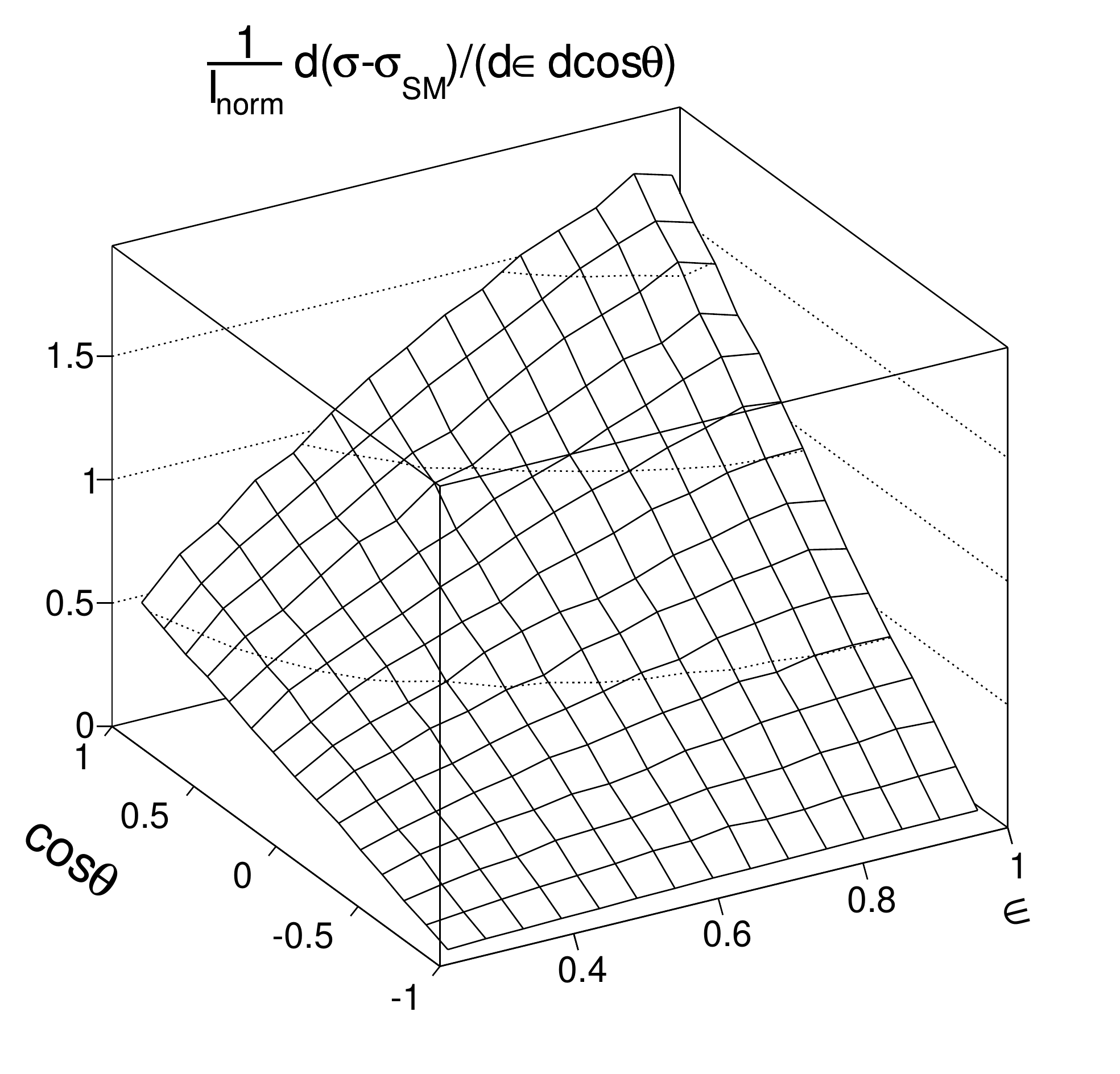}
		\end{minipage}
		\begin{minipage}[t]{.325\linewidth}
			\centering
			\includegraphics[width=6cm,height=6cm]{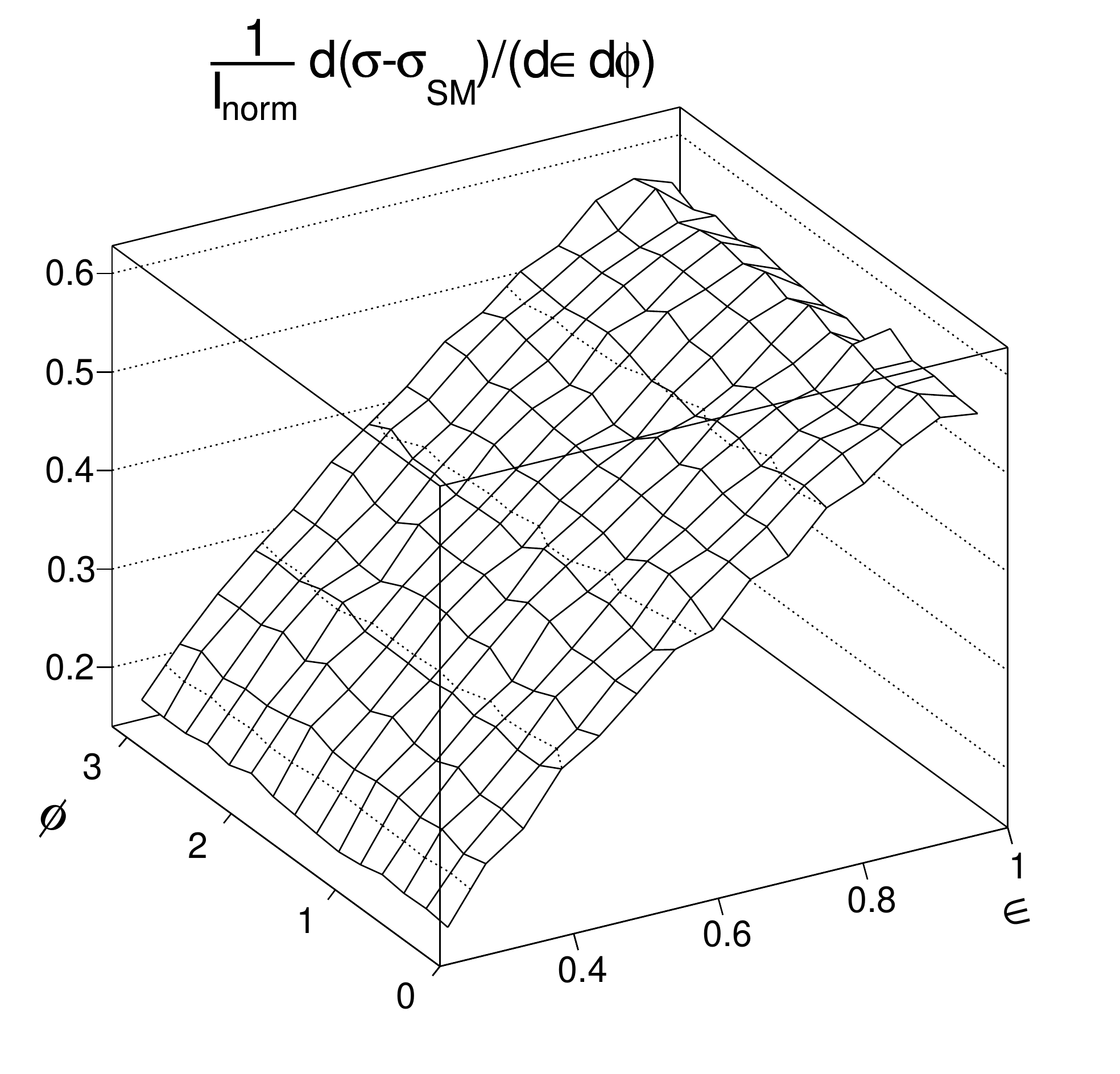}
		\end{minipage}
		\begin{minipage}[t]{.325\linewidth}
			\centering
			\includegraphics[width=6cm,height=6cm]{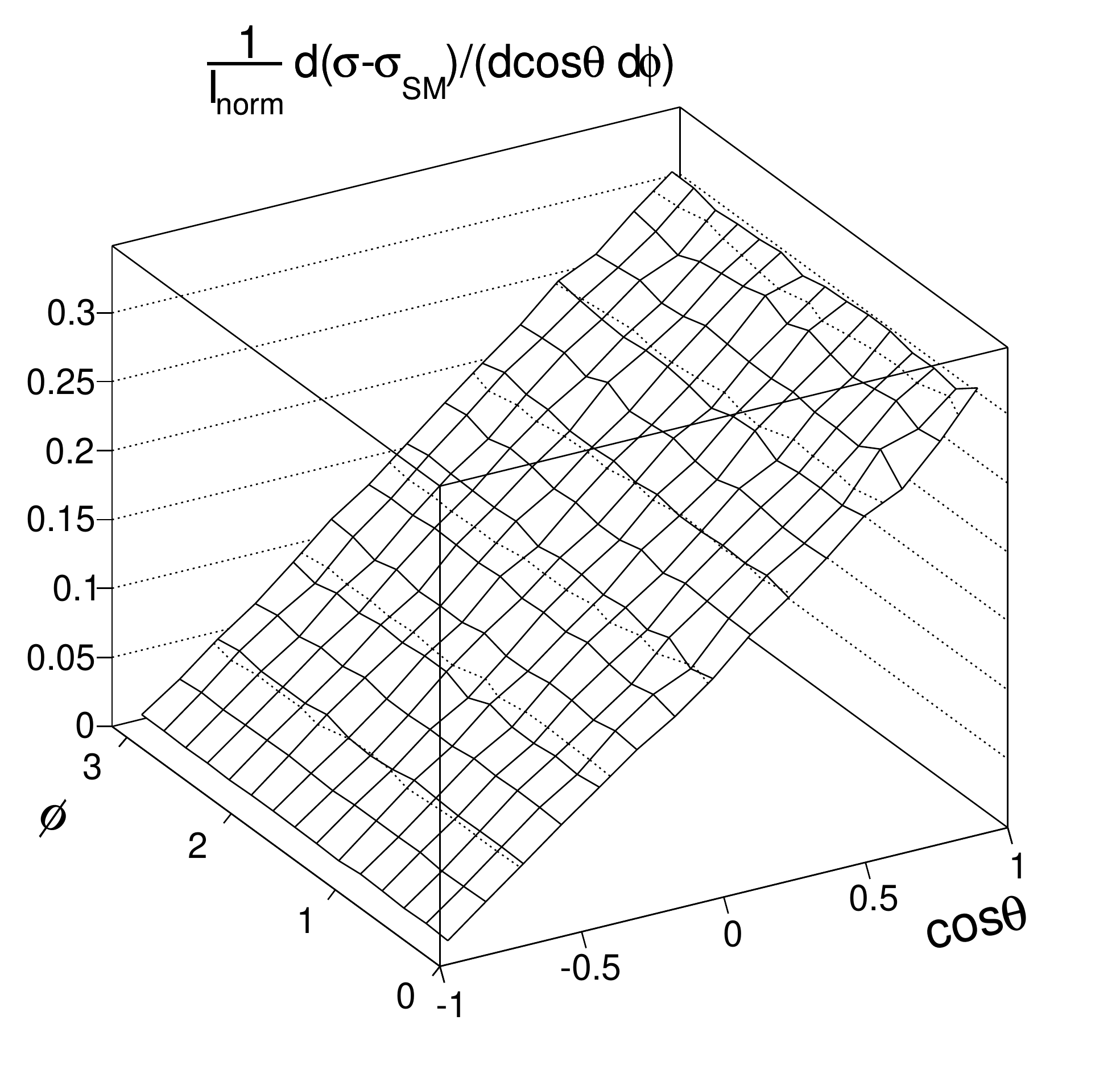}
		\end{minipage}
	\\				
	\end{center}
	\caption{ \label{pic8} \footnotesize  Scenario $Ref_{LT}$ = 0.057. The upper figures show plots of the normalized double-differential t-quark decay partial width $\frac{d(\Gamma-\Gamma_{SM})}{d\epsilon~\cdot~d\cos\theta}$, ~$\frac{d(\Gamma-\Gamma_{SM})}{d\epsilon~\cdot~d\phi}$, and $\frac{d(\Gamma-\Gamma_{SM})}{d\cos\theta~\cdot~d\phi}$ built from formulas (\ref{twidth1}), (\ref{twidth2}), (\ref{twidth3}). The middle figures show plots of the normalized double-differential cross sections $\frac{d(\sigma-\sigma_{SM})}{d\epsilon~\cdot~d\cos\theta}$, ~$\frac{d(\sigma-\sigma_{SM})}{d\epsilon~\cdot~d\phi}$, and $\frac{d(\sigma-\sigma_{SM})}{d\cos\theta~\cdot~d\phi}$ built from formulas (\ref{totalcrossec1}), (\ref{totalcrossec2}), (\ref{totalcrossec3}). The lower figures show plots of the normalized double-differential cross sections built from Monte Carlo events.}
\end{figure}

For the third scenario, we set $Ref_{LT}$ not equal to 0, and the remaining anomalous couplings are equal to 0.
In this case, the last two terms of the expression (\ref{diffcrossec}) are zero, and there is no dependence on the angle $\phi$ in the first and second terms. The dependence of both terms on $\cos\theta$ is the same as in the SM, but the energy distributions are different. The second term of formula (\ref{diffcrossec}) slightly deviates the shapes of differential cross sections (Fig.~\ref{pic8} middle left and middle central) from the corresponding shapes for the differential width (Fig.~\ref{pic8} upper left and upper central).

\begin{figure}
	\begin{center}
		\begin{minipage}[t]{.325\linewidth}
			\centering
			\includegraphics[width=6cm,height=6cm]{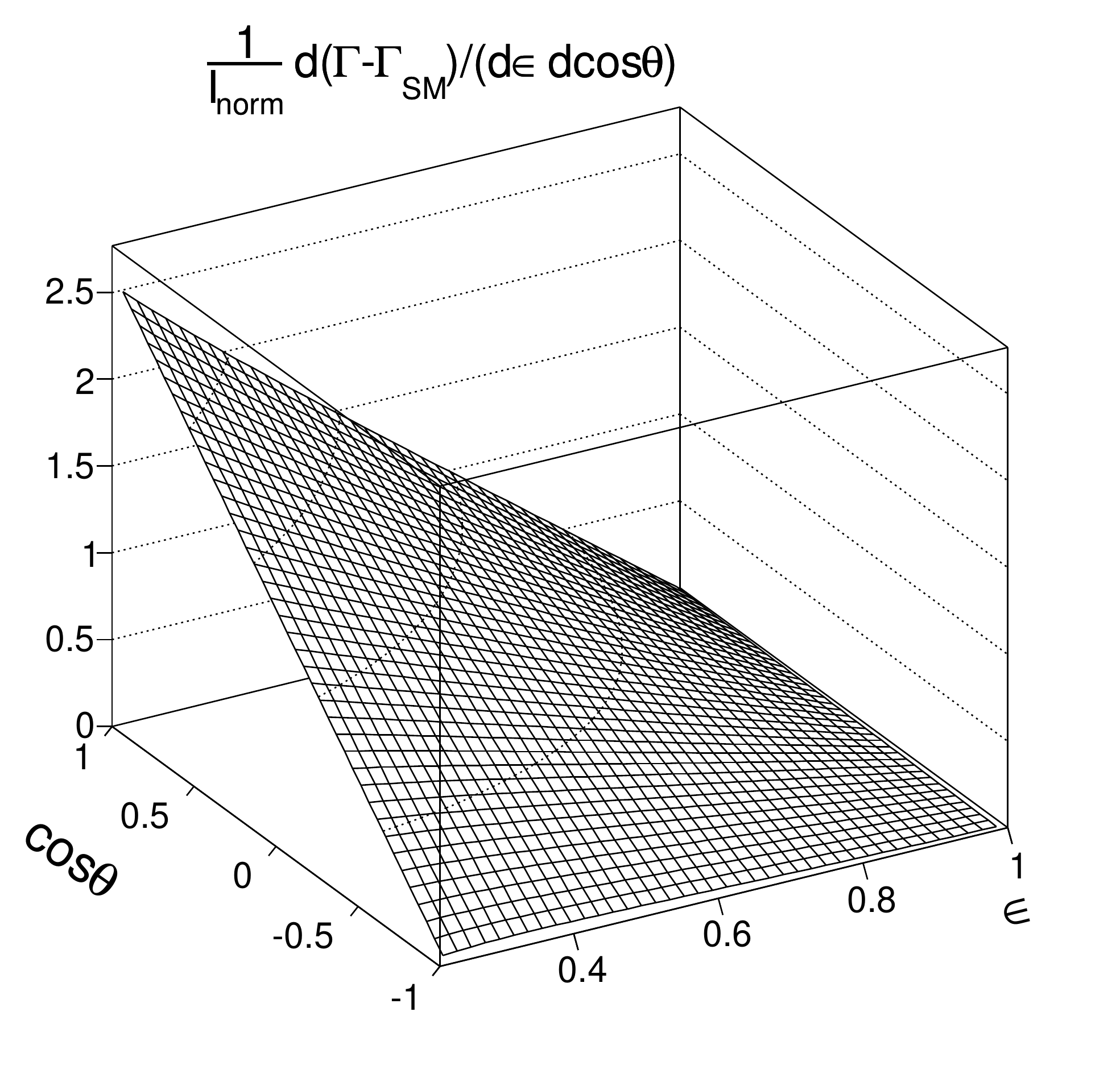}
		\end{minipage}
		\begin{minipage}[t]{.325\linewidth}
			\centering
			\includegraphics[width=6cm,height=6cm]{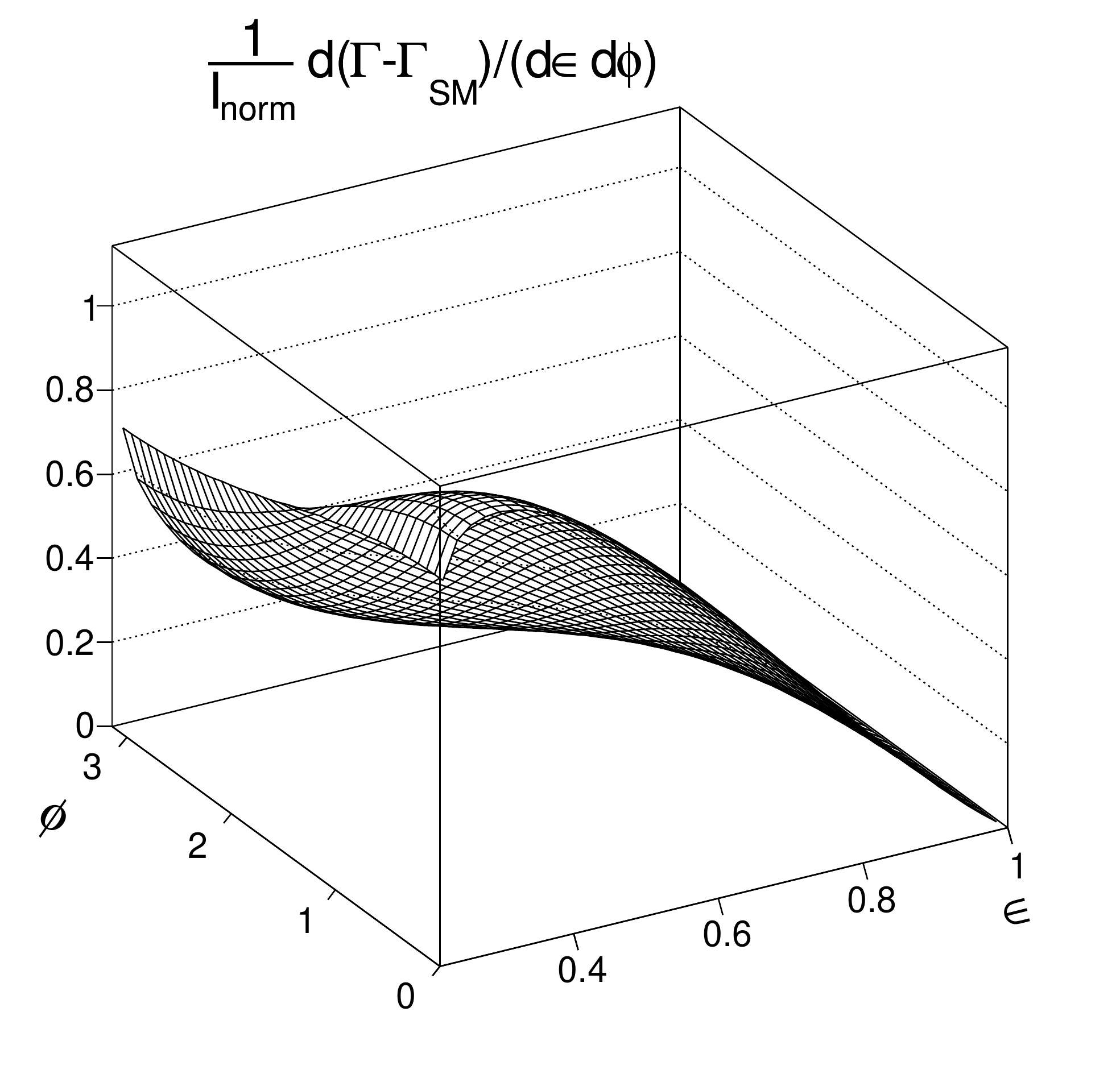}
		\end{minipage}
		\begin{minipage}[t]{.325\linewidth}
			\centering
			\includegraphics[width=6cm,height=6cm]{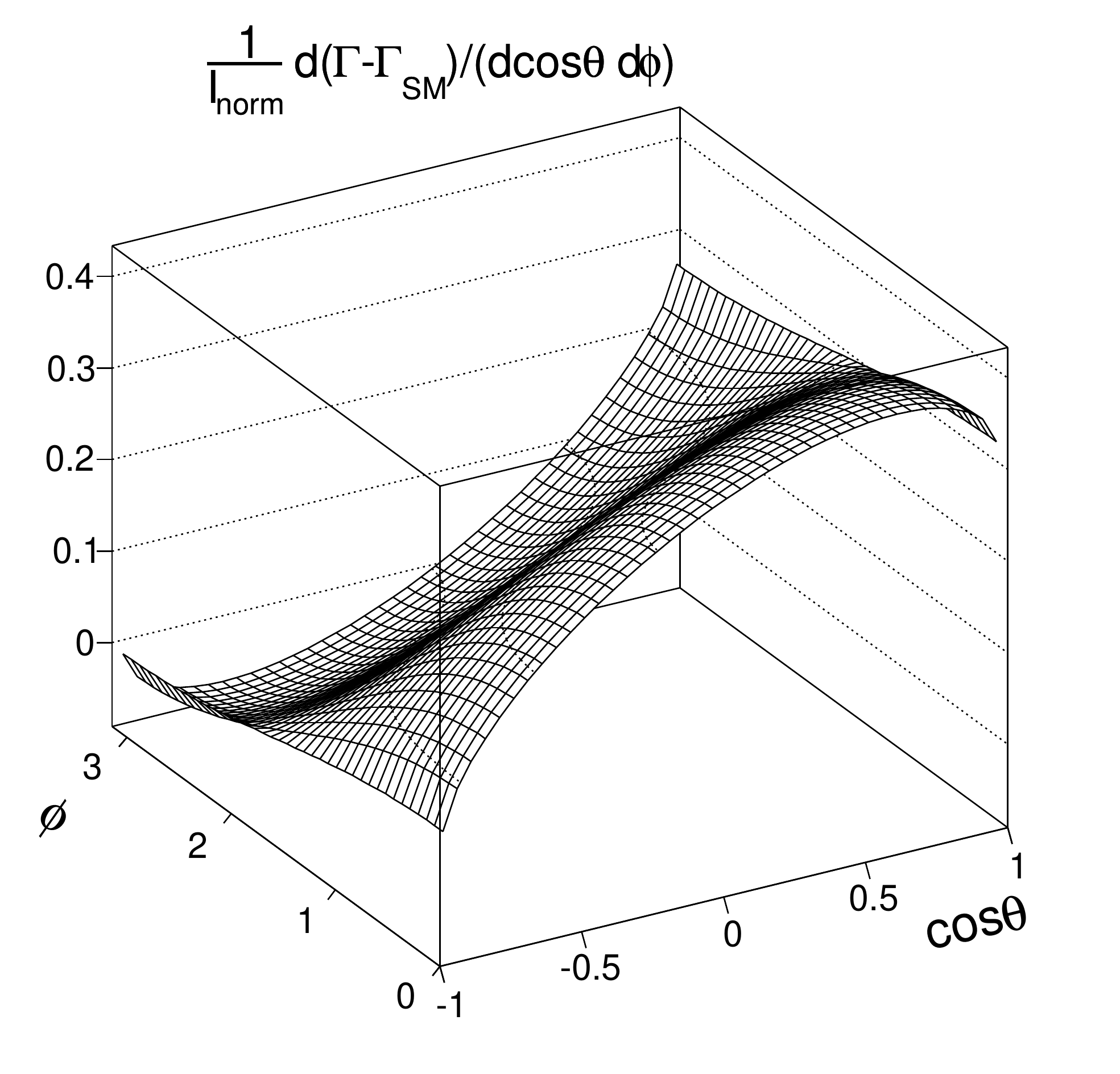}
		\end{minipage}
	\\
		\begin{minipage}[t]{.325\linewidth}
			\centering
			\includegraphics[width=6cm,height=6cm]{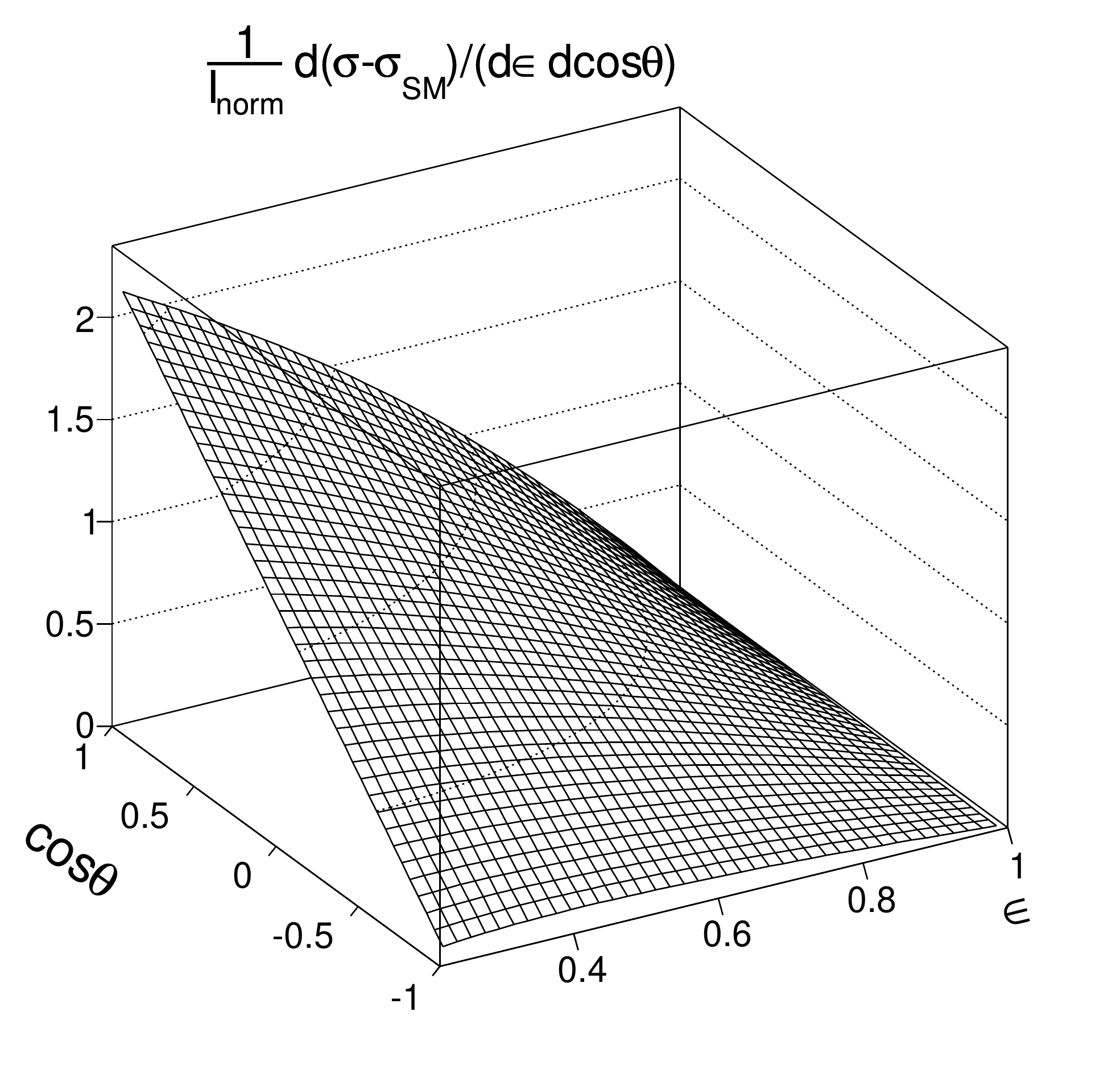}
		\end{minipage}
		\begin{minipage}[t]{.325\linewidth}
			\centering
			\includegraphics[width=6cm,height=6cm]{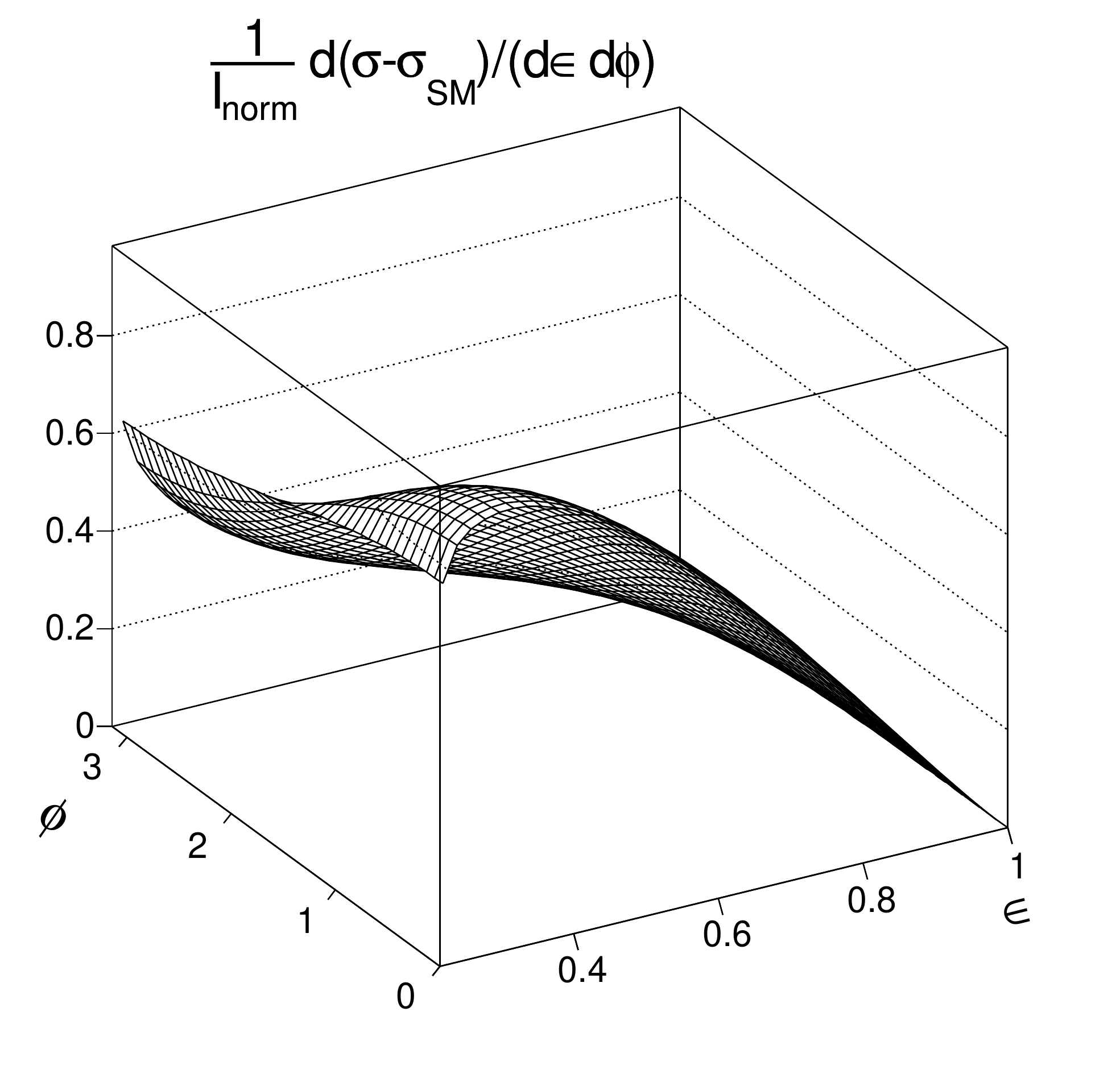}
		\end{minipage}
		\begin{minipage}[t]{.325\linewidth}
			\centering
			\includegraphics[width=6cm,height=6cm]{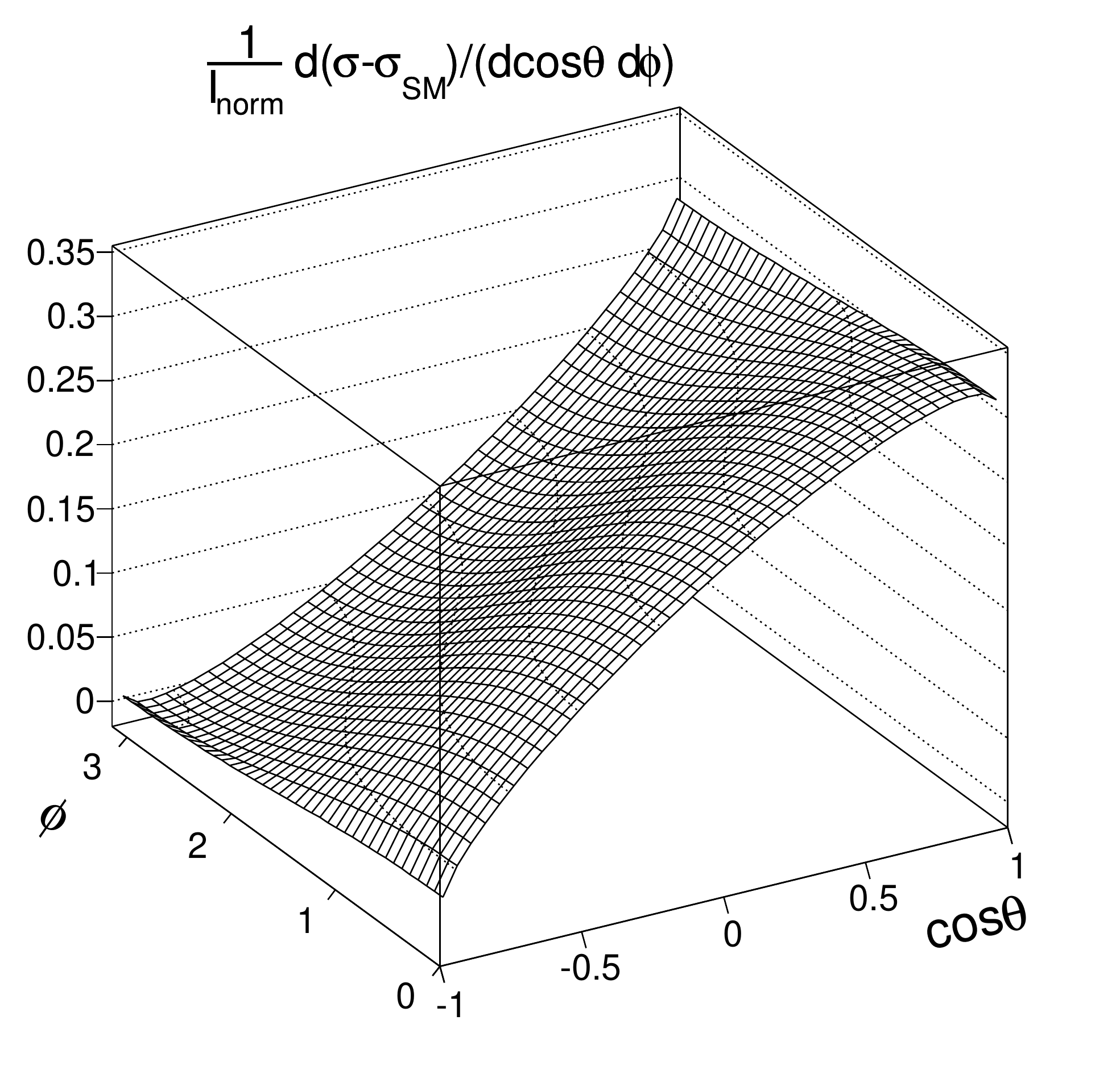}
		\end{minipage}
	\\
		\begin{minipage}[t]{.325\linewidth}
			\centering
			\includegraphics[width=6cm,height=6cm]{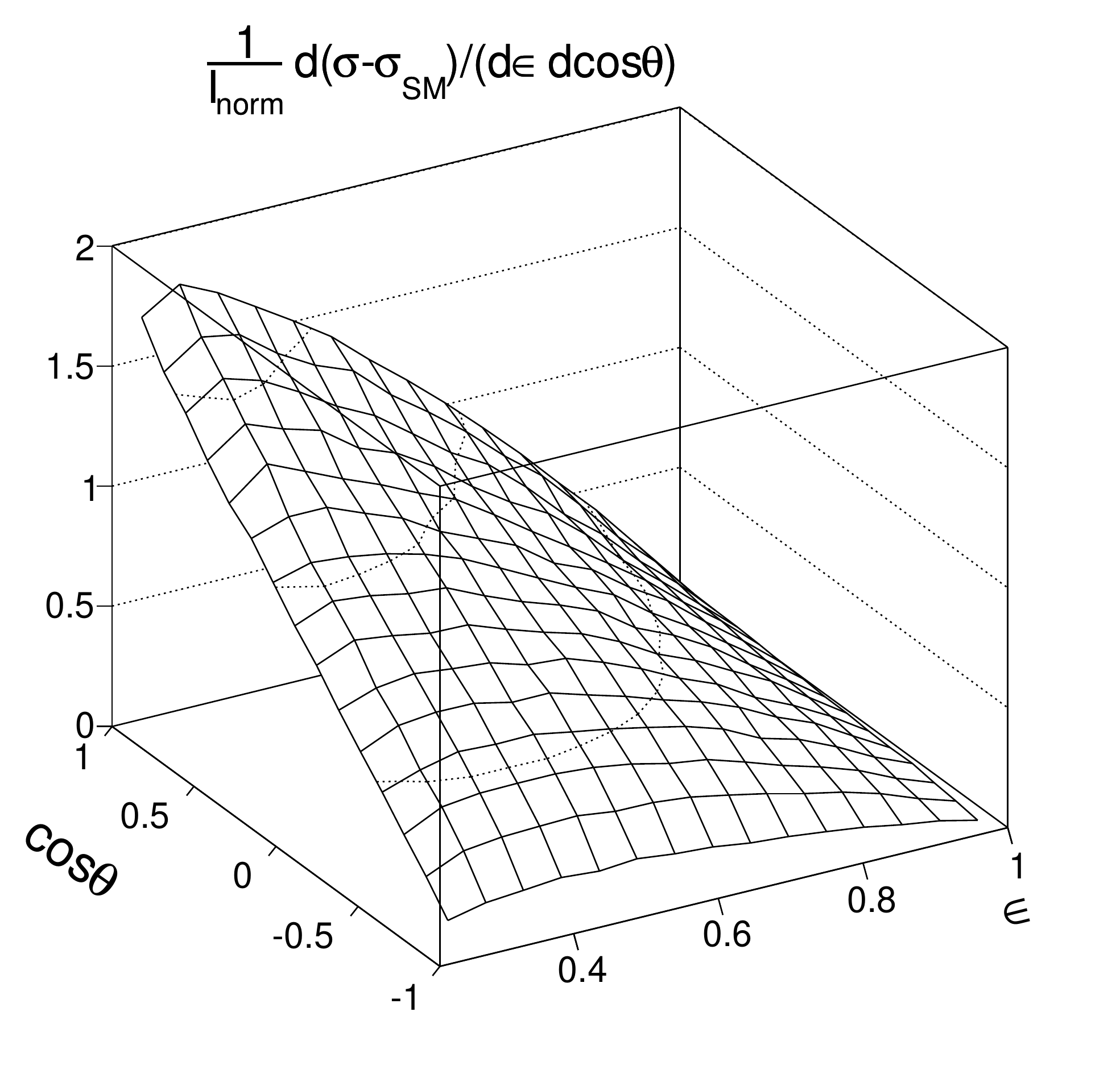}
		\end{minipage}
		\begin{minipage}[t]{.325\linewidth}
			\centering
			\includegraphics[width=6cm,height=6cm]{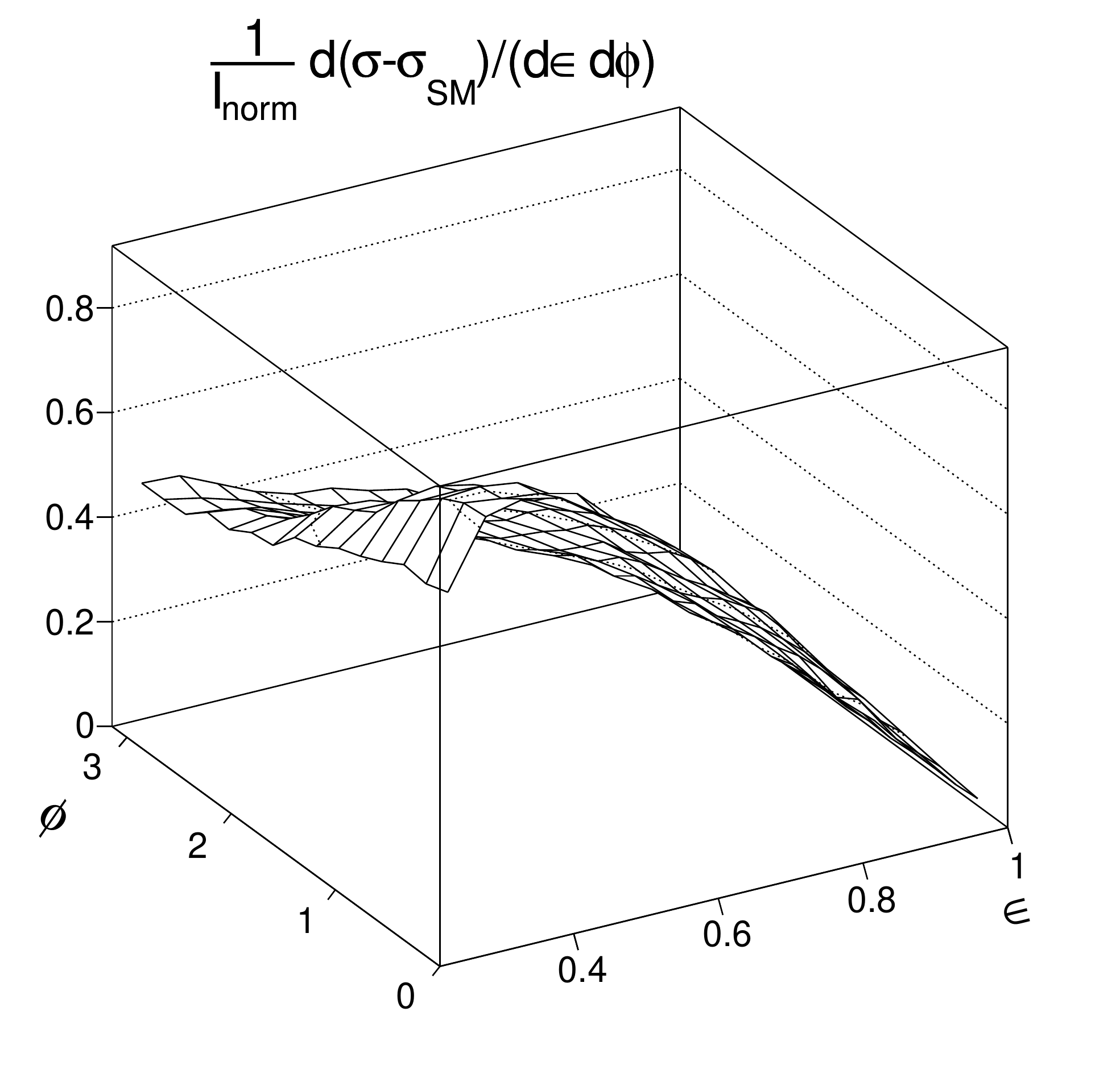}
		\end{minipage}
		\begin{minipage}[t]{.325\linewidth}
			\centering
			\includegraphics[width=6cm,height=6cm]{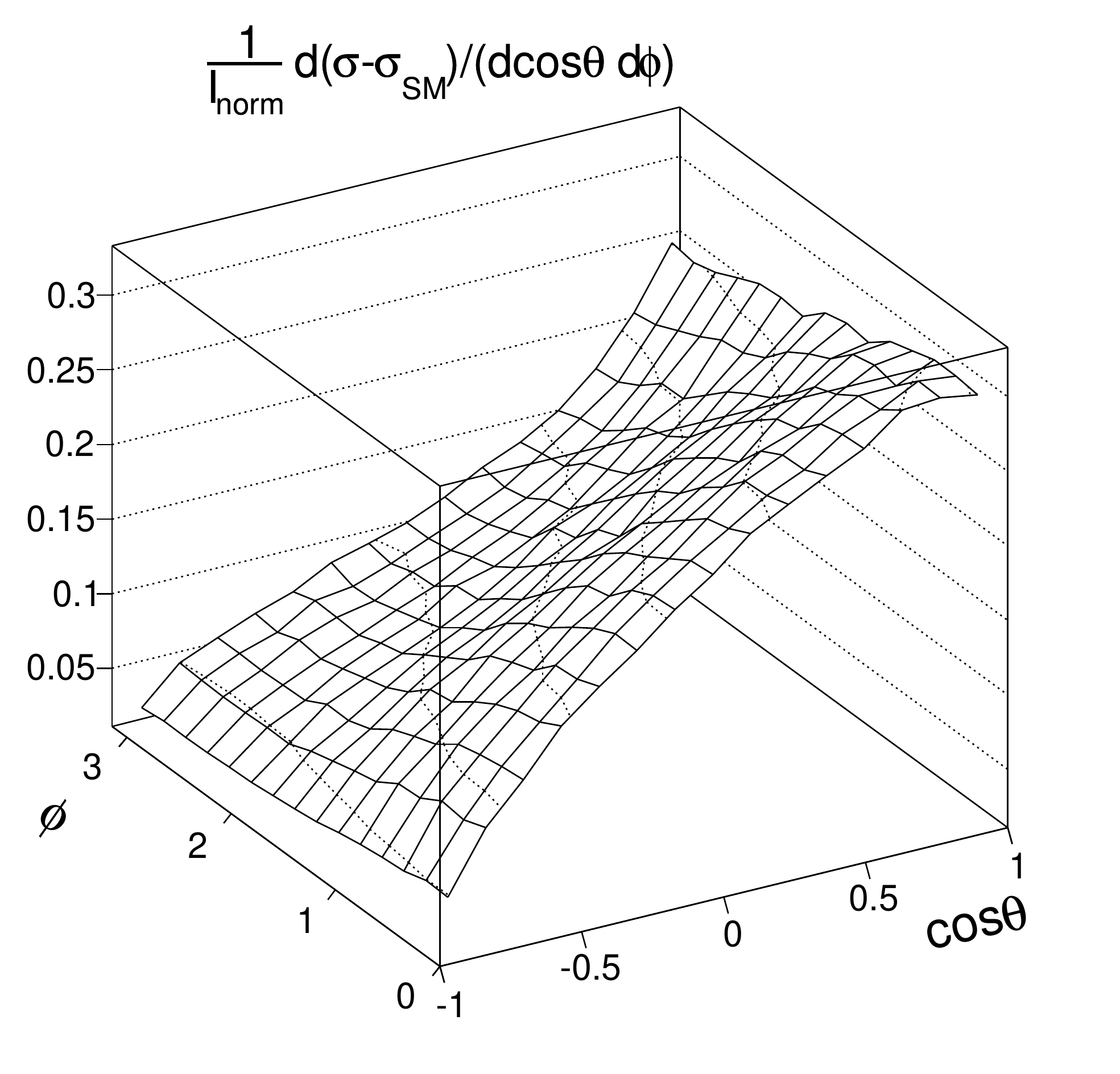}
		\end{minipage}
     \\			
	\end{center}
	\caption{ \label{pic9} \footnotesize  Scenario $Ref_{RT}$ = 0.048. The upper figures show plots of the normalized double-differential t-quark decay partial width $\frac{d(\Gamma-\Gamma_{SM})}{d\epsilon~\cdot~d\cos\theta}$, ~$\frac{d(\Gamma-\Gamma_{SM})}{d\epsilon~\cdot~d\phi}$, and $\frac{d(\Gamma-\Gamma_{SM})}{d\cos\theta~\cdot~d\phi}$ built from formulas (\ref{twidth1}), (\ref{twidth2}), (\ref{twidth3}). The middle figures show plots of the normalized double-differential cross sections $\frac{d(\sigma-\sigma_{SM})}{d\epsilon~\cdot~d\cos\theta}$, ~$\frac{d(\sigma-\sigma_{SM})}{d\epsilon~\cdot~d\phi}$, and $\frac{d(\sigma-\sigma_{SM})}{d\cos\theta~\cdot~d\phi}$ built from formulas (\ref{totalcrossec1}), (\ref{totalcrossec2}), (\ref{totalcrossec3}). The lower figures show plots of the normalized double-differential cross sections built from Monte Carlo events.}
\end{figure}

For the fourth scenario, we set $Ref_{RT}$ nonzero and positive, and the remaining anomalous couplings are equal to 0.
For this scenario, the formula for the differential cross section (\ref{diffcrossec}) is the most complex of all the listed scenarios. The first part of formula (\ref{diffcrossec}) contains quadratic anomalous terms, as well as the leading linear interference term, which mainly determines the shape of the differential cross sections. The second, third and fourth terms of the formula (\ref{diffcrossec}) contain quadratic anomalous terms, which only slightly affect the shape of the differential cross sections (Fig.~\ref{pic9} middle). 

\begin{figure}
	\begin{center}
		\begin{minipage}[t]{.325\linewidth}
			\centering
			\includegraphics[width=6cm,height=6cm]{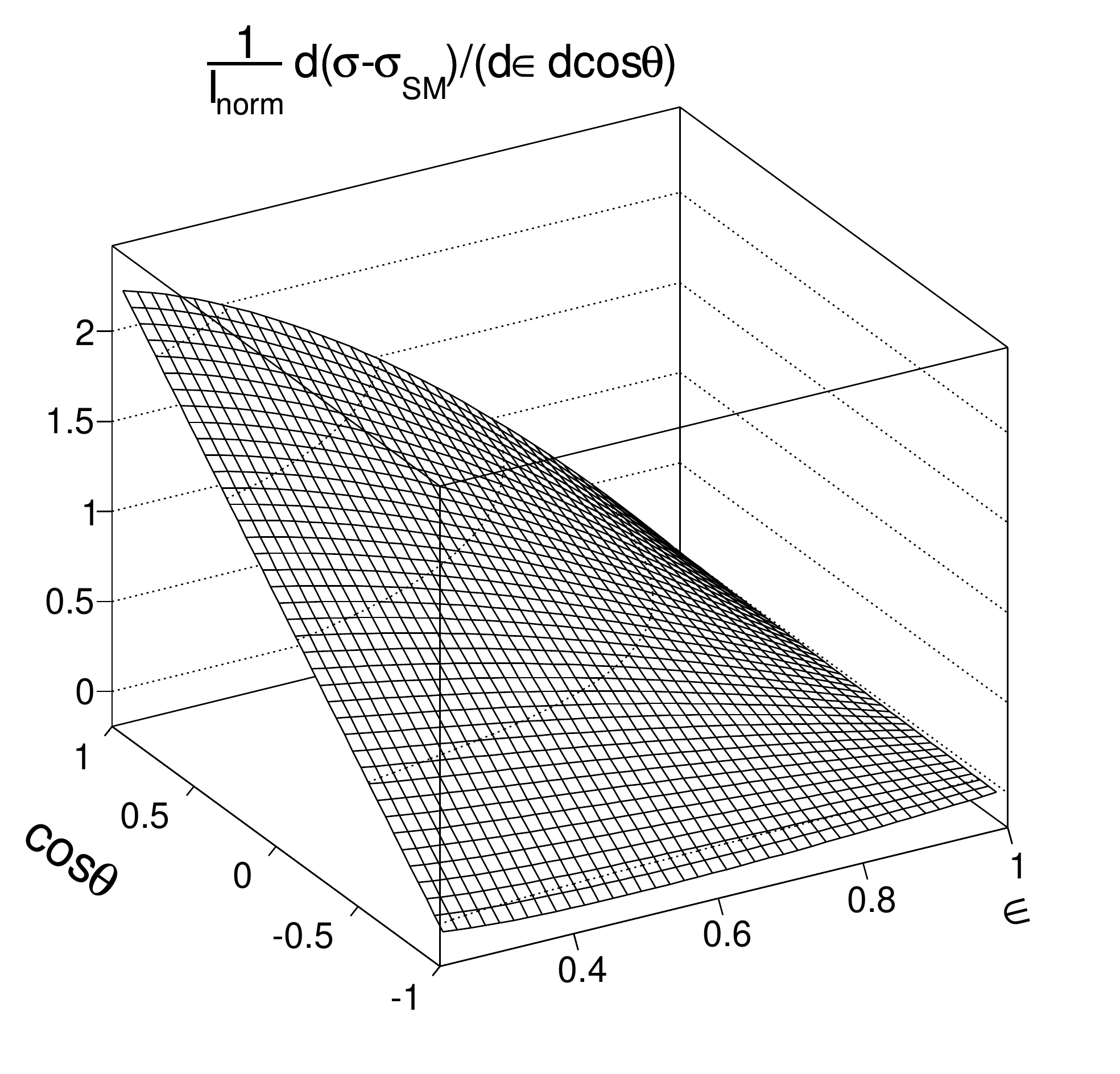}
		\end{minipage}
		\begin{minipage}[t]{.325\linewidth}
			\centering
			\includegraphics[width=6cm,height=6cm]{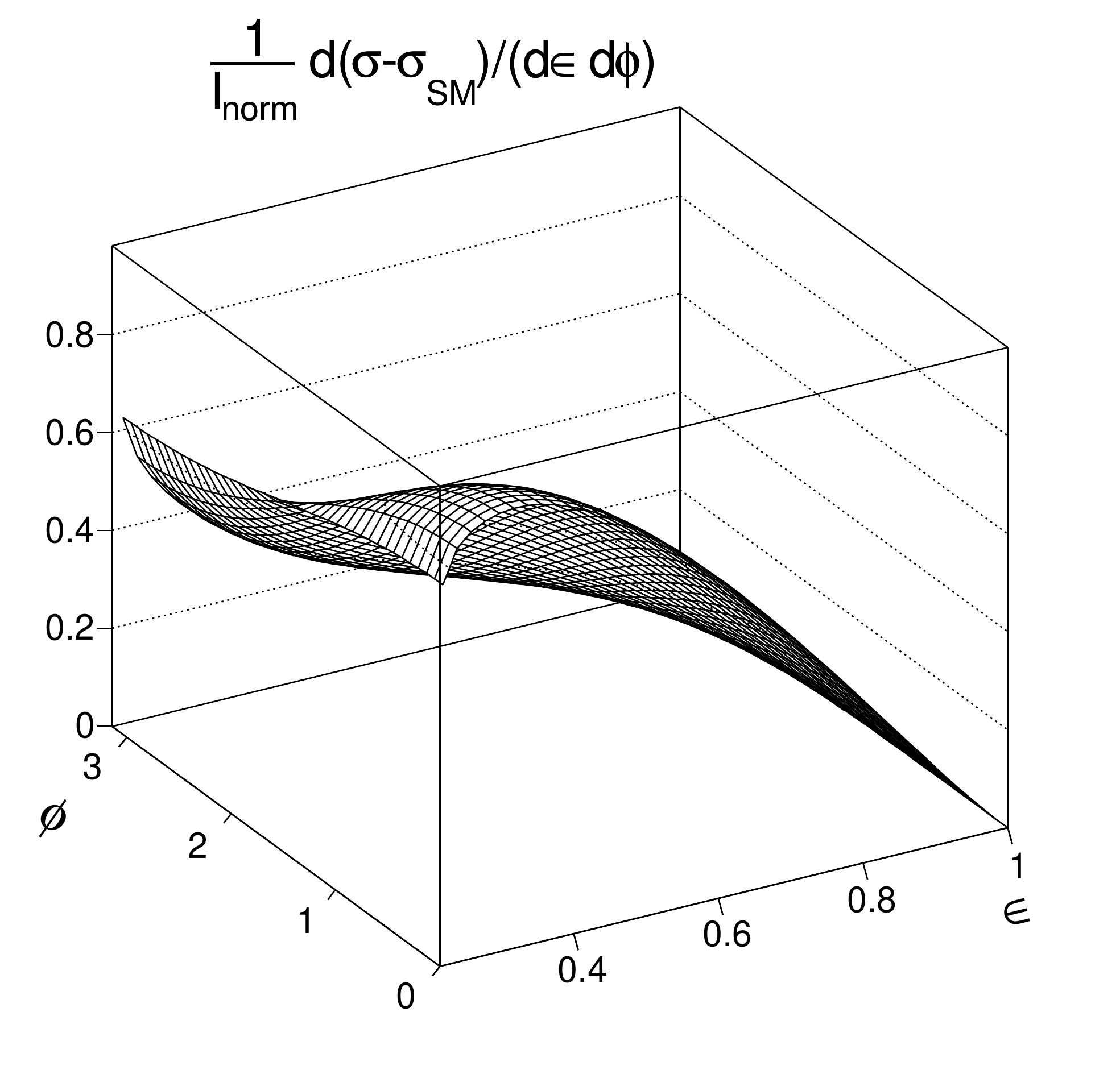}
		\end{minipage}
		\begin{minipage}[t]{.325\linewidth}
			\centering
			\includegraphics[width=6cm,height=6cm]{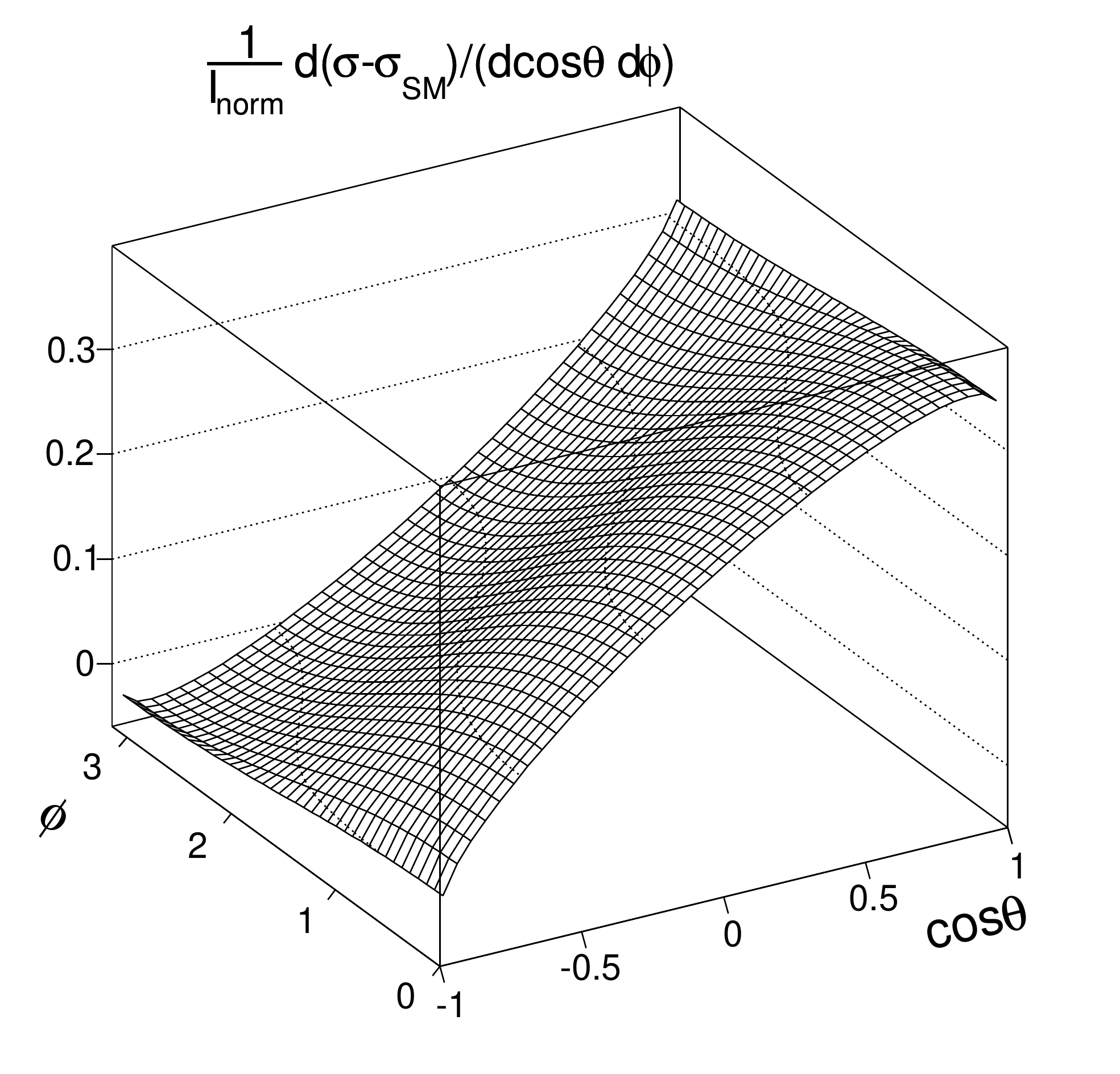}
		\end{minipage}
	\end{center}
	\caption{ \label{pic10} \footnotesize  Scenario $Ref_{RT}$ = -0.048. Plots of the normalized double-differential cross sections $\frac{d(\sigma-\sigma_{SM})}{d\epsilon~\cdot~d\cos\theta}$, ~$\frac{d(\sigma-\sigma_{SM})}{d\epsilon~\cdot~d\phi}$, and $\frac{d(\sigma-\sigma_{SM})}{d\cos\theta~\cdot~d\phi}$. }
\end{figure}

For the last scenario, $Ref_{RT}$ is set to be nonzero and negative, and the remaining anomalous couplings are equal to 0. The overall picture (Fig.~\ref{pic10}) is very similar to the previous case, but one needs to keep in mind that deviations from the prediction of the Standard Model have the opposite sign, and this sign is not displayed on normalized distributions. Despite the similarities, Fig.~\ref{pic10} is not identical to (Fig.~\ref{pic9} middle). Differences can be observed from the bottom of the left and right plots. In the case of the ($Re f_{RT}\ne 0$) scenario, the differential cross section of the process includes linear and quadratic anomalous terms. Although the relative contribution of the quadratic terms is small, they have an influence on the total sum, which will be different for the case of positive and negative couplings. In the general case, the differential scattering cross section for this scenario can be written as:\\
$d\sigma=d\sigma_{SM} + f_{RT}\cdot dS_1 + f_{RT}^2\cdot dS_2$.
The normalized difference of the differential scattering cross section with anomalous couplings and the scattering cross section of the Standard Model is\\
$\frac{d(\sigma-\sigma_{SM})}{\sigma-\sigma_{SM}} = 
\frac{f_{RT}\cdot dS_1 + f_{RT}^2\cdot dS_2}{f_{RT}\cdot S_1 + f_{RT}^2\cdot S_2} = \frac{dS_1 + f_{RT}\cdot dS_2}{S_1 + f_{RT}\cdot S_2}$.
The leading term of this expression dS1/S1 is independent of anomalous couplings and is crucial for the distribution shape. A term of the following order is proportional to anomalous coupling and depends on its sign. This term defines the small differences between the corresponding pictures of Fig.~\ref{pic9} middle and Fig.~\ref{pic10}. 

%=========================================================================
%=========================================================================
%=========================================================================
%=========================================================================
%=========================================================================
%=========================================================================
\begin{figure}
	\begin{center}
		\begin{minipage}[t]{.325\linewidth}
			\centering
			\includegraphics[width=6cm,height=6cm]{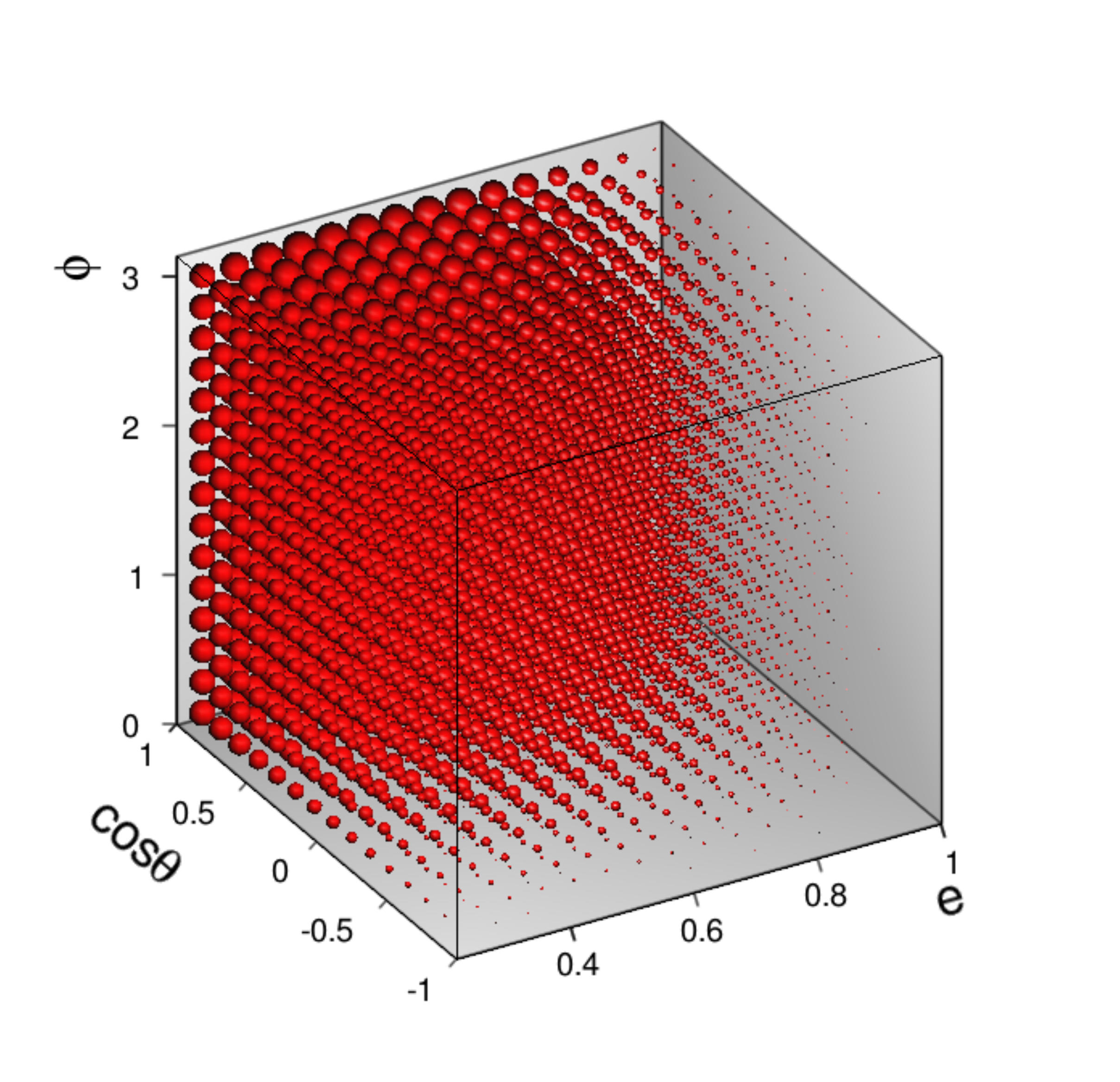}
			\begin{center}
				{\it $Re f_{LV}= 1.03$}
			\end{center}
		\end{minipage}
		\begin{minipage}[t]{.325\linewidth}
			\centering
			\includegraphics[width=6cm,height=6cm]{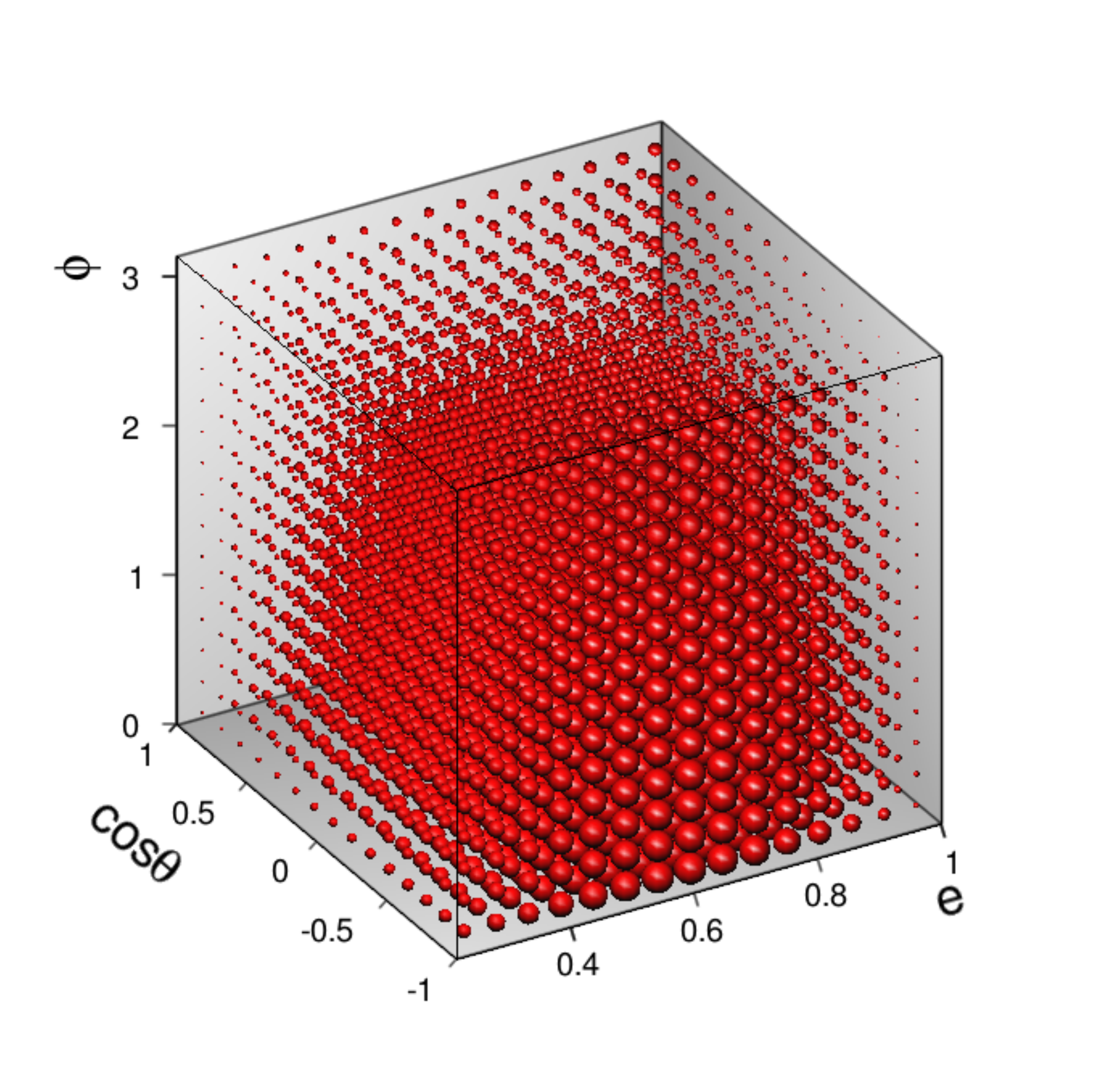}
			\begin{center}
				{\it $Re f_{RV}= 0.16$}
			\end{center}
		\end{minipage}
		\begin{minipage}[t]{.325\linewidth}
			\centering
			\includegraphics[width=6cm,height=6cm]{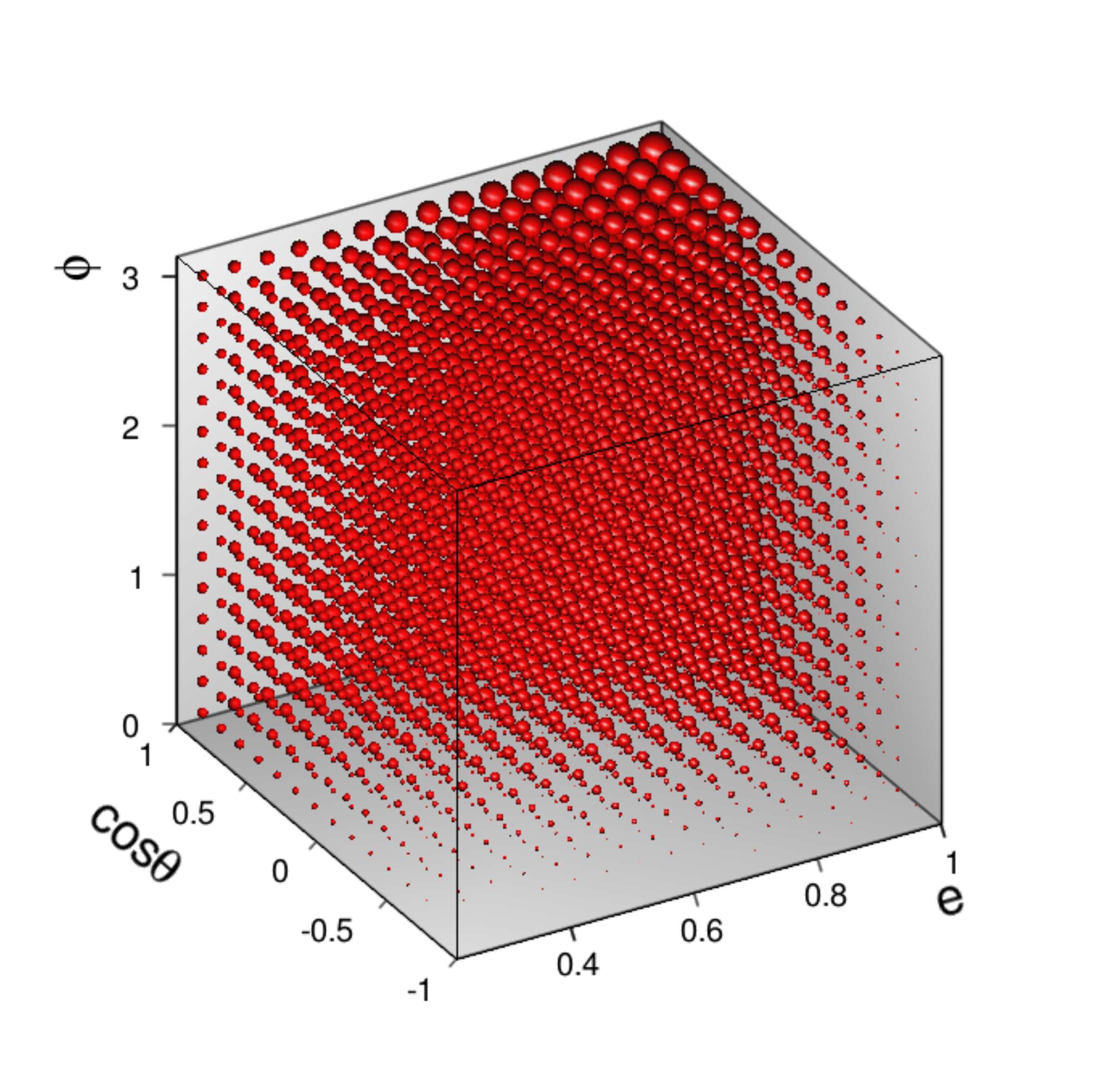}
			\begin{center}
				{\it $Re f_{LT}= 0.057$}
			\end{center}
		\end{minipage}
		\begin{minipage}[t]{.325\linewidth}
			\centering
			\includegraphics[width=6cm,height=6cm]{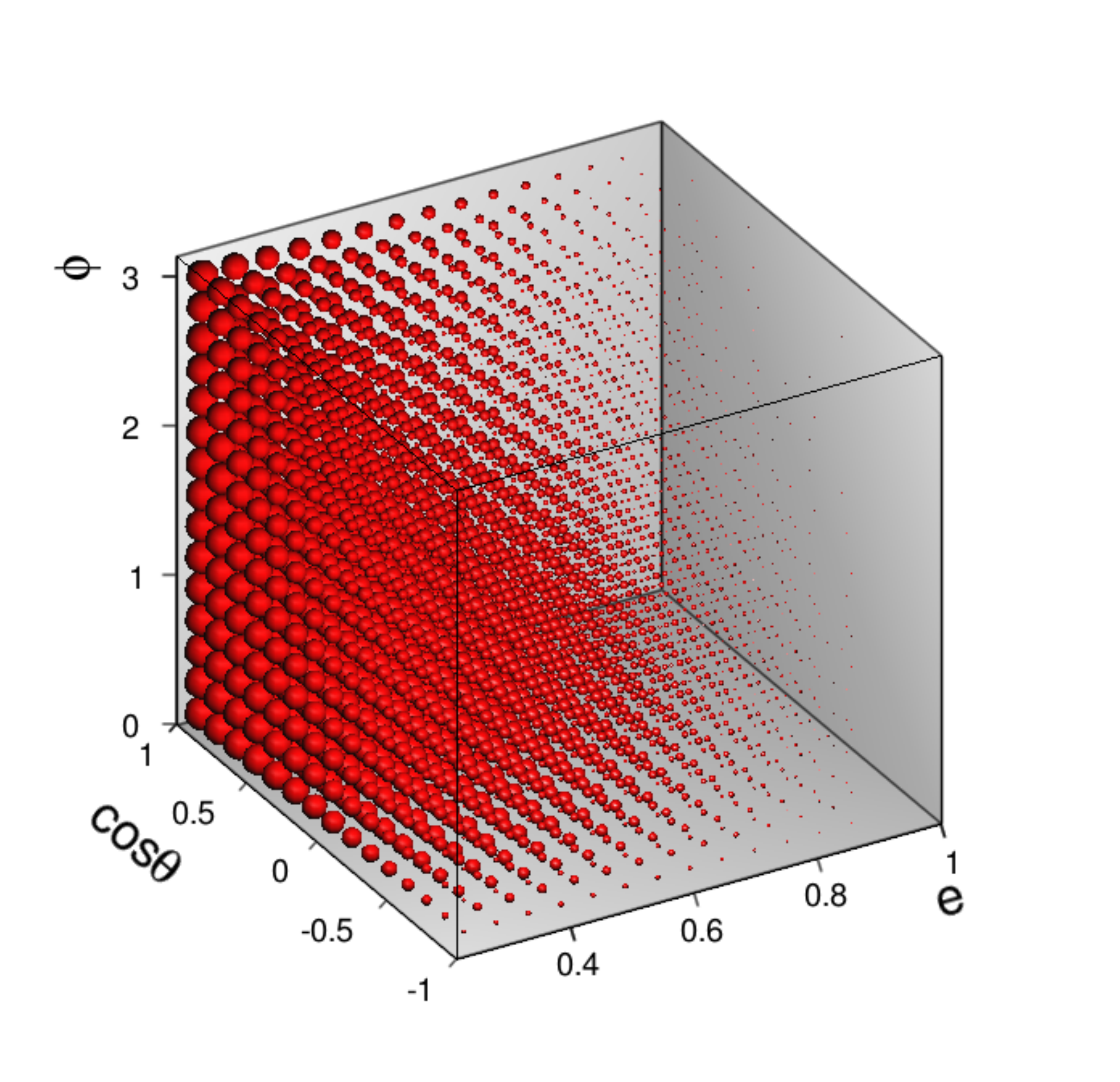}
			\begin{center}
				{\it $Re f_{RT}= 0.048$}
			\end{center}
		\end{minipage}
		\begin{minipage}[t]{.325\linewidth}
			\centering
			\includegraphics[width=6cm,height=6cm]{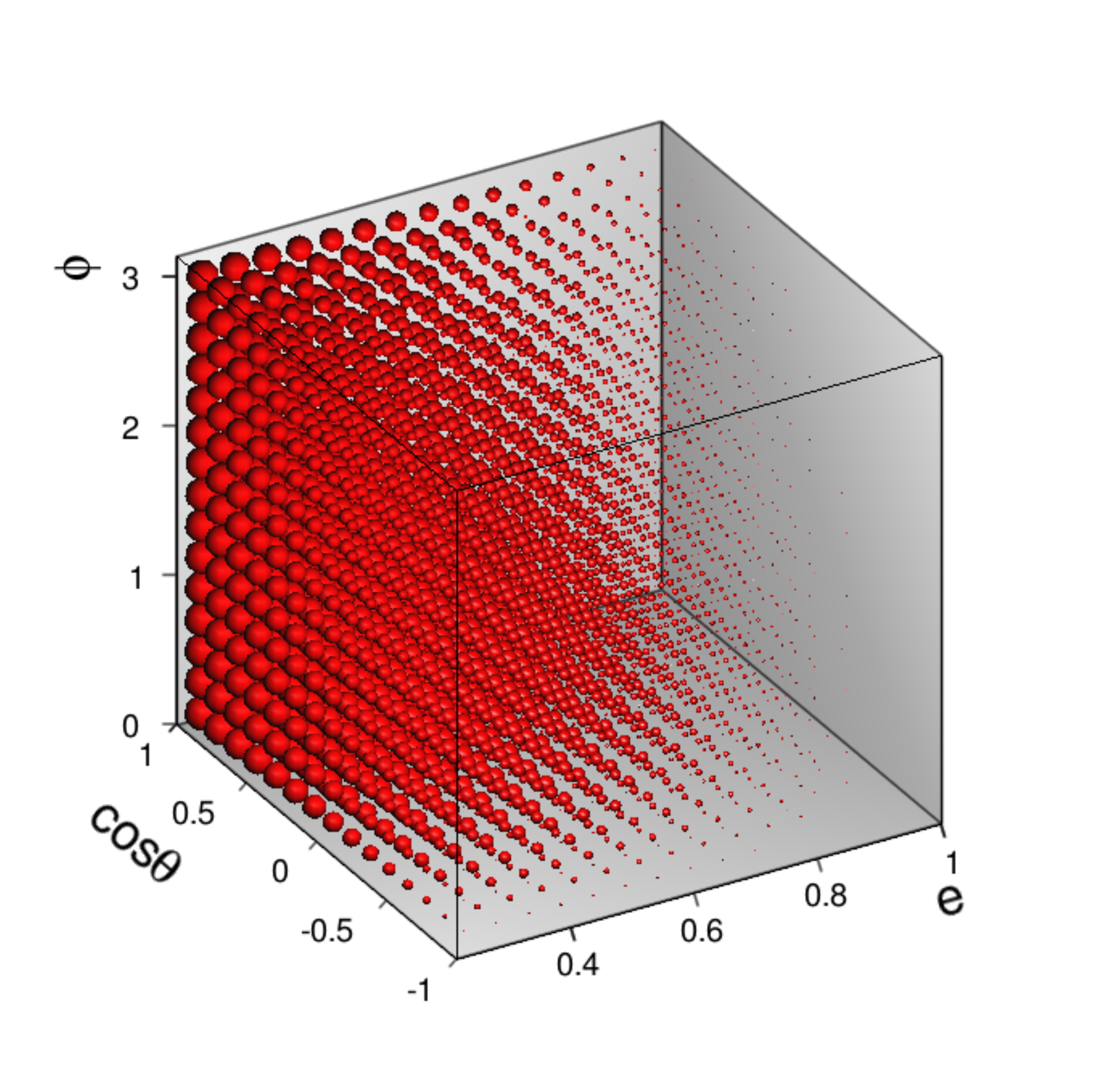}
			\begin{center}
				{\it $Re f_{RT}=-0.048$}
			\end{center}
		\end{minipage}
	\end{center}
	\caption{ \label{pic11} \footnotesize Plots of probability density $\frac{d(\sigma-\sigma_{SM})}{d\epsilon~\cdot~d\cos\theta~\cdot~d\phi}$ in 3D space ($\epsilon$, $\cos\theta$, $\phi$) for the scenarios $Re f_{LV}= 1.03$,~~$Re f_{RV}= 0.16$,~~$Re f_{LT}= 0.057$,~~$Re f_{RT}= 0.048$,~~and~~$Re f_{RT}=-0.048$.}
\end{figure}
%=========================================================================
One can see good agreement of the plots obtained using the analytical formula (\ref{totalcrossec}) with the corresponding Monte Carlo distributions for all listed scenarios. This comparison confirms the correctness of the analytical calculations performed. It should be noted that the presence of an additional subprocess $\bar{d} b \rightarrow l^+, \nu_{l}, b, \bar{u}$ does not have a noticeable effect on the shape of the distributions. It is clearly seen that in all scenarios listed above, except for the second one, the shape of the differential cross section is determined mainly by the differential width of the t-quark decay. But in the case of scenario ($Ref_{RV}\ne 0$), the second and fourth terms of formula (\ref{diffcrossec}) significantly change the shape of the differential cross sections.

As the main conclusion, it can be noted that the corresponding shapes of the surfaces in the coordinate space ($\epsilon$, $\cos\theta$) and ($\epsilon$, $\phi$) are very different for various scenarios. The most spectacular differences can be observed in coordinate space ($\epsilon$, $\cos\theta$). This allows one to separate all scenarios from each other.  
At the same time, the plots in the space ($\cos\theta$, $\phi$) are quite similar for the listed scenarios (except for the $Ref_{RV}\ne 0$ case) and do not allow them to be uniquely identified. 

In addition to the two-dimensional distributions, we draw plots showing the probability density $\frac{d(\sigma-\sigma_{SM})}{d\epsilon~\cdot~d\cos\theta~\cdot~d\phi}$ in 3D space ($\epsilon$, $\cos\theta$, $\phi$) for the cases listed (Fig.~\ref{pic11}). The size of each cell in the figures is proportional to the probability density. One can see that the areas of maximum density for the different scenarios are located in different places of the 3D cube ($\epsilon$, $\cos\theta$, $\phi$). 
Although the three-dimensional areas are almost the same for cases ($Ref_{RT}$ = 0.048) and ($Ref_{RT}$ = -0.048), 
the corresponding deviations from the Standard Model due to anomalous contributions for these scenarios have opposite signs.

Now we use the obtained analytical expressions to extract the values of anomalous coupling applying a fitting procedure.  For completeness of the MC simulation, we also included additional subprocess $\bar{d} b \rightarrow l^+, \nu_{l}, b, \bar{u}$ with a $\bar{d}$ quark in the initial state. In the case of the SM, the contribution to the rate from such events is about $13\%$, while the impact on the distribution shapes is practically the same.
We generate SM Monte Carlo event samples for different values of the integral luminosity and collision energy of the LHC collider: 30 fb$^{-1}$ at 13 TeV, 300 fb$^{-1}$ and 3000 fb$^{-1}$ at 14 TeV. For these events, we constructed 2D histograms $\frac{d\sigma_{SM}}{d\epsilon \cdot d\cos\theta}$.
Using the method of maximum likelihood, we fit histograms built from these SM MC events. 
We used formula (\ref{totalcrossec1}) as a fitting function. The values of the coupling $Re f_ {LV}$ and one of the nonzero anomalous couplings of the above scenarios were used as fitting parameters.
Applying the fitting function to the Standard Model MC events, we expect to extract $Re f_{LV}$ values close to 1 and $Re f_{LR}$, $Re f_{LT}$, $Re f_{RT}$ values close to 0. It should be noted that in this study, we are not interested in the values themselves but in the prediction of the accuracy of their measurement $\delta~Re f_{LV}$, $\delta~Re f_{RV}$, $\delta~Re f_{LT}$, $\delta~Re f_{RT}$. We did not use information about the latest experimental limits and did not set boundaries for finding parameter values when fitting.
For fitting, we use the MINUIT algorithm \cite{James:1975dr} build into the ROOT package \cite{Brun:1997pa}. The results of the two-parametric fitting of the two-dimensional histogram are given in Table~\ref{table1}.

Similar to the two-dimensional case, we fit the distribution of the SM events in 3D space ($\epsilon$,~$\cos\theta$,~$\phi$). Table~\ref{table2} shows the results of this fitting. The values of coupling $Re f_{LV}$ and one of the nonzero anomalous couplings of the above scenarios were used as fitting parameters.

\begin{table}
\begin{center}
\begin{tabular}{|l|l|l|l|} \hline
$L,~fb^{-1}$ & $\delta~Re f_{LV}$ / $\delta~Re f_{RV}$  & \hskip 2mm $\delta~Re f_{LV}$ / $\delta~Re f_{LT}$ & \hskip 2mm $\delta~Re f_{LV}$ / $\delta~Re f_{RT}$ \\ \hline
30 & $6.9\cdot10^{-4}$ / $1.2\cdot10^{-2}$ & $8.1\cdot10^{-4}$ / $5.0\cdot10^{-3}$ & $1.6\cdot10^{-3}$ / $2.5\cdot10^{-3}$ \\  
300 & $1.9\cdot10^{-4}$ / $4.7\cdot10^{-3}$ & $2.5\cdot10^{-4}$ / $1.9\cdot10^{-3}$ & $5.6\cdot10^{-4}$ / $8.7\cdot10^{-4}$ \\
3000 & $5.9\cdot10^{-5}$ / $6.8\cdot10^{-4}$ & $8.0\cdot10^{-5}$ / $8.0\cdot10^{-4}$ & $1.6\cdot10^{-4}$ / $2.4\cdot10^{-4}$ \\ \hline
\end{tabular}
\end{center}
\caption{\label{table1} The accuracy of measuring the two anomalous parameters by fitting in the 2D coordinate space ($\epsilon$,~$\cos~\theta$).}
\end{table}

\begin{table}
\begin{center}
\begin{tabular}{|l|l|l|l|} \hline
$L,~fb^{-1}$ & $\delta~Re f_{LV}$ / $\delta~Re f_{RV}$  & \hskip 2mm $\delta~Re f_{LV}$ / $\delta~Re f_{LT}$ & \hskip 2mm $\delta~Re f_{LV}$ / $\delta~Re f_{RT}$ \\ \hline
30 & $6.1\cdot10^{-4}$ / $8.2\cdot10^{-3}$ & $6.8\cdot10^{-4}$ / $6.4\cdot10^{-3}$ & $1.3\cdot10^{-3}$ / $1.8\cdot10^{-3}$ \\  
300 & $1.8\cdot10^{-4}$ / $3.1\cdot10^{-3}$ & $2.5\cdot10^{-4}$ / $2.3\cdot10^{-3}$ & $4.0\cdot10^{-4}$ / $5.6\cdot10^{-4}$ \\
3000 & $5.9\cdot10^{-5}$ / $7.1\cdot10^{-4}$ & $8.0\cdot10^{-5}$ / $8.1\cdot10^{-4}$ & $1.2\cdot10^{-4}$ / $1.7\cdot10^{-4}$ \\ \hline
\end{tabular}
\end{center}
\caption{\label{table2} The accuracy of measuring the two anomalous parameters by fitting in the 3D coordinate space ($\epsilon$, $\cos\theta$, $\phi$).}
\end{table}

It can be seen that fitting in 3D space ($\epsilon$,~$\cos\theta$,~$\phi$) made it possible to improve the accuracy of measuring the couplings $f_{LV}$, $f_{RV}$, and $f_{RT}$. But the accuracy of measuring the coupling $f_ {LT}$ was slightly worse than the results of fitting in two-dimensional space ($\epsilon$,~$\cos\theta$).
Despite the fact that the 3D histogram contains more information than the 2D histogram, the fitting in the case of 3D distribution is technically more complex. Here, much depends on the histogram binning and some other settings. By optimizing these parameters and also using information on the latest experimental limits, it is possible to further increase the measurement accuracy of the couplings. It can be seen that the predicted accuracy of measuring of anomalous couplings is much higher (for $Ref_ {LV}$ the accuracy is 50 times higher, for $Ref_ {RV}$ it is 20 times higher, for $Ref_ {LT}$ it is 12 times higher, and for $Ref_ {RT}$ it is 30 times higher) than the current available experimental accuracy \cite{Khachatryan:2016sib} with the corresponding value of the integral luminosity of the LHC. Of course, our simulation corresponds to the ideal case where we can accurately restore the t quark system and do not take into account the effects of the detector response. However, this illustration demonstrates the potential for increasing the measurement accuracy using the proposed method.

\section{Numerical illustration, imaginary couplings}
\label{sec:Numimagi}
In the previous examples, we considered cases where only $\phi$-even terms of formula (\ref{totalcrossec}) were involved.
To use the remaining terms proportional to $\sin\phi$, we must consider a scenario with nonzero anomalous imaginary couplings.
It should be noted that in the above simulation, we had in mind that the $\phi$ angle is restored by formula (\ref{cosphi}). With this approach, events corresponding to the angles $\phi$ and ($2\pi-\phi$) are counted as events with the same $\phi$. Therefore, integration of the differential cross section (\ref{totalcrossec}) over (\ref{cosphi}) from 0 to $\pi$ is equivalent to integration over angle $\phi$ from 0 to $2\pi$. This made sense in the considered cases where the $\phi$-even terms of formula (\ref{totalcrossec}) were involved. In this case, the analysis was simplified without loss of information but not in the case of the scenarios where terms proportional to $\sin\phi$ are involved. To reflect the contribution of such terms, it is necessary to carefully separate the events corresponding to angle $\phi$ in the range of 0 to $\pi$, and the events corresponding to $\phi$ ranging from $\pi$ to $2\pi$ by using triple product $T = (\bold p_{e^+}\times\bold p_b)\cdot \bold s$. If $T>0$: $\phi$ $\in$ (0, $\pi$). If $T<0$: $\phi$ $\in$ ($\pi$, $2\pi$).

\begin{figure}
	\begin{center}
		\begin{minipage}[t]{.325\linewidth}
			\centering
			\includegraphics[width=6cm,height=6cm]{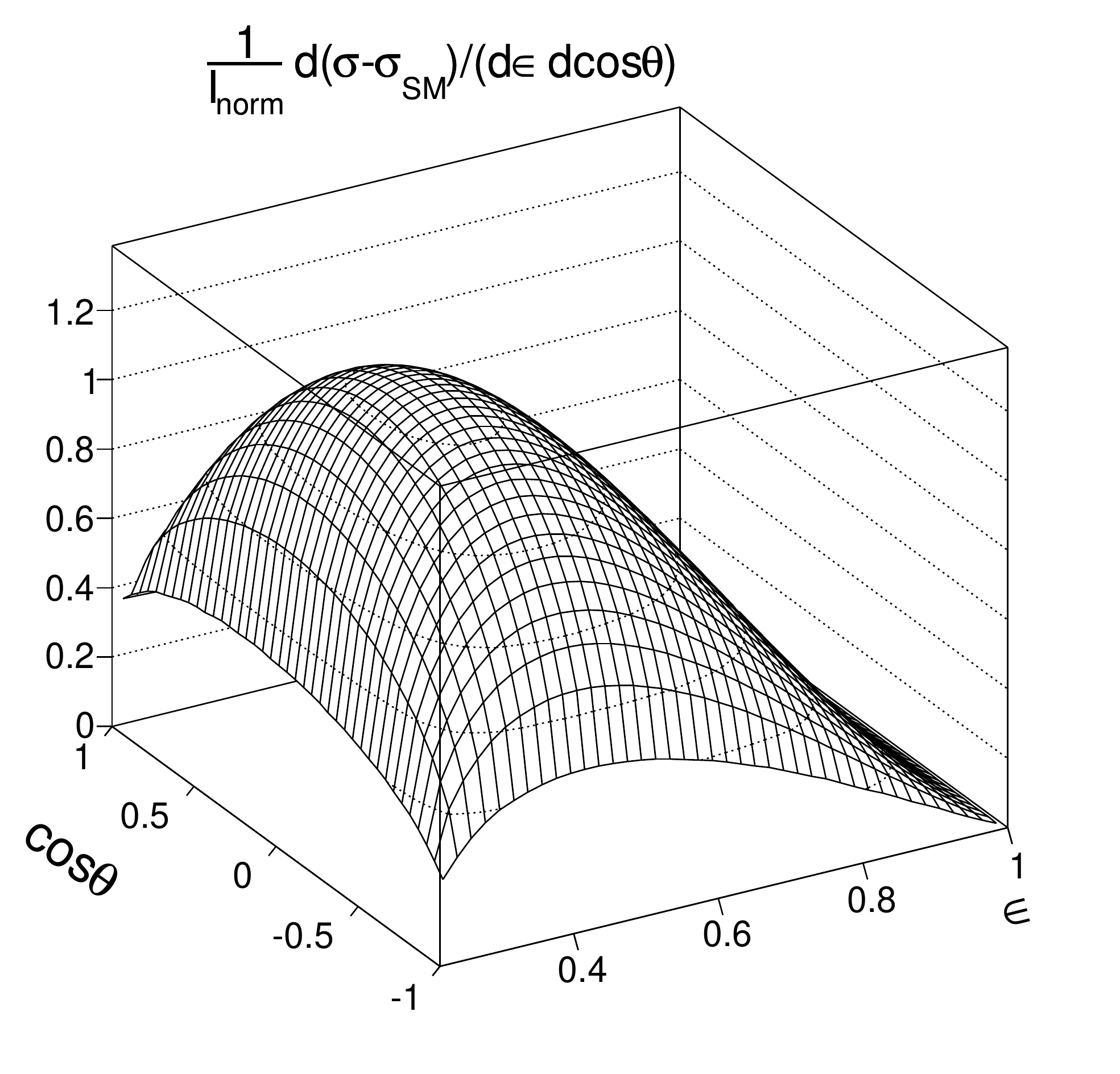}
		\end{minipage}
		\begin{minipage}[t]{.325\linewidth}
			\centering
			\includegraphics[width=6cm,height=6cm]{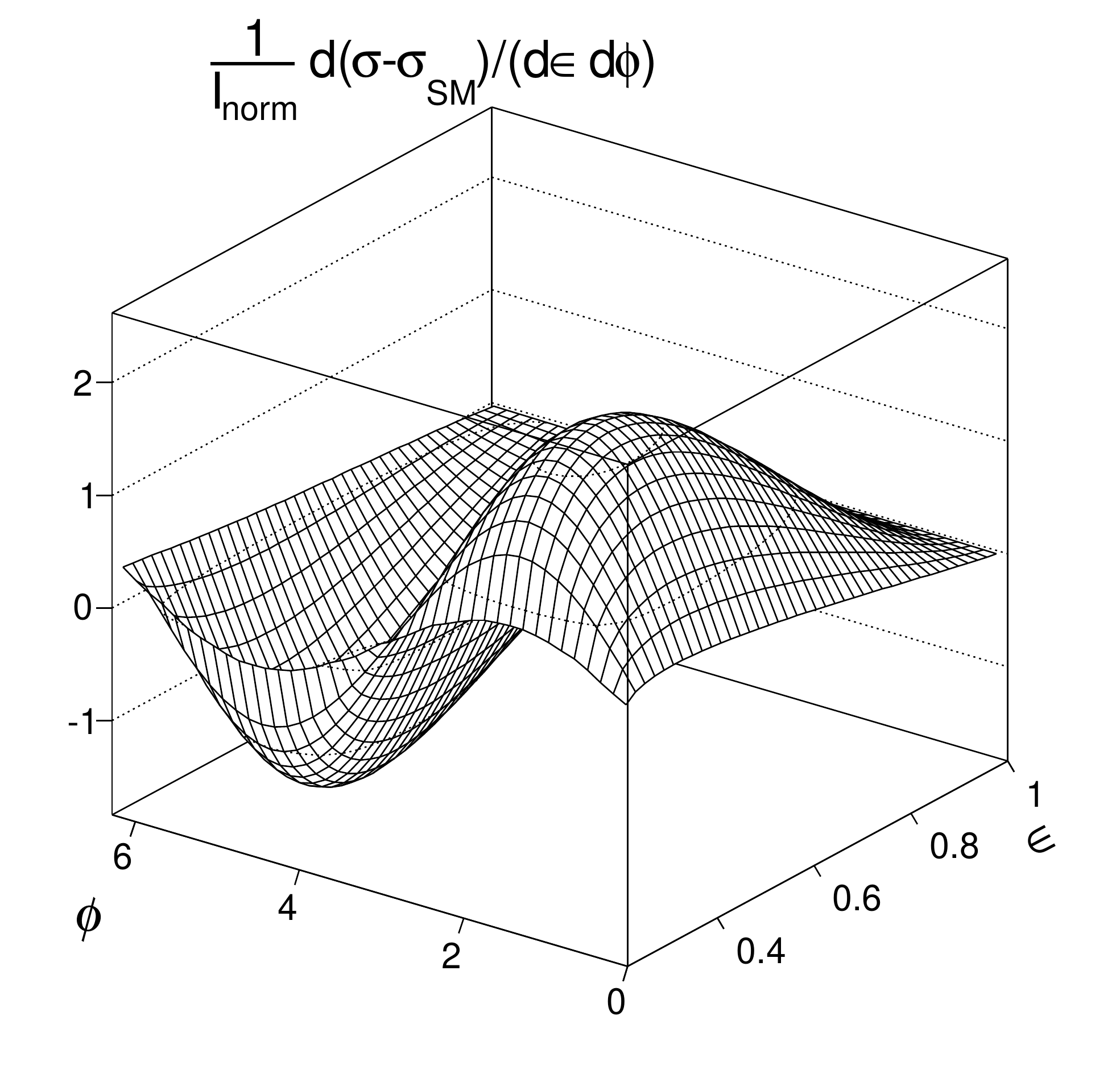}
		\end{minipage}
		\begin{minipage}[t]{.325\linewidth}
			\centering
			\includegraphics[width=6cm,height=6cm]{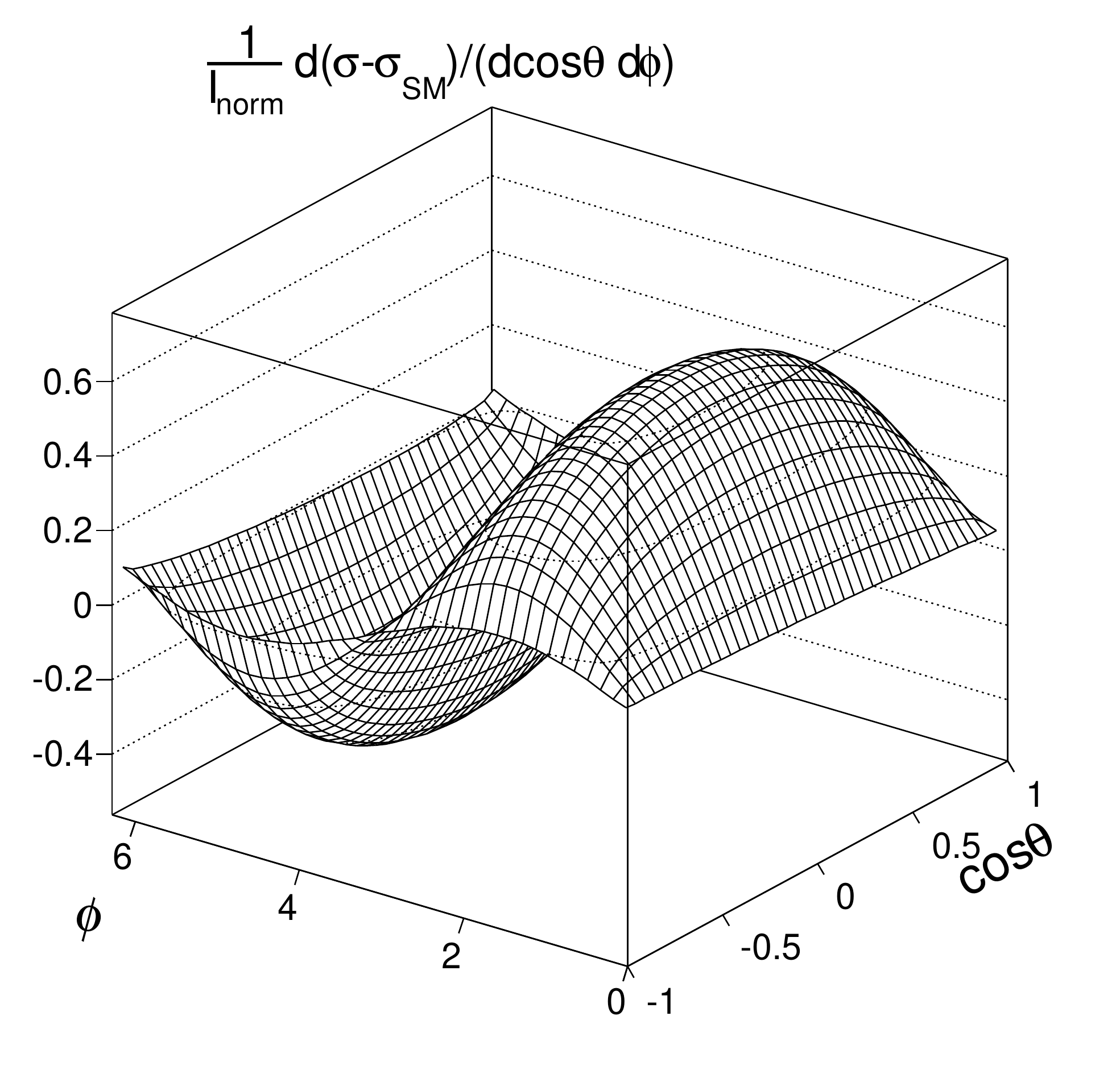}
		\end{minipage}
		\begin{minipage}[t]{.325\linewidth}
			\centering
			\includegraphics[width=6cm,height=6cm]{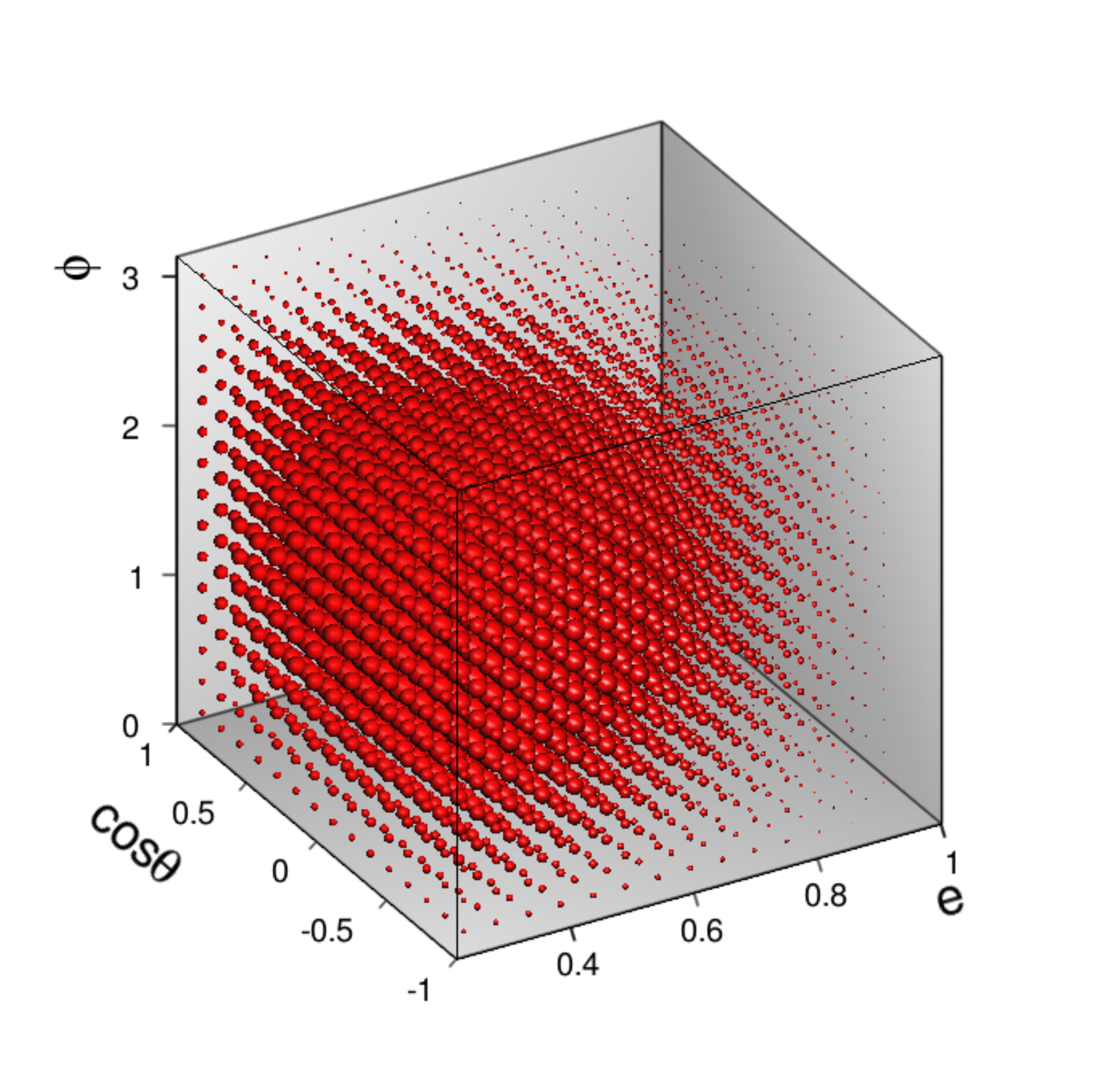}
		\end{minipage}
		\begin{minipage}[t]{.325\linewidth}
			\centering
			\includegraphics[width=6cm,height=6cm]{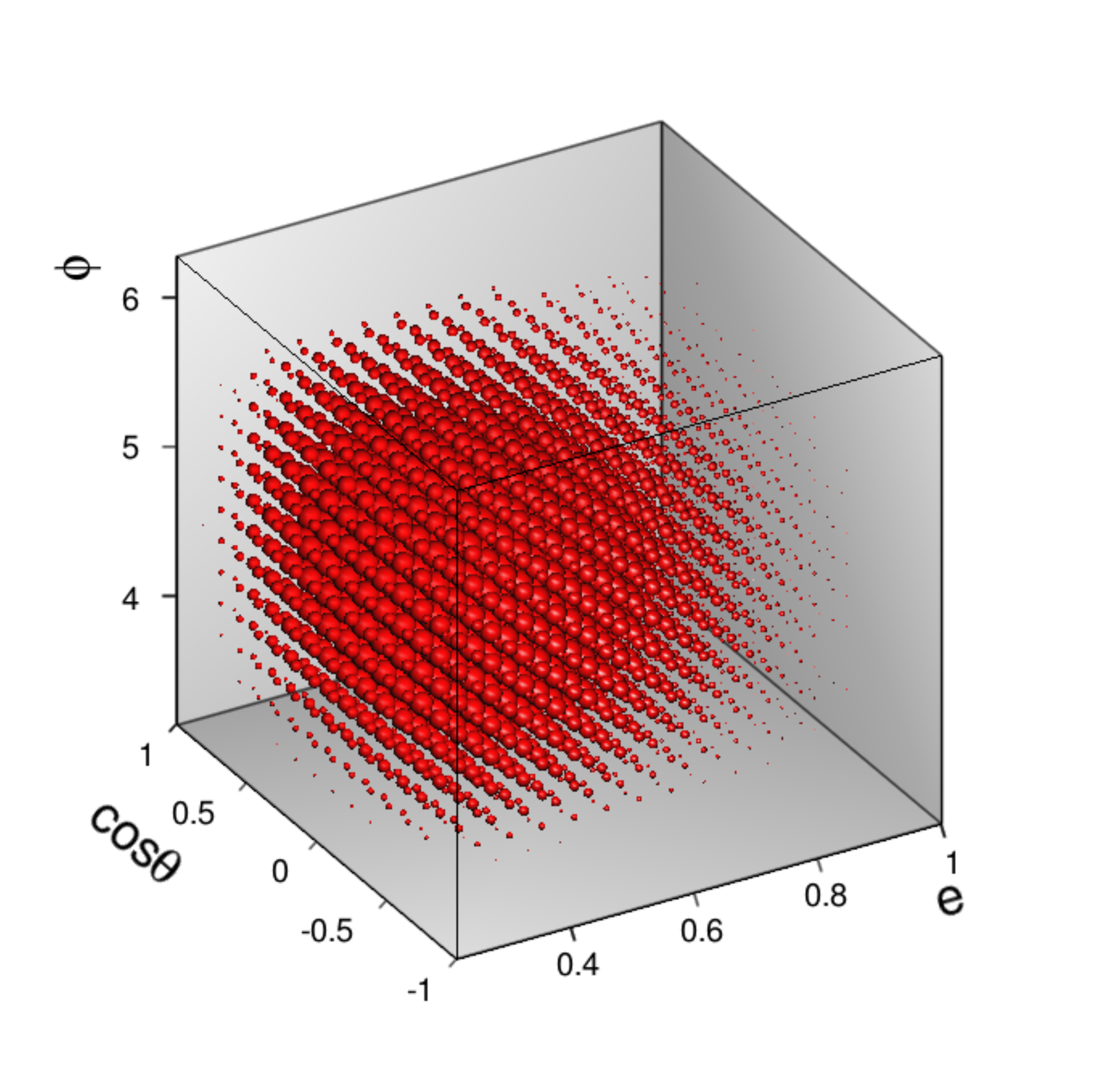}
		\end{minipage}	
	\end{center}
	\caption{ \label{pic12} \footnotesize Scenario $Imf_{RT}$ = - 0.048. The upper figures show plots of the normalized double-differential cross sections $\frac{d(\sigma-\sigma_{SM})}{d\epsilon~\cdot~d\cos\theta}$, ~$\frac{d(\sigma-\sigma_{SM})}{d\epsilon~\cdot~d\phi}$, and $\frac{d(\sigma-\sigma_{SM})}{d\cos\theta~\cdot~d\phi}$. The lower figures show the plots of probability density $\frac{d(\sigma-\sigma_{SM})}{d\epsilon~\cdot~d\cos\theta~\cdot~d\phi}$ in 3D space ($\epsilon$, $\cos\theta$, $\phi$). The lower left figure corresponds to the $\phi$ range (0, $\pi$), while the lower right figure corresponds to the $\phi$ range ($\pi$, $2\pi$). }
\end{figure}

As the first numerical illustration with imaginary couplings, we draw plots of double-differential cross sections (Fig.\ref{pic12}, upper plots) and plots of the probability density in 3D phase space (Fig.\ref{pic12}, lower plots) for the scenario where $Imf_{RT}$ is not equal to 0, and the remaining anomalous couplings are equal to 0. For a correct comparison with the case of a real coupling, we set the value of the imaginary coupling $Imf_{RT}$=-0.048 similar to the corresponding real one.
The upper plots of Fig.\ref{pic12} show $\frac{d(\sigma-\sigma_{SM})}{d\epsilon~\cdot~dcos\theta}$ corresponding to $\phi$ ranging from 0 to $\pi$, while $\frac{d(\sigma-\sigma_{SM})}{d\epsilon~\cdot~d\phi}$ and $\frac{d(\sigma-\sigma_{SM})}{dcos\theta~\cdot~d\phi}$ corresponding to $\phi$ ranging from 0 to $2\pi$. 
It can be seen that the deviation from the prediction of the  Standard Model due to anomalous contributions $(\sigma-\sigma_{SM})$ on the intervals of the angle $\phi$, $(0, \pi)$ and $(\pi, 2\pi)$ differ in sign for this scenario. 
One can also notice a slight difference in the absolute values of the deviations from the SM corresponding to different intervals of $\phi$ (Fig.\ref{pic12}, lower plots). 
This is due to the fact that the linear anomalous term from the first part of (\ref{diffcrossec}) is an odd function of $\phi$ and, when changing the interval of $\phi$, it changes sign, while the other quadratic anomalous terms from (\ref{diffcrossec}) are even and do not change. Therefore, the total sum of the even and odd terms will differ at different intervals, 
but since the linear term dominates, and the contribution of quadratic terms is substantially suppressed, this difference is small. 

\begin{figure}[htbp!]
	\begin{center}
		\begin{minipage}[t]{.325\linewidth}
			\centering
			\includegraphics[width=5.5cm,height=5.5cm]{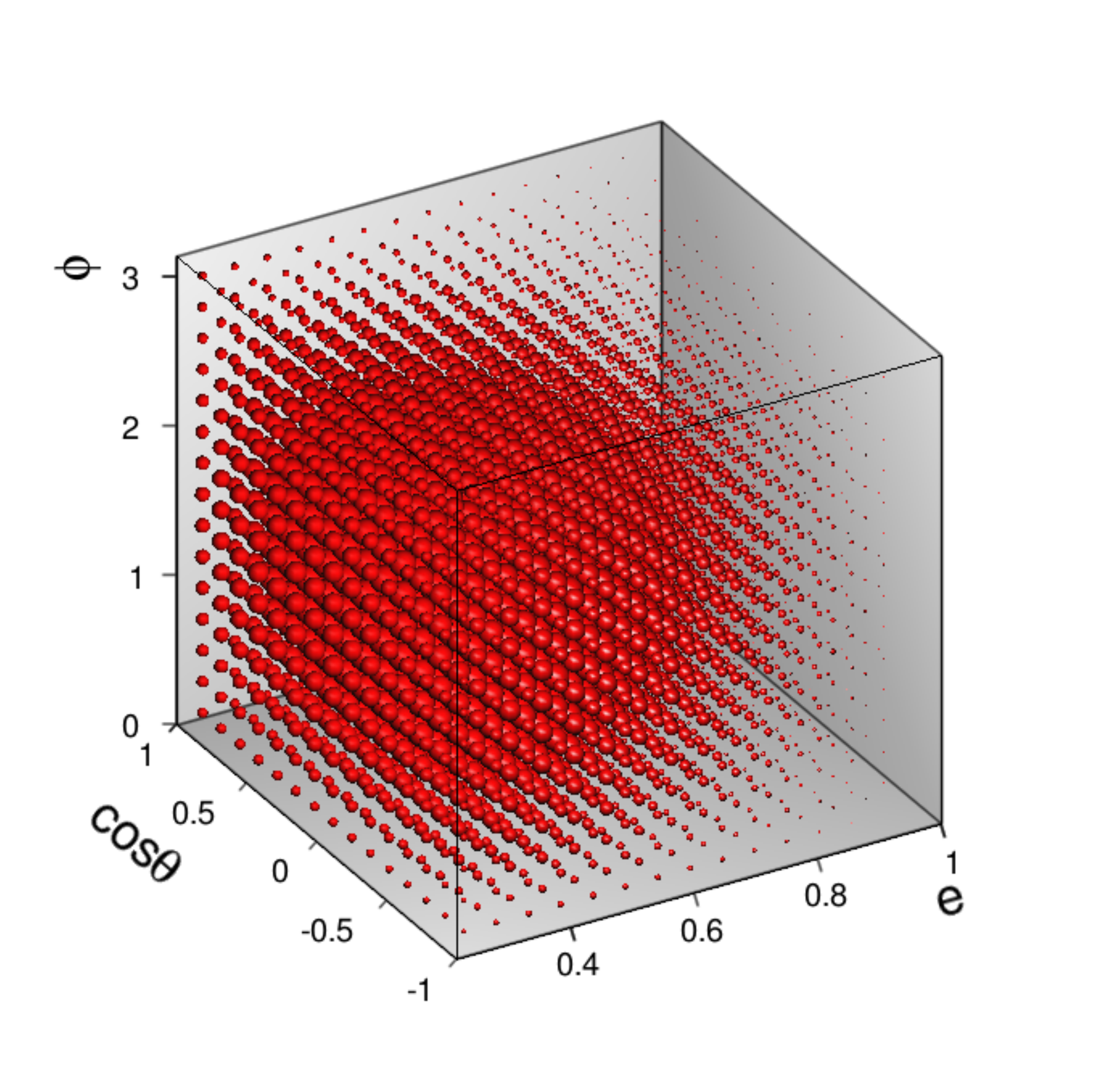}
		\end{minipage}
		\begin{minipage}[t]{.325\linewidth}
			\centering
			\includegraphics[width=5.5cm,height=5.5cm]{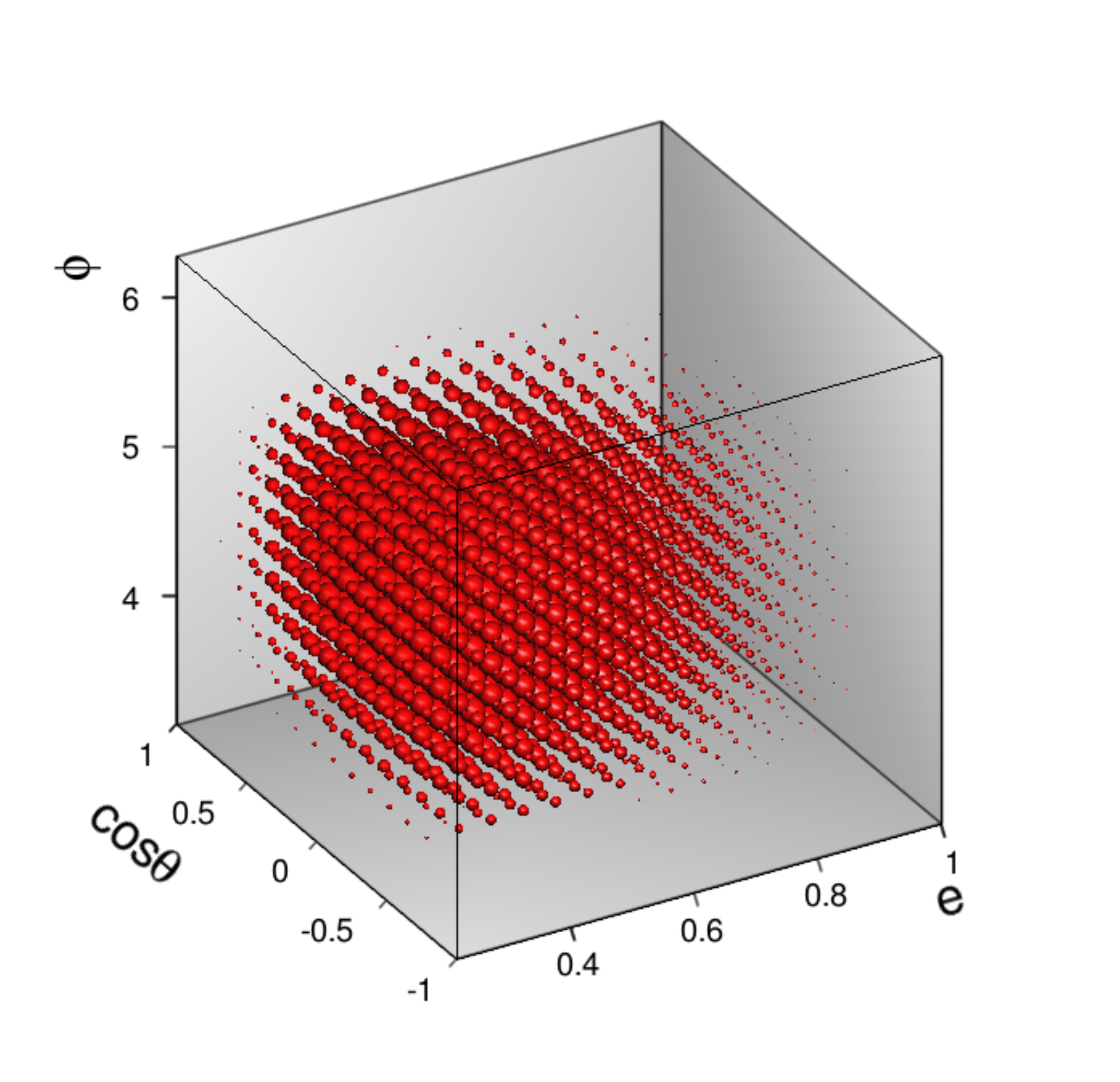}
		\end{minipage}	
	\end{center}
	\caption{ \label{pic13} \footnotesize Scenario ($Imf_{LV}$ = 0.03, $Imf_{RT}$ = - 0.048). Plots of probability density $\frac{d(\sigma-\sigma_{SM})}{d\epsilon~\cdot~d\cos\theta~\cdot~d\phi}$ in 3D space ($\epsilon$, $\cos\theta$, $\phi$). The left figure corresponds to the $\phi$ range (0, $\pi$), while the right figure corresponds to the $\phi$ range ($\pi$, $2\pi$). }
\end{figure}

For another example, when ($Imf_{LV}$ and $Imf_{RT}\ne 0$) the joint contribution of the quadratic terms becomes somewhat larger, and the difference in the absolute values of the deviations at different intervals increases (Fig.\ref{pic13}).

\begin{figure}[htbp!]
	\begin{center}
		\begin{minipage}[t]{.325\linewidth}
			\centering
			\includegraphics[width=6cm,height=6cm]{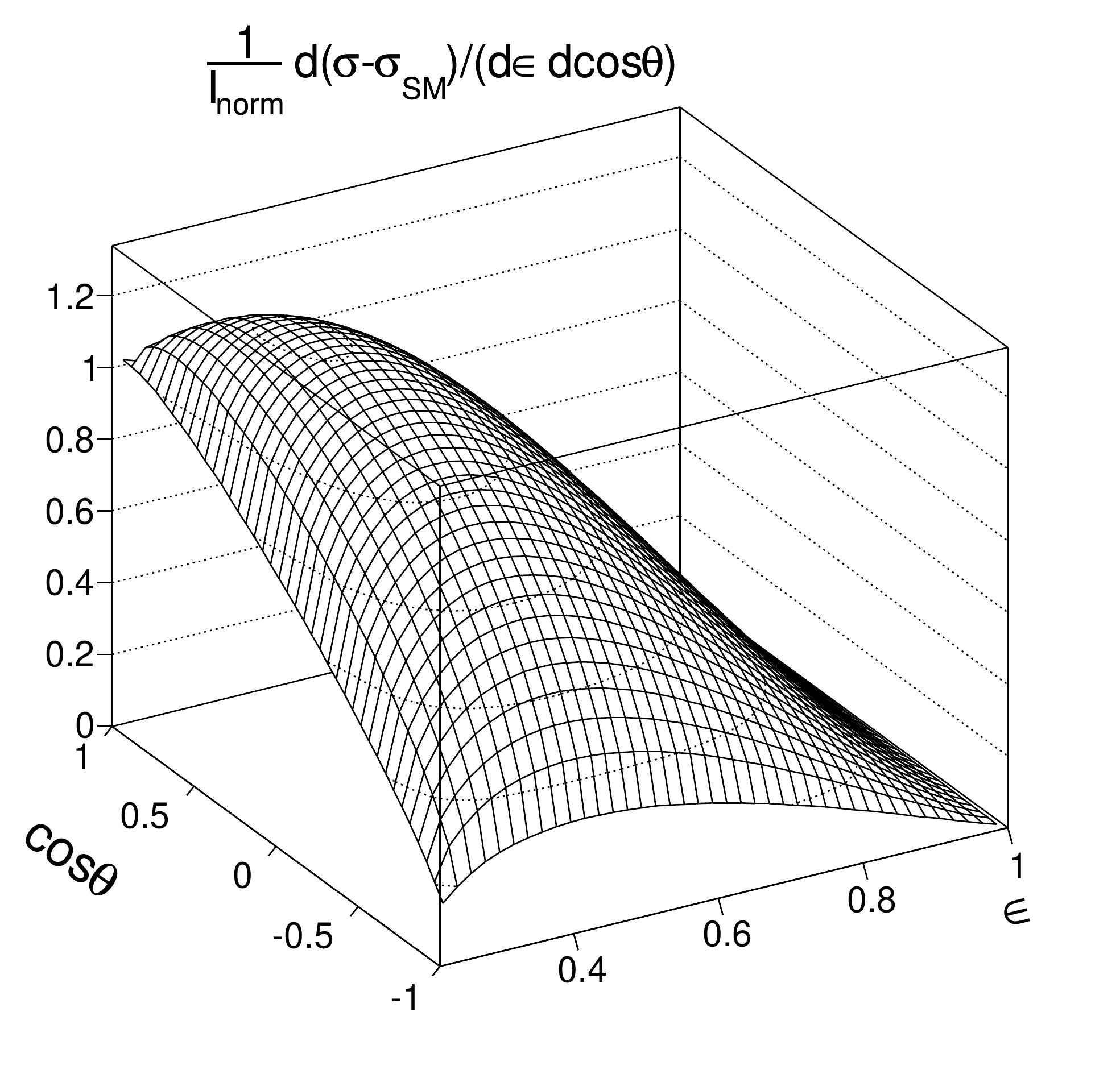}
		\end{minipage}
		\begin{minipage}[t]{.325\linewidth}
			\centering
			\includegraphics[width=6cm,height=6cm]{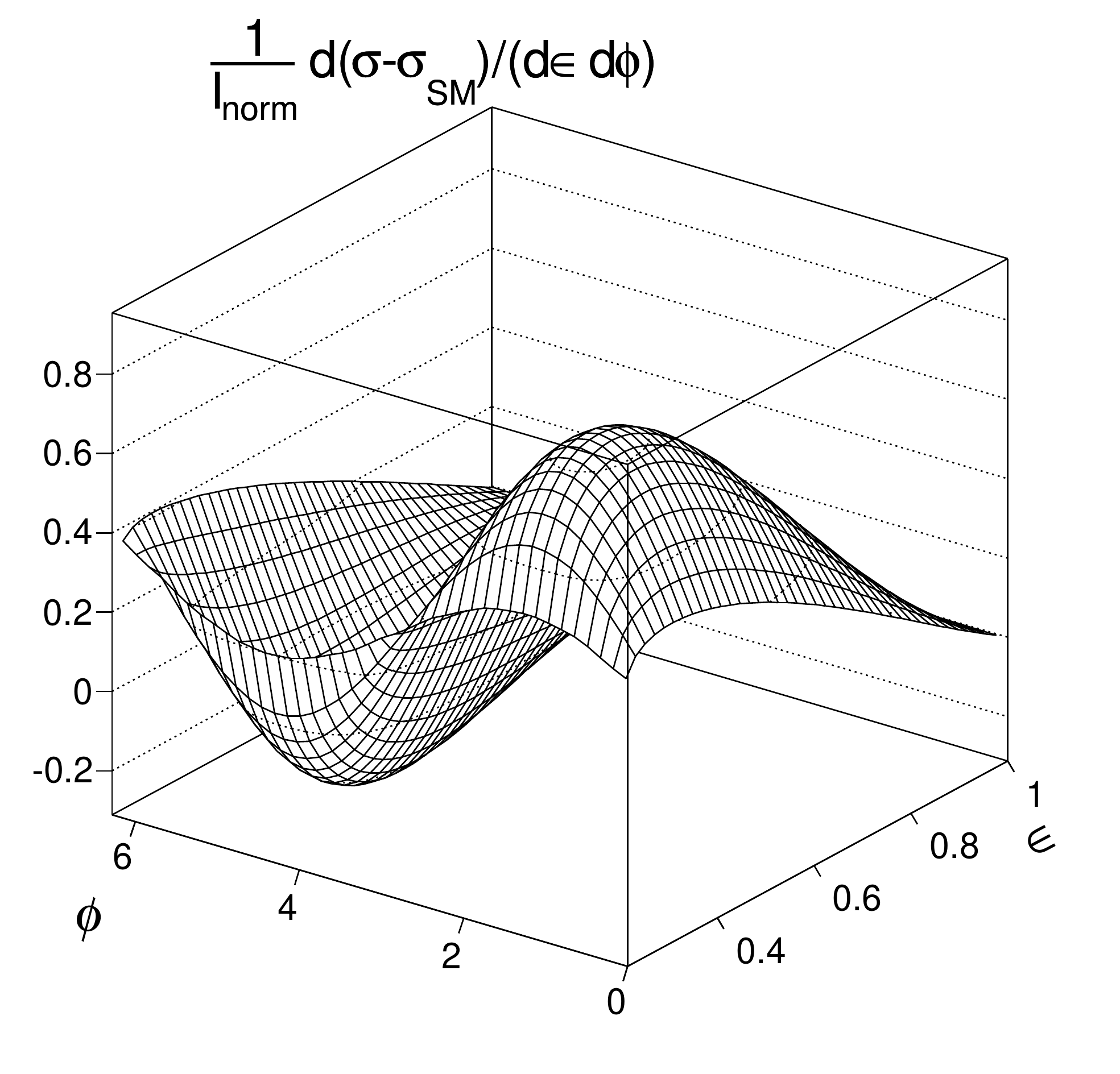}
		\end{minipage}
		\begin{minipage}[t]{.325\linewidth}
			\centering
			\includegraphics[width=6cm,height=6cm]{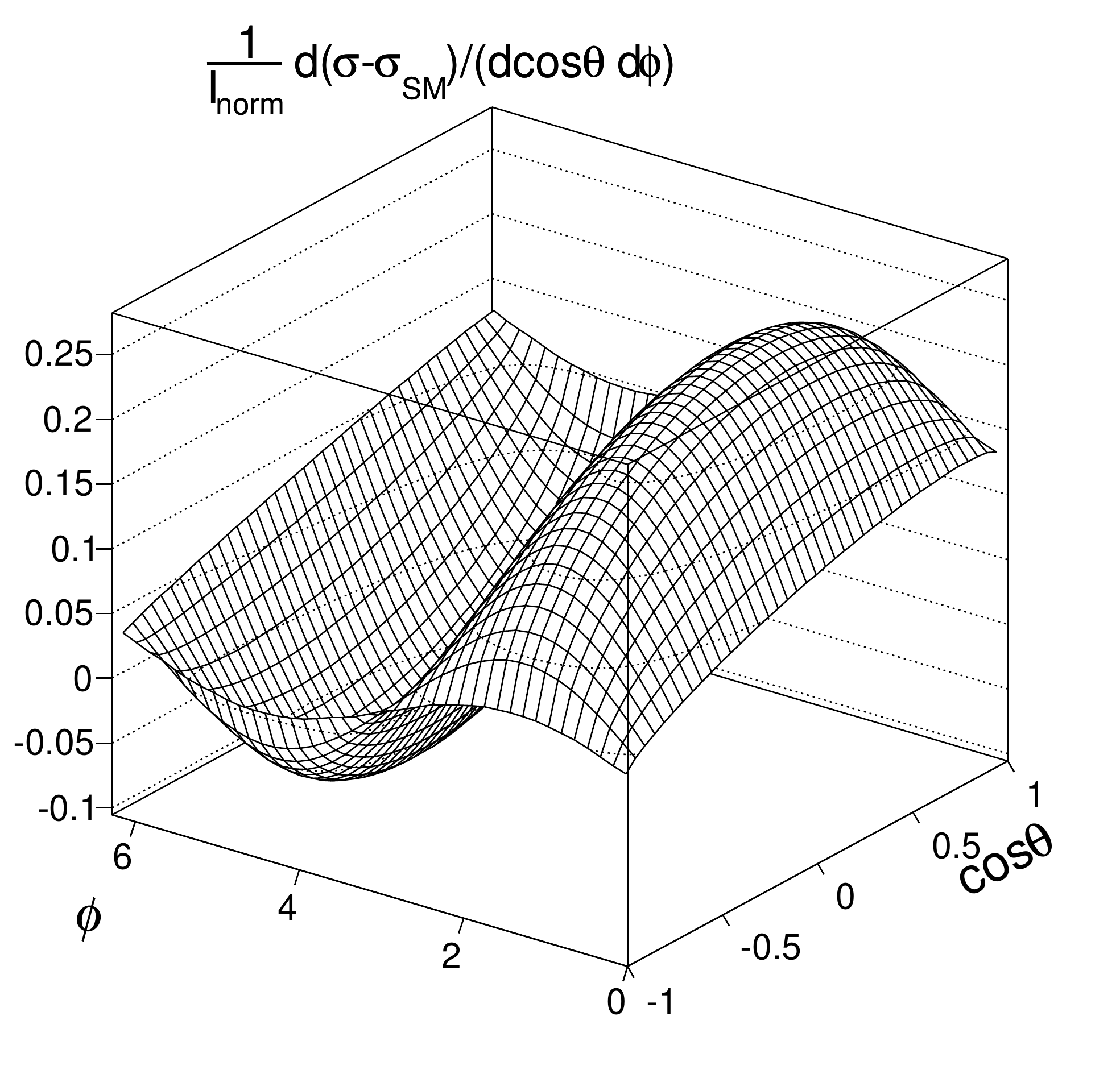}
		\end{minipage}
		\begin{minipage}[t]{.325\linewidth}
			\centering
			\includegraphics[width=6cm,height=6cm]{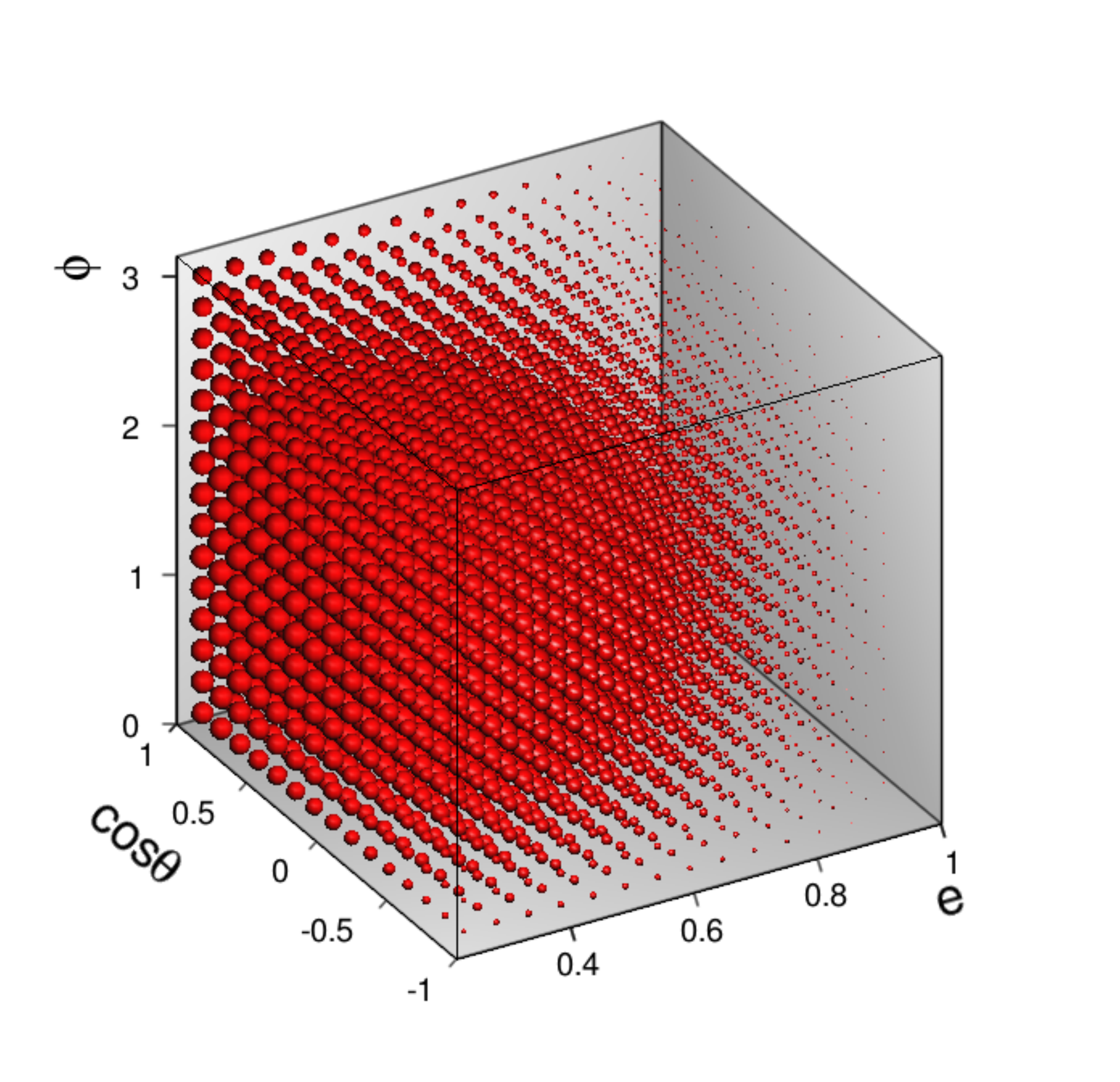}
		\end{minipage}
		\begin{minipage}[t]{.325\linewidth}
			\centering
			\includegraphics[width=6cm,height=6cm]{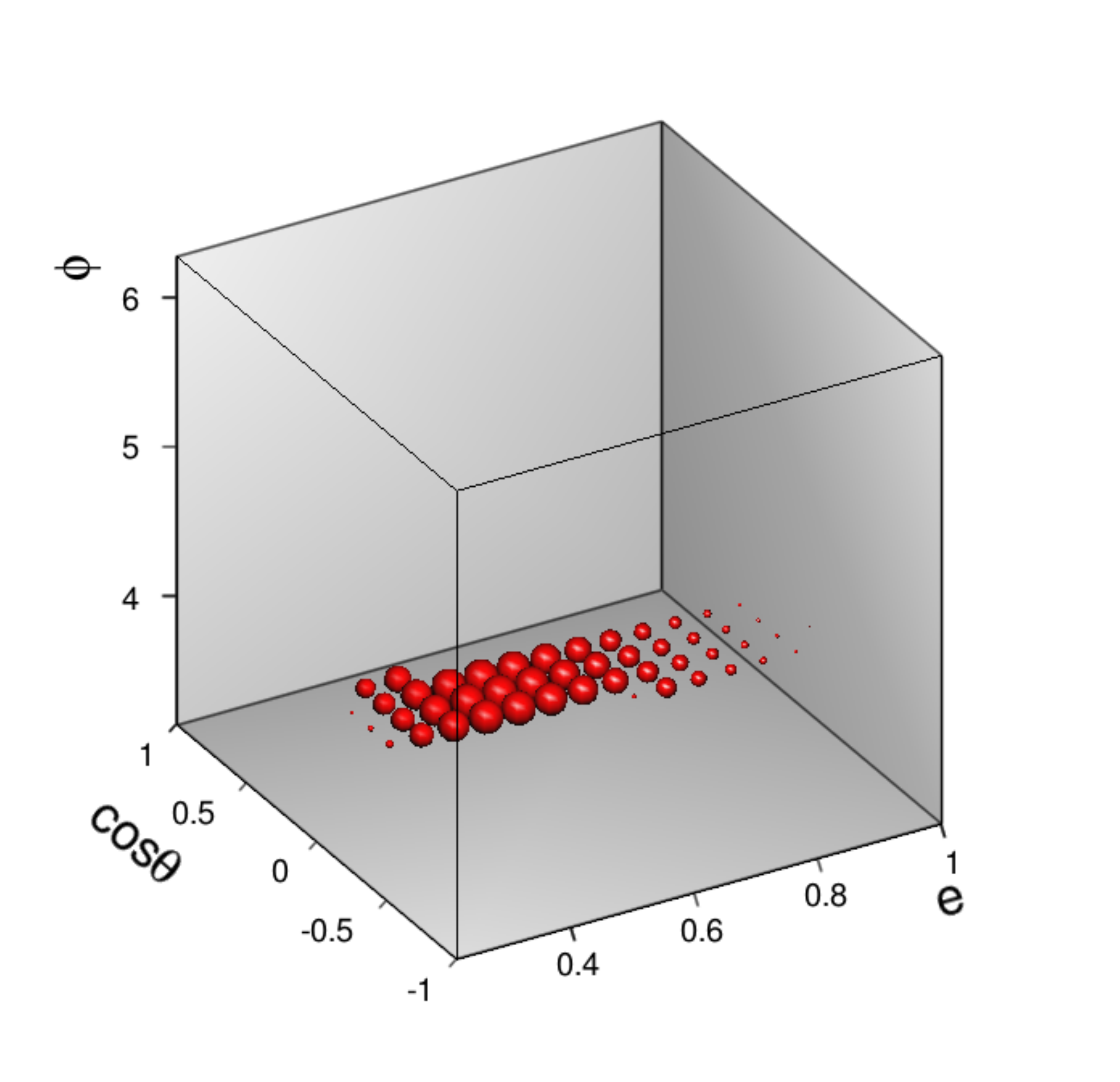}
		\end{minipage}	
	\end{center}
	\caption{ \label{pic14} \footnotesize Scenario ($Ref_{RT}$ = 0.00876, $Imf_{RT}$ = - 0.048). The upper figures show plots of the normalized double-differential cross sections $\frac{d(\sigma-\sigma_{SM})}{d\epsilon~\cdot~d\cos\theta}$, ~$\frac{d(\sigma-\sigma_{SM})}{d\epsilon~\cdot~d\phi}$, and $\frac{d(\sigma-\sigma_{SM})}{d\cos\theta~\cdot~d\phi}$. The lower figures show plots of probability density $\frac{d(\sigma-\sigma_{SM})}{d\epsilon~\cdot~d\cos\theta~\cdot~d\phi}$ in 3D space ($\epsilon$, $\cos\theta$, $\phi$). The lower left figure corresponds to the $\phi$ range (0, $\pi$), while the lower right figure corresponds to the $\phi$ range ($\pi$, $2\pi$). }
\end{figure}

It is also interesting to consider the case ($Ref_{RT}$ and $Imf_{RT}\ne 0$) when both linear terms of formula (\ref{twidth0}) are involved in the game, but one of them proportional to $Ref_{RT}$ is even, and the other one proportional to $Imf_{RT}$ is an odd function of $\phi$. It is possible to choose a combination of coupling values in which the even and odd components amplify each other in one interval $\phi$, $(0,\pi)$ and fully compensate each other in another interval $\phi$, $(\pi, 2\pi)$, as shown in Fig.\ref{pic14}.

\begin{figure}[htbp!]
	\begin{center}
		\begin{minipage}[t]{.325\linewidth}
			\centering
			\includegraphics[width=6cm,height=6cm]{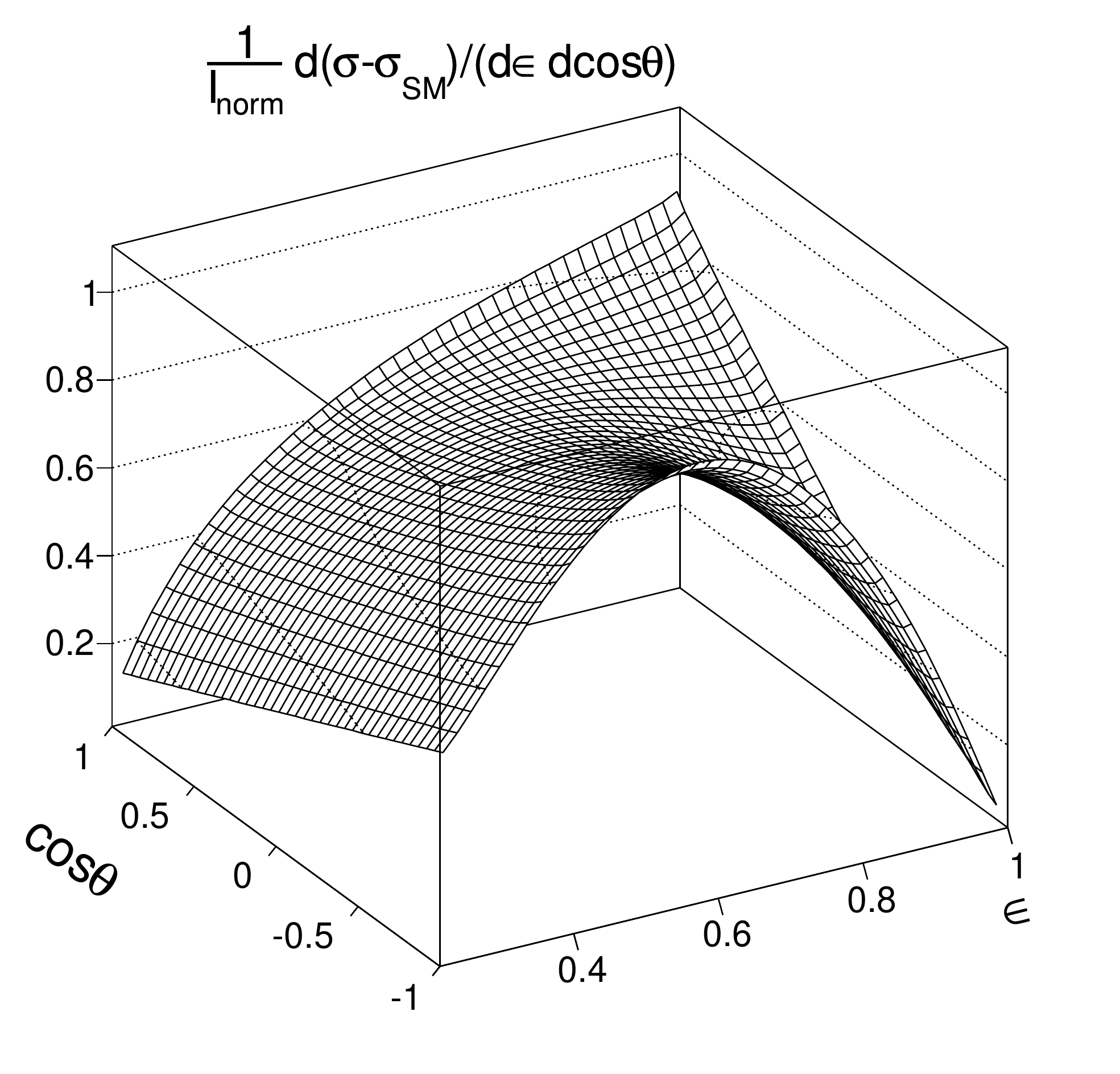}
		\end{minipage}
		\begin{minipage}[t]{.325\linewidth}
			\centering
			\includegraphics[width=6cm,height=6cm]{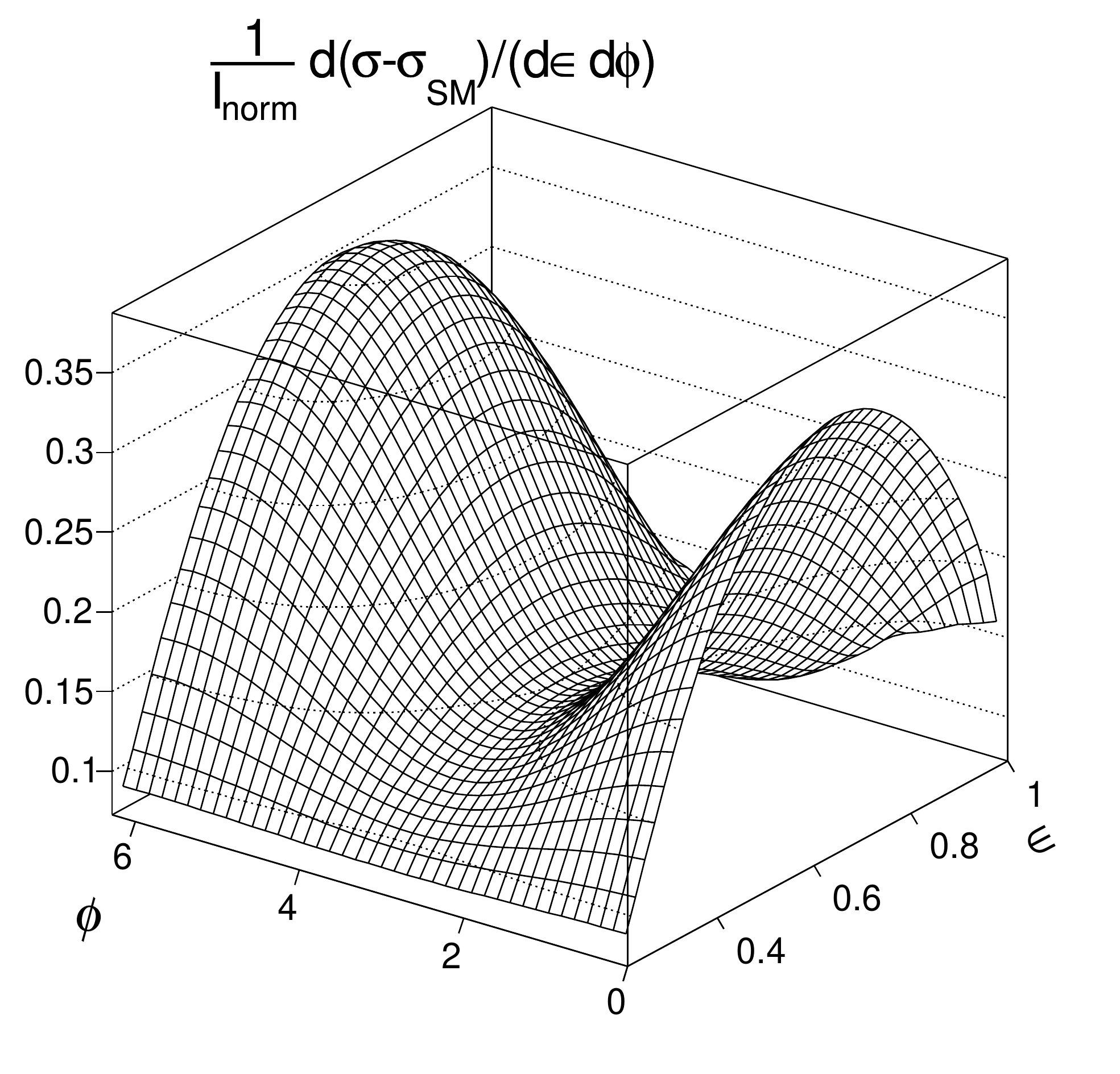}
		\end{minipage}
		\begin{minipage}[t]{.325\linewidth}
			\centering
			\includegraphics[width=6cm,height=6cm]{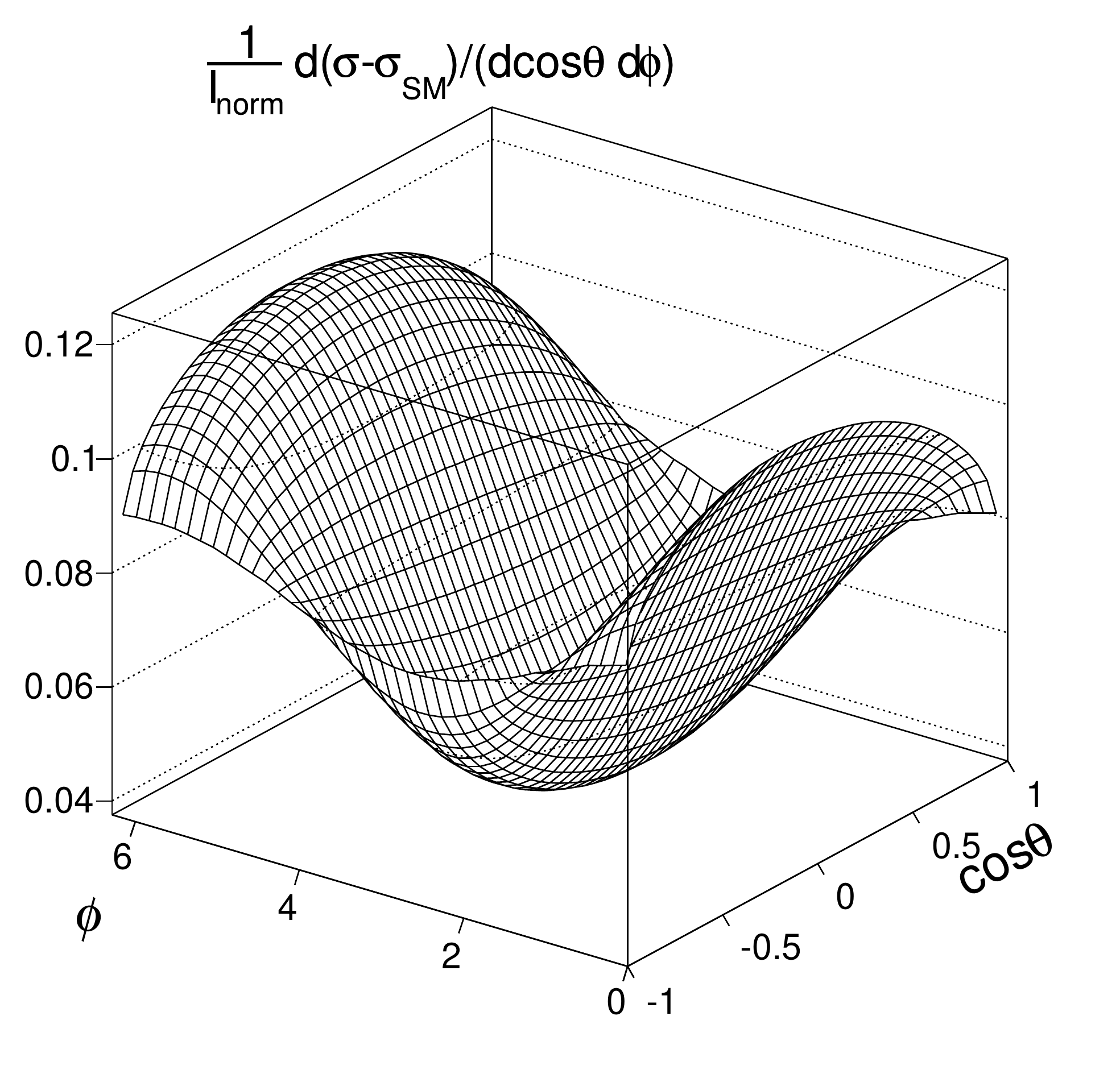}
		\end{minipage}
		\begin{minipage}[t]{.325\linewidth}
			\centering
			\includegraphics[width=6cm,height=6cm]{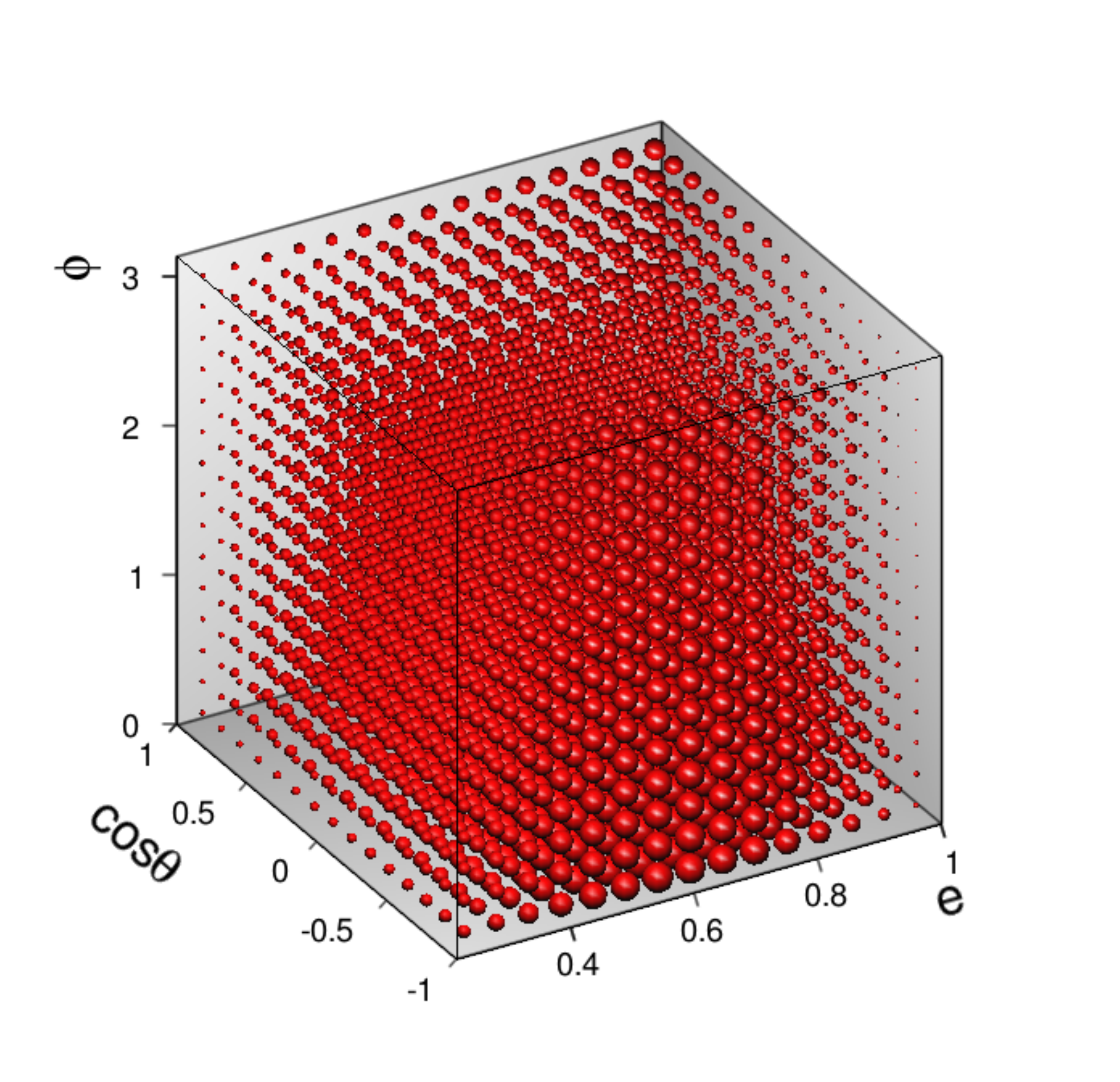}
		\end{minipage}
		\begin{minipage}[t]{.325\linewidth}
			\centering
			\includegraphics[width=6cm,height=6cm]{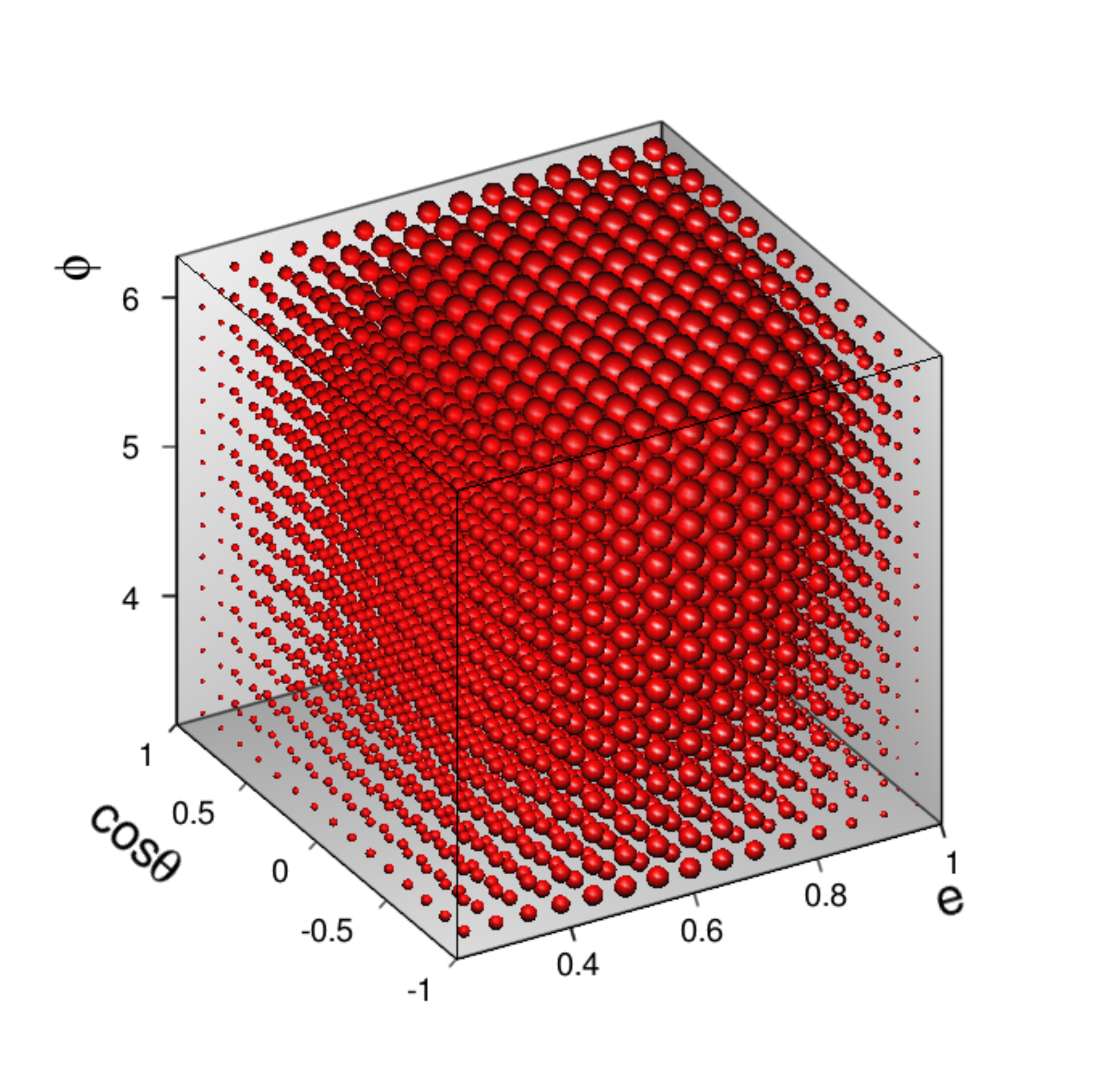}
		\end{minipage}	
	\end{center}
	\caption{ \label{pic15} \footnotesize Scenario ($Ref_{LT}$ = 0.057, $Imf_{RV}$ = 0.16). The upper figures show plots of the normalized double-differential cross sections $\frac{d(\sigma-\sigma_{SM})}{d\epsilon~\cdot~d\cos\theta}$, ~$\frac{d(\sigma-\sigma_{SM})}{d\epsilon~\cdot~d\phi}$, and $\frac{d(\sigma-\sigma_{SM})}{d\cos\theta~\cdot~d\phi}$. The lower figures show plots of probability density $\frac{d(\sigma-\sigma_{SM})}{d\epsilon~\cdot~d\cos\theta~\cdot~d\phi}$ in 3D space ($\epsilon$, $\cos\theta$, $\phi$). The lower left figure corresponds to the $\phi$ range (0, $\pi$), while the lower right figure corresponds to the $\phi$ range ($\pi$, $2\pi$). }
\end{figure}

Finally, we consider a scenario ($Ref_{LT}$ and $Imf_{RV}\ne 0$) where linear anomalous terms are absent, and even and odd components are represented by quadratic terms only (Fig.\ref{pic15}). It can be seen that in this case the even terms dominate; therefore, the deviation from the SM does not change sign on all intervals of the $\phi$ angle. However, the contribution of the odd components is manifested in the asymmetry of the deviations at different intervals of the $\phi$ angle.

As the main conclusion of this part, we note that various combinations of real and imaginary couplings belonging to the terms of formula (\ref{twidth0}) that are odd with respect to the $\phi$ angle can manifest themselves in the form of asymmetries of differential cross sections at different intervals of $\phi$: $(0, \pi)$ and $(\pi, 2\pi)$. The shapes of such distributions differ significantly from the corresponding distributions for cases of pure real couplings, which will allow us to experimentally detect and identify imaginary couplings.

Similar to the considered case of real couplings, we apply the method of fitting the Monte Carlo events with the obtained analytical formula  to estimate the accuracy of measuring imaginary anomalous couplings. Accuracy values for the joint measurement of $Ref_{LV}$ and $Imf_{RT}$ couplings by fitting in 3D space ($\epsilon$,~$\cos\theta$,~$\phi$) are given in Table~\ref{table3}.
%==============================
\begin{table}
	\begin{center}
		\begin{tabular}{|l|l|} \hline
			$L,~fb^{-1}$ & $\delta~Re f_{LV}$ / $\delta~Im f_{RT}$\\ \hline
			30 & $7.5\cdot10^{-4}$ / $3.4\cdot10^{-3}$\\  
			300 & $2.4\cdot10^{-4}$ / $1.0\cdot10^{-3}$\\
			3000 & $7.1\cdot10^{-5}$ / $3.3\cdot10^{-4}$\\ \hline
		\end{tabular}
	\end{center}
	\caption{\label{table3} The accuracy of measuring the two anomalous parameters by fitting in the 3D coordinate space ($\epsilon$,~$\cos\theta$,~$\phi$).}
\end{table}
%=========================================================================
%=========================================================================
%=========================================================================
%=========================================================================
%=========================================================================

\newpage

\section{Conclusions}
\label{sec:Conclusions} 
We obtained an analytical expression for the differential width of the three-particle decay of a polarized t quark in its rest frame as a function of the energy of a charged lepton and two angles of orientation of the quantization axis of the t-quark spin. The expression was presented in the most general form for the case of real and imaginary vector and tensor anomalous \textit{Wtb} couplings. The parts of this expression containing the contribution of the SM and its interference with the anomalous contributions are fully consistent with the published results.
We showed that the expression for the differential width of the t quark can be divided into eight kinematically different terms corresponding to possible combinations of anomalous parameters. These eight terms have different dependences on distribution variables and we suggest using them as basis functions when fitting experimental data and extracting anomalous parameter values. Also, these eight analytic functions can be used as multidimensional variables for analysis based on the neural network method. This makes it possible to most effectively separate various anomalous contributions.

In addition, expressions for the differential width of the t quark were obtained as functions of various combinations of two variables: the energy of the charged lepton and one of the t-quark spin orientation angles, as well as two spin orientation angles. 
Using the obtained analytical expressions, we constructed various two-dimensional plots corresponding to different anomalous scenarios. The most noticeable differences appear in the shape of the surfaces of two-dimensional distributions, where one of the variables is the energy of a charged lepton, and the other is one of the t-quark spin orientation angles. At the same time, the dependence of the t-quark width only on two angles is less informative and does not effectively separate one anomalous component of the width from the other.

Also, we showed that the formulas obtained for the differential width of the t quark can be used to derive the differential cross section for the full process of the production and decay of the t quark ($u b \to d, b, \nu, l^+$) where the optimal direction of the quantization axis of the t quark in its rest frame is the direction of the d quark from the t-quark production.

To verify the obtained analytical results, we performed a numerical simulation of the full t-channel processes of t-quark production and decay at the LHC collider for various scenarios with real and imaginary anomalous couplings. The values of the recent experimental upper limits were taken as the values of the anomalous parameters.
Based on the Monte Carlo generators created by CompHEP, various two-dimensional distributions with respect to the energy of the charged lepton and the orientation angles of the t-quark spin in its rest frame were constructed. Also, for the considered scenarios, using the obtained analytical expressions for the t-quark production and decay differential cross sections, the corresponding two-dimensional plots were constructed. A comparison of the shapes of Monte Carlo distributions with the plots obtained from the formulas showed their full agreement with each other and confirmed the correctness of the analytical calculations performed.

In addition, using the obtained analytical expressions, we estimated the accuracy of extracting the values of the anomalous \textit{Wtb} couplings for different levels of the integral luminosity of the LHC collider using fitting methods. The predicted accuracy values are much higher (for $Ref_ {LV}$, the accuracy is 50 times higher; for $Ref_ {RV}$, the accuracy is 20 times higher; for $Ref_ {LT}$, the accuracy is 12 times higher; for $Ref_ {RT}$, the accuracy is 30 times higher) than the current experimental accuracy for the same integral luminosity. Despite the ideal nature of our theoretical experiment, it showed the potential for improving the accuracy of measurements using the method we proposed in addition to the experimental methods already used.

\section{Acknowledgements}
The research was supported by the Russian Science Foundation [http://dx.doi.org/10.13039/501100006769] Grant No. 16-12-10280. The authors are grateful to L. Dudko and Y. Kurihara for useful discussions and critical remarks.

\appendix

\section{Derivation of a symbolic expression for fully differential single top quark production cross section}
We explicitly express scalar products,
containing the t-quark spin vector,
through 4-momentum components
in the t-quark decay matrix element (\ref{melement}).
The matrix element of the polarized t-quark decay takes the form

\begin{align}\label{decme}
|M|^2_{t \to b \nu e^+}~=&~~~~F_L^{dec}\cdot(1+\cos\theta_{e^+s}) + F_R^{dec}\cdot (1-\cos\theta_{\nu s}) \\ \nonumber&
+ F_{int1}^{dec}\cdot (1+\cos\theta_{e^+ s}-\cos\theta_{\nu e^+}-\cos\theta_{\nu s})
+F_{int2}^{dec}\cdot\sin\theta_{be^+}\sin\phi_{e^+s}\sin\theta_{e^+s} 
\end{align}
where
\begin{align}
&F_L^{dec}~=~K^{dec}\cdot\left(|f_{LV}|^2\cdot(p_b p_{\nu})
+ |f_{LT}|^2\cdot\frac{2}{m_W^2} \cdot (p_b p_{e^+})(p_{\nu} p_{e^+})\right)\cdot m_t\cdot E_{e^+}~, \\
\nonumber &
F_R^{dec}~=~K^{dec}\cdot\left(|f_{RV}|^2\cdot(p_b p_{e^+})
+ |f_{RT}|^2\cdot\frac{2}{m_W^2} \cdot (p_b p_{\nu})(p_{\nu} p_{e^+})\right)\cdot m_t\cdot E_{\nu}~, \\
\nonumber &
F_{int1}^{dec}~=~K^{dec}\cdot(Re f_{LV}\cdot Re f_{RT} + Im f_{LV}\cdot Im f_{RT})\cdot\frac{2}{m_W}
\cdot (p_b p_{\nu})\cdot m_t\cdot E_{\nu} E_{e^+} \\
\nonumber &
~~~~~~~~+ K^{dec}\cdot(Re f_{LT} \cdot Re f_{RV} + Im f_{LT}\cdot Im f_{RV})\cdot\frac{2}{m_W} \cdot (p_b p_{e^+}) \cdot m_t\cdot E_{\nu} E_{e^+}~,\\
\nonumber &
F_{int2}^{dec}~=~K^{dec}\cdot(Re f_{LV}\cdot Im f_{RT} - Im f_{LV}\cdot Re f_{RT})\cdot\frac{-2}{m_W}\cdot\left( \frac{m_t^2}{2}-(p_{e^+} p_t) \right)\cdot m_t\cdot E_{b} E_{e^+} \\
\nonumber &
~~~~~~~~+ K^{dec}\cdot(Re f_{LT}\cdot Im f_{RV}  - Im f_{LT}\cdot Re f_{RV}  )\cdot\frac{-2}{m_W} \cdot\left((p_b p_t) + (p_{e^+} p_t) - \frac{m_t^2}{2}\right)\cdot m_t\cdot E_{b} E_{e^+}~,\\
\nonumber &
K^{dec}~=~\frac{g^4}{\left(2(p_{\nu} p_{e^+})-m_W^2\right)^2 + \Gamma_W^2 m_W^2}~.
\end{align}
Changing $e^+$ to $d$ and $\nu$ to $u$ in (\ref{decme}) and using crossing symmetry, we write in the same manner the matrix element of the t-channel production of the polarized t quark.
In this case, the components of the 4-momentum are also written in the t-quark rest frame

\begin{align}
|M|^2_{ub\to td}~=&~~~~F_L^{prod}\cdot(1+\cos\theta_{ds}) + F_R^{prod}\cdot (1-\cos\theta_{u s}) \\ \nonumber&
+ F_{int1}^{prod}\cdot (1+\cos\theta_{d s}-\cos\theta_{u d}-\cos\theta_{u s})
+F_{int2}^{prod}\cdot\sin\theta_{bd}\sin\phi_{ds}\sin\theta_{ds} 
\end{align}
where
\begin{align}
&F_L^{prod}~=~K^{prod}\cdot\left(|f_{LV}|^2\cdot(p_b p_{u})
+ |f_{LT}|^2\cdot\frac{2}{m_W^2} \cdot (p_b p_{d})(p_{u} p_{d})\right)\cdot m_t\cdot E_d~, \\
\nonumber &
F_R^{prod}~=~K^{prod}\cdot\left(|f_{RV}|^2\cdot(p_b p_{d})
+ |f_{RT}|^2\cdot\frac{2}{m_W^2} \cdot (p_b p_{u})(p_{u} p_{d})\right)\cdot m_t\cdot E_{u}~, \\
\nonumber &
F_{int1}^{prod}~=~K^{prod}\cdot(Re f_{LV}\cdot Re f_{RT} + Im f_{LV}\cdot Im f_{RT})\cdot\frac{2}{m_W}
\cdot (p_b p_{u})\cdot m_t\cdot E_{u} E_d \\
\nonumber &
~~~~~~~~+ K^{prod}\cdot(Re f_{LT} \cdot Re f_{RV} + Im f_{LT}\cdot Im f_{RV})\cdot\frac{2}{m_W} \cdot (p_b p_{d}) \cdot m_t\cdot E_{u} E_d~,\\
\nonumber &
F_{int2}^{prod}~=~K^{prod}\cdot(Re f_{LV}\cdot Im f_{RT} - Im f_{LV}\cdot Re f_{RT})\cdot\frac{-2}{m_W}\cdot\left( \frac{m_t^2}{2}-(p_{d} p_t) \right)\cdot m_t\cdot E_{b} E_d \\
\nonumber &
~~~~~~~~+ K^{prod}\cdot(Re f_{LT}\cdot Im f_{RV}  - Im f_{LT}\cdot Re f_{RV}  )\cdot\frac{-2}{m_W} \cdot\left((p_b p_t) + (p_{d} p_t) - \frac{m_t^2}{2}\right)\cdot m_t\cdot E_{b} E_d~,\\ \nonumber &
K^{prod}~=~\frac{g^4\cdot V_{ud}^2}{\left(2(p_{u} p_{d})-m_W^2\right)^2}~.
\end{align}
The matrix element of the complete process of single top production with its subsequent decay has the form

\begin{align}
&|M|^2_{u b \to d b \nu e^+}~~=~~D_t\cdot|\sum\limits_{s}M_{u b \to d t}\cdot M_{t \to b \nu e^+}|^2
\end{align}
where:
\begin{align}
&D_t~=~\frac{1}{\left((p_b + p_{\nu} + p_{e^+})^2-m_t^2\right)^2 + \Gamma_t^2 m_t^2}
\end{align}
After squaring the amplitude of the complete process,
the remaining nonzero parts can be expressed in terms of the matrix elements of polarized production and polarized decay of the t quark:

\begin{align}\label{totalme}
|M|^2_{u b \to d b \nu e^+}~=~&\sum\limits_{s}~~\big[
~~\big(~F_L^{prod}\cdot(1+\cos\theta_{ds}) + F_R^{prod}\cdot (1-\cos\theta_{u s}) \\ \nonumber&
~~~~~+ F_{int1}^{prod}\cdot (1+\cos\theta_{d s}-\cos\theta_{u d}-\cos\theta_{u s})
+F_{int2}^{prod}\cdot\sin\theta_{bd}\sin\phi_{ds}\sin\theta_{ds}~\big)  \\ \nonumber&
~~~~\times \\ \nonumber&
~~~~~D_t\cdot\big(~F_L^{dec}\cdot(1+\cos\theta_{e^+s}) + F_R^{dec}\cdot (1-\cos\theta_{\nu s}) \\ \nonumber&
~~~~~+ F_{int1}^{dec}\cdot (1+\cos\theta_{e^+ s}-\cos\theta_{\nu e^+}-\cos\theta_{\nu s})
+F_{int2}^{dec}\cdot\sin\theta_{be^+}\sin\phi_{e^+s}\sin\theta_{e^+s}~\big)~~] \\ \nonumber&
\end{align}
We chose the direction of the d-quark three-momentum as the quantization axis of the t-quark spin 
and we explicitly write out two components of expression (\ref{totalme}). The t-quark spin projection is directed along the d-quark momentum in the first component and against the d-direction in the second one:

\begin{align}\label{totalme1}
|M|^2_{u b \to d b \nu e^+}~&=~\big(~~F_L^{prod}\cdot 2 + F_R^{prod}\cdot (1-\cos\theta_{u d}) 
+ F_{int1}^{prod}\cdot 2\cdot(1-\cos\theta_{u d})
~~\big)  \\ \nonumber&
~\times D_t\cdot
\big(~~F_L^{dec}\cdot(1+\cos\theta_{e^+d}) + F_R^{dec}\cdot (1-\cos\theta_{\nu d}) \\ \nonumber&
~~~~~+ F_{int1}^{dec}\cdot (1+\cos\theta_{e^+ d}-\cos\theta_{\nu e^+}-\cos\theta_{\nu d})
+F_{int2}^{dec}\cdot\sin\theta_{be^+}\sin\phi_{e^+d}\sin\theta_{e^+d}~~\big) \\ \nonumber&
+\\ \nonumber&
~~~~~F_R^{prod}\cdot (1+\cos\theta_{u d}) \\ \nonumber&
~\times D_t\cdot
\big(~~F_L^{dec}\cdot(1-\cos\theta_{e^+d}) + F_R^{dec}\cdot (1+\cos\theta_{\nu d}) \\ \nonumber&
~~~~~+ F_{int1}^{dec}\cdot (1-\cos\theta_{e^+ d}-\cos\theta_{\nu e^+}+\cos\theta_{\nu d})
+F_{int2}^{dec}\cdot\sin\theta_{be^+}\sin\phi_{e^+d}\sin\theta_{e^+d}~~\big)
\end{align}
In order to separate the factorized part equal to the cross section multiplied by the width, we add and subtract the term $F_R^{prod}\cdot 1$ in the first factor of the first term of (\ref{totalme1}). Then, collecting similar terms proportional to $F_R^{prod}\cdot(1+\cos\theta_{ud})$, we get

\begin{align}\label{totalme2}
&|M|^2_{u b \to d b \nu e^+}~=~\big(~~F_L^{prod}\cdot 2 + F_R^{prod}\cdot 2 
+ F_{int1}^{prod}\cdot 2\cdot(1-\cos\theta_{u d})
~~\big)  \\ \nonumber&
~~~~~\times
D_t\cdot\big(~~F_L^{dec}\cdot(1+\cos\theta_{e^+d}) + F_R^{dec}\cdot (1-\cos\theta_{\nu d}) \\ \nonumber&
~~~~~~~~~~~~~~+ F_{int1}^{dec}\cdot (1+\cos\theta_{e^+ d}-\cos\theta_{\nu e^+}-\cos\theta_{\nu d})
+F_{int2}^{dec}\cdot\sin\theta_{be^+}\sin\phi_{e^+d}\sin\theta_{e^+d}~~\big) \\ \nonumber&
-F_R^{prod}\cdot 2\cdot (1+\cos\theta_{u d})
\times D_t\cdot 
\big(~F_L^{dec}\cdot\cos\theta_{e^+d} - F_R^{dec}\cdot\cos\theta_{\nu d}
+ F_{int1}^{dec}\cdot (\cos\theta_{e^+ d}-\cos\theta_{\nu d})
~\big)
\end{align}
The first term corresponds to the factorized product of the unpolarized t-quark production matrix element and the matrix element of the polarized t-quark decay. The second term consists of the factorized product of the part of the unpolarized t-quark production matrix element $F_R^{prod}$ multiplied by the factor $(1+\cos\theta_{ud})$ and the part of the polarized t-quark decay width. Also, both terms are multiplied by the square of the t-quark propagator denominator. We tested this expression analytically using the CompHEP package \cite{Boos:2004kh}. Writing down expression (\ref{totalme2}) in a compact form, we obtain

\begin{align}
&|M|^2_{u b \to d b \nu e^+}~~=~~\big(\sum\limits_{s}|M|^2_{u b \to d t}\big)\times D_t\cdot \big(|M|^2_{t \to b \nu e^+}\big)_{polar}\\ \nonumber&
-\big(\sum\limits_{s}|M_R|^2_{u b \to d t}\big)\cdot (1+\cos\theta_{u d})
\times D_t\cdot 
\big(~F_L^{dec}\cdot\cos\theta_{e^+d} - F_R^{dec}\cdot\cos\theta_{\nu d}
+ F_{int1}^{dec}\cdot (\cos\theta_{e^+ d}-\cos\theta_{\nu d})
~\big)
\end{align}
Using the narrow-width approximation of the t quark, we integrate this matrix element over all phase space variables, except for the positron energy $E_{e^+}$ and its orientation angles ($\cos\theta$ and $\phi$) with respect to the d-quark direction. It should be noted that after integration the factor $(1+\cos\theta_{ud})$ becomes equal to 2. As an integration result, we obtain the differential cross section of the t-channel production of the polarized t quark, with its subsequent decay 

\begin{align}\label{completecs}
&\frac{d\sigma(\hat{s})_{u b \to d b \nu e^+}}{d\epsilon\cdot d\cos\theta\cdot d\phi}~~=~~
\sigma(\hat{s})_{ub\to td}\cdot\frac{1}{\Gamma_t}~\cdot\frac{d\Gamma_{t \to b \nu e^+}}{d\epsilon\cdot d\cos\theta\cdot d\phi}\\ \nonumber \\ \nonumber &
- \sigma_{R}(\hat{s})_{ub\to td}\cdot\frac{1}{\Gamma_t}~\cdot\frac{\alpha^2\cdot m_t^3}{64\cdot \pi\cdot \sin^4{\Theta_W}\cdot \Gamma_W\cdot m_W}\cdot \big[ \\ \nonumber &
+ |f_{LV}|^2~~\cdot~~(1-\epsilon)\cdot \epsilon \cdot\cos\theta\\ \nonumber &
+ |f_{LT}|^2~~\cdot~~(\epsilon-r^2)\cdot \epsilon \cdot\cos\theta\\ \nonumber &
+ |f_{RT}|^2~~\cdot~~(1-\epsilon)\cdot\left(\frac{2r\cdot c(\epsilon)}{\epsilon}\cdot\sin\theta\cos\phi + \left(\frac{2r^2}{\epsilon}+\epsilon-r^2-1\right)\cdot\cos\theta\right)\\ \nonumber &
+ |f_{RV}|^2~~\cdot~~(\epsilon-r^2)\cdot\left(\frac{2r\cdot\c(\epsilon)}{\epsilon}\cdot\sin\theta\cos\phi + \left(\frac{2r^2}{\epsilon}+\epsilon-r^2-1\right)\cdot\cos\theta\right)\\ \nonumber &
+ (Re f_{LV}\cdot Re f_{RT} + Im f_{LV}\cdot Im f_{RT})~~\cdot~~(1-\epsilon)\cdot 2\cdot\left(c(\epsilon)\cdot\sin\theta\cos\phi + r\cdot\cos\theta\right) \\ \nonumber &
+ (Re f_{LT}\cdot Re f_{RV} + Im f_{LT}\cdot Im f_{RV})~~\cdot~~(\epsilon-r^2)\cdot 2\cdot\left(c(\epsilon)\cdot\sin\theta\cos\phi + r\cdot\cos\theta\right)
~~~\big]
\end{align}
\\

where~~
$c(\epsilon) = \sqrt{(1-\epsilon)(\epsilon-r^2)},~~~\epsilon = 2E_{e^+}/m_t,~~~\epsilon_{max} = 1,~~~\epsilon_{min} = r^2,~~~r=m_W/m_t$,
$\sigma_{ub\to td}$ is the cross section of the unpolarized t-quark production (\ref{prodcross}) and $(\sigma_R)_{ub\to td}$ is a part of this cross section that is proportional to $|f_{RV}|^2$ and $|f_{RT}|^2$, $\Gamma_t$ is the total decay width of the t quark, taking into account the anomalous couplings and all decay modes, $\frac{d\Gamma_{t \to b \nu e^+}}{d\epsilon~\cdot~d\cos\theta~\cdot~d\phi}$ is the differential partial width of the polarized t-quark decay (\ref{twidth0}), and $\theta$ and $\phi$ are orientation angles of the positron with respect to direction of the d-quark momentum.
\\

This expression was obtained for the first time and includes terms of all orders of magnitude of anomalous couplings. Once again, we note that this expression was obtained in the approximation of the t-quark narrow width and $m_b$ = 0. The denominator of the expression is equal to the total width of the unpolarized decay of the t quark, which is constant and does not contain functional dependences that affect the shape of the differential distributions, but only changes their normalization. To study spin correlations, we use differential cross section normalized to the full integral. In this case, the dependence of the denominator of the expression on anomalous couplings does not play a role.
To simplify the analysis, we leave in the numerator of expression (\ref{completecs}) only terms up to the second order of magnitude of the anomalous couplings and we get formula (\ref{totalcrossec}):

\begin{align}
&\frac{d\sigma(\hat{s})_{u b \to d b \nu e^+}}{d\epsilon\cdot d\cos\theta\cdot d\phi}~~=~~\frac{1}{\Gamma_t}~\cdot\big[~~
\sigma(\hat{s})_{ub\to td}\cdot\frac{d\Gamma_{t \to b \nu e^+}}{d\epsilon\cdot d\cos\theta\cdot d\phi}\nonumber& 
\\ \nonumber&
- \sigma_{R}(\hat{s})_{ub\to td}\cdot\frac{\alpha^2\cdot m_t^3\cdot V_{tb}^2}{64\cdot \pi\cdot \sin^4{\Theta_W}\cdot \Gamma_W\cdot m_W}\cdot(1-\epsilon)\cdot\epsilon \cdot\cos\theta~~\big]\\ \nonumber&
\end{align}
We left the first term of the sum (\ref{totalcrossec}) unchanged to simplify the notation, but we mean that we omit in it the terms of the third and fourth order of magnitude of anomalous couplings. 

Now, for the first term of expression (\ref{totalcrossec}), we write out explicitly the terms only up to the second order of magnitude of anomalous couplings:

\begin{align}\label{factorcs}
&\sigma_{ub\to td}\cdot\frac{d\Gamma_{t \to b \nu e^+}}{d\epsilon\cdot d\cos\theta\cdot d\phi}~~=~~
\big(\sigma_{SM}+(\sigma-\sigma_{SM})\big)_{ub\to td}\cdot\frac{d\big(\Gamma_{SM}+(\Gamma-\Gamma_{SM})\big)_{t \to b \nu e^+}}{d\epsilon\cdot d\cos\theta\cdot d\phi}~=~
\\ \nonumber&
\big(\sigma_{SM}+(\sigma-\sigma_{SM})\big)_{ub\to td}\cdot \frac{d(\Gamma_{SM})_{t \to b \nu e^+}}{d\epsilon\cdot d\cos\theta\cdot d\phi}
~+~(\sigma_{SM})_{ub\to td}\cdot\frac{d(\Gamma-\Gamma_{SM})_{t \to b \nu e^+}}{d\epsilon\cdot d\cos\theta\cdot d\phi}
\\ \nonumber&
+~(\sigma-\sigma_{SM})_{ub\to td}\cdot \frac{d(\Gamma-\Gamma_{SM})_{t \to b \nu e^+}}{d\epsilon\cdot d\cos\theta\cdot d\phi}~\approx~\\ \nonumber&
 \sigma_{ub\to td}\cdot \frac{d(\Gamma_{SM})_{t \to b \nu e^+}}{d\epsilon\cdot d\cos\theta\cdot d\phi}
~+~(\sigma_{SM})_{ub\to td}\cdot\frac{d(\Gamma-\Gamma_{SM})_{t \to b \nu e^+}}{d\epsilon\cdot d\cos\theta\cdot d\phi}
~+~(\sigma_{RT})_{ub\to td}\cdot \frac{d(\Gamma_{RT})_{t \to b \nu e^+}}{d\epsilon\cdot d\cos\theta\cdot d\phi}
\end{align}
\\

where $(\sigma_{RT})_{ub\to td}$ is part of the cross section of the unpolarized t-quark production that is proportional to  $V_{tb}\cdot Ref_{RT}$, $(\Gamma_{RT})_{t \to b \nu e^+}$ is a part of the top partial width that is proportional to $V_{tb}\cdot Ref_{RT}$ and $V_{tb}\cdot Imf_{RT}$, $(\sigma_{SM})_{ub\to td}$ is the SM cross section of the unpolarized t-quark production, and $(\Gamma_{SM})_t$ is the SM total decay width of the t quark.
\\

Substituting expression (\ref{factorcs}) into (\ref{totalcrossec}) and subtracting expression 
$(\sigma_{SM})_{ub\to td}\cdot\frac{1}{(\Gamma_{SM})_t}\cdot\frac{d(\Gamma_{SM})_{t \to b \nu e^+}}{d\epsilon\cdot d\cos\theta\cdot d\phi}$ from (\ref{totalcrossec}), we obtain the expression for the difference of the differential cross section with anomalous couplings and the Standard Model differential cross section:

\begin{align}
\frac{d(\sigma-\sigma_{SM})_{u b \to d b \nu e^+}}{d\epsilon\cdot d\cos\theta\cdot d\phi}&=
\frac{1}{\Gamma_t}\cdot\big[~~
(\sigma_{SM})_{ub\to td}\cdot\frac{d(\Gamma-\Gamma_{SM})_{t \to b \nu e^+}}{d\epsilon\cdot d\cos\theta\cdot d\phi}\\ \nonumber&
+ \left(\sigma_{ub\to td}-r_{\Gamma}\cdot(\sigma_{SM})_{ub\to td}\right)\cdot \frac{d(\Gamma_{SM})_{t \to b \nu e^+}}{d\epsilon\cdot d\cos\theta\cdot d\phi}\\ \nonumber&
+(\sigma_{RT})_{ub\to td}\cdot \frac{d(\Gamma_{RT})_{t \to b \nu e^+}}{d\epsilon\cdot d\cos\theta\cdot d\phi}\\ \nonumber&
- (\sigma_{R})_{ub\to td}\cdot \frac{\alpha^2\cdot m_t^3}{64\cdot \pi\cdot \sin^4{\Theta_W}\cdot \Gamma_W\cdot m_W}\cdot V_{tb}^2\cdot(1-\epsilon)\cdot\epsilon \cdot\cos\theta~~\big]
\end{align}
where $r_{\Gamma}=\frac{\Gamma_t}{(\Gamma_{SM})_t}$.

Normalizing this expression by the value of the full integral, we obtain expression (\ref{diffcrossec}).

\end{document}